\documentclass[letterpaper]{article} 
\usepackage[]{aaai25}  
\usepackage{times}  
\usepackage{helvet}  
\usepackage{courier}  
\usepackage[hyphens]{url}  
\usepackage{graphicx} 
\usepackage{natbib}  
\usepackage{caption} 
\usepackage{algorithm}
\usepackage{algorithmic}

\usepackage{xcolor}
\newcommand{\answerYes}[1]{\textcolor{blue}{#1}} 
 
\newcommand{\answerNA}[1]{\textcolor{gray}{#1}}

\usepackage{newfloat}
\usepackage{listings}
\DeclareCaptionStyle{ruled}{labelfont=normalfont,labelsep=colon,strut=off} 
\floatstyle{ruled}
\newfloat{listing}{tb}{lst}{}
\floatname{listing}{Listing}
\title{Forced Migration and Information-Seeking Behavior on Wikipedia: \\
Insights from the Ukrainian Refugee Crisis}

\author {
    Carolina Coimbra Vieira\textsuperscript{\rm 1 \rm 2 \rm *},
    Ebru Sanlit\"{u}rk\textsuperscript{\rm 1},
    Emilio Zagheni\textsuperscript{\rm 1}
}
\affiliations {
    \textsuperscript{\rm 1}Max Planck Institute for Demographic Research 
    \textsuperscript{\rm 2}Max Planck Institute for Software Systems\\
    \rm * \textit{coimbravieira@demogr.mpg.de}
}

\usepackage{caption}
\usepackage{subcaption}
\usepackage{booktabs}
\usepackage{longtable}
\usepackage{multirow}
\usepackage{mwe}
\usepackage[percent]{overpic}
\usepackage[T1]{fontenc}
\usepackage{tikz}
\usepackage{overpic}

\begin{document}

\maketitle

\begin{abstract} 
Gathering information about where to migrate to is an important component of the migration process, particularly in the context of forced migration, when individuals must make rapid decisions under conditions of uncertainty. In this study, we investigate the relationship between forced migration and online information-seeking behavior on Wikipedia. Focusing on the 2022 Russian invasion of Ukraine, we analyze how the resulting refugee crisis, which led to over six million Ukrainians fleeing to countries across Europe, affected information-seeking behavior on Wikipedia, as measured by views of articles about European cities. We compared changes in the number of views of Ukrainian-language Wikipedia articles, used as a proxy for information-seeking by Ukrainians, with those in four other language editions of Wikipedia. Our results show that views of Ukrainian-language Wikipedia articles about European cities are more strongly correlated with the number of Ukrainian refugees who applied for temporary protection in European countries than views of articles in other languages. As Poland and Germany emerged as the main destinations of Ukrainian refugees, we investigated further and found that the number of applications for temporary protection in Polish and German cities is also more strongly correlated with views of their respective Wikipedia articles in Ukrainian. Finally, we analyzed the temporal relationship between refugee flows to Poland and information-seeking on Wikipedia. Our findings reveal that refugee border crossings preceded the increase in views of Ukrainian-language Wikipedia articles about Polish cities, suggesting that information-seeking surged after displacement. This highlights the reactive nature of information-seeking behavior during forced migration, in contrast to the pre-departure planning behavior that is characteristic of regular labor migration. However, while official applications for protection often lagged behind border crossings by at least a couple of weeks, Wikipedia activity increased almost immediately after border crossings. Thus, information-seeking on Wikipedia can offer a near real-time indicator of emerging migration patterns during crises.
\end{abstract}

\section{Introduction}
\label{sec:intro}

``At the end of 2022, 108.4 million people worldwide were forcibly displaced as a result of persecution, conflict, violence, human rights violations, and events seriously disturbing public order.''\footnote{\url{https://www.unhcr.org/global-trends}}
Despite these large numbers, persons in need of international protection 
-- refugees, asylum-seekers, displaced persons, other persons in need of international protection, and stateless persons -- constitute a distinct migrant subgroup that is often not fully captured by traditional data~\cite{robinson1998importance}. 

Studies on forced migration, such as those observing patterns of movement and estimating forced migration flows, are important for enabling international organizations, governments, and humanitarian agencies to allocate resources more effectively and to provide timely protection. 
A growing body of literature focuses on repurposing digital trace data to quantify sudden displacement, observe patterns of population movement, and nowcast forced migration flows
~\cite{gonzalez2024have,leasure2023nowcasting}. 
Digital traces represent an alternative data source for migration studies, particularly in contexts where timely and granular information is critical. 
 
Timely information is also crucial for forced migrants, as forced displacement increases the need for quick and reliable access to information throughout the migration journey.
In the context of standard labor migration, the preparation for migration and the gathering of information typically take time, as studies linking online information-seeking behavior to migration flows have shown~\cite{bohme2020searching,wladyka2017queries}.
In contrast, in the context of forced displacement, this preparation happens much more rapidly. Since forced migrants often flee imminent danger, their need for information extends beyond the pre-migration stage and into the peri- and post-migration phases.
Recent studies using digital trace data to examine forced migration patterns suggest that the search for migration-related information intensifies after crises~\cite{sanliturk2024search,anastasiadou2024war}.
Survey-based studies further show that forced migrants frequently use smartphones and the internet to search for information during their journey and after their arrival at their destination~\cite{merisalo2020digital}.
As an example, Wikipedia has been identified as an important information source for Middle Eastern asylum-seekers in Germany~\cite{zimmer2020age}.

Given the importance of online sources of information for forced migrants, we investigate the relationship between online information-seeking behavior and forced migration flows. As a case study, we focus on the Ukrainian refugee crisis that began on February 24, 2022, following Russia’s invasion of Ukraine.
We examine the use of Wikipedia as an online source of information by Ukrainian refugees. In addition to being the largest and most widely used free online encyclopedia, Wikipedia is also appealing from a data collection perspective. Unlike Google Trends data, which provide only relative search popularity indices, Wikipedia provides publicly accessible data on the absolute number of daily pageviews dating back to 2015, making it a valuable resource for analyzing users' interests over time. 
Previous research has shown that Wikipedia readership reflects real-world events and trending topics~\cite{miz2020trending}, and it has been used to monitor, forecast, and assess information-seeking behavior related to health crises and natural disasters~\cite{ribeiro_sudden_nodate,jemielniak2021wikiproject,tizzoni2020impact}.
However, to the best of our knowledge, this is the first study to assess the relationship between information-seeking behavior on Wikipedia and refugee flows. We address the following research questions: \\
\textbf{RQ1:} How did the Ukrainian refugee crisis affect information-seeking behavior on Wikipedia? \\
{\textbf{RQ2:} What was the temporal relationship between information-seeking behavior on Wikipedia and Ukrainian refugee flows?

According to the United Nations High Commissioner for Refugees (UNHCR),\footnote{\url{https://www.unhcr.org/emergencies/ukraine-emergency}} more than 5.9 million refugees left Ukraine for destinations across Europe up to April 2024. For statistical purposes, the UNHCR uses the term refugees to broadly refer to all individuals who have left Ukraine due to the war. We employ this usage throughout this paper. The majority of Ukrainian refugees initially sought refuge in neighboring countries. In particular, Poland and Germany emerged as primary destinations of Ukrainian refugees.
We focus our analysis on Ukrainian refugees in Poland and Germany, examining patterns of information-seeking behavior in Polish and German cities. Poland is particularly relevant, not only as a major destination during the early stages of the Ukrainian refugee crisis~\cite{duszczyk2022war}, but also because official data on border crossings are available from Polish authorities. Building on these data, we analyze the temporal relationship between Wikipedia-based information-seeking behavior and refugee flows into Poland.

Our methodology leverages data from Wikipedia Pageviews to assess how the number of views of Wikipedia articles about cities, used as proxies for potential destinations, varied over time across different language editions, which serve as proxies for countries or regions of origin, in response to migration events.
We collected the daily number of views in Ukrainian (the official language of Ukraine), Russian, and English of Wikipedia articles about European capitals. In addition, to account for views by domestic populations in Poland and Germany, we included daily views in the Polish- and German-language editions of Wikipedia. Finally, to assess the null effect on information-seeking behavior related to cities less affected by refugee flows, we collected daily views in Ukrainian, Russian, and English of Wikipedia articles about five of the most populous capitals worldwide.
To account for fluctuations in the overall popularity of Wikipedia across different language editions, we computed the proportion of views for each article relative to the total number of views of Wikipedia in that language during the same time period.

We found a positive correlation between the proportion of views of Ukrainian-language Wikipedia articles about capitals of countries in the European Union (EU) and the stocks of Ukrainian refugees in those countries by year since 2022. Wikipedia articles about the capitals of Poland and Germany, for instance, were consistently among the five most viewed articles in the Ukrainian-language Wikipedia in the years analyzed (2022-2024). Within Poland, we also observed a strong correlation between views of Ukrainian-language Wikipedia articles about the most populous Polish cities and the numbers of Ukrainian refugees registered in those locations. The same pattern was observed for the German context, where we noticed a stronger correlation in 2022 between the numbers of Ukrainian refugees under temporary protection in the most populous German cities and views of Wikipedia articles about those locations in Ukrainian than in other Wikipedia languages.

We further explored the temporal dynamics of information-seeking as measured by views of Wikipedia articles about Polish cities in relation to the timing of Ukrainian refugee border crossings into Poland, according to data from the UNHCR. 
We found a consistent positive association between article views and the daily number of refugees crossing the border. 
Additionally, we applied Granger causality analysis to examine the lag structure and direction of the relationship between refugee flows and Wikipedia readership. The results of the analysis showed that border crossings by Ukrainian refugees into Poland Granger-caused increases in views of Ukrainian-language Wikipedia articles about Polish cities, with an average optimal lag of around eight days. This indicates that spikes in information-seeking behavior typically followed refugee arrivals by just over a week, rather than preceded them, underscoring the reactive nature of information-seeking behavior during forced migration.

Our results open up new avenues for understanding the relationship between information-seeking behavior and forced migration flows, highlighting the role of Wikipedia as an online source of information during crises. This study makes several key contributions.
First, we demonstrate how real-world crises leading to refugee flows influence online information-seeking behavior, as reflected in Wikipedia views. Second, we introduce Wikipedia data as a novel and timely source for analyzing information-seeking in the context of forced migration. By leveraging the increase in views of Wikipedia articles during crises, we show that forced migration follows a distinct temporal pattern. During forced migration, individuals often leave the origin location first and seek information later, whereas during traditional forms of migration, such as labor migration, individuals tend to engage in more preparatory information searches. This insight sheds light on how people seek information in times of crisis and uncertainty, including during forced migration, and how the behavior of such people differs from the pre-departure planning typical of regular labor migrants. Importantly, the migration process did not end at the moment of crossing the border. 
While official applications for protection often lagged border crossings by at least a couple of weeks, Wikipedia activity increased almost immediately after border crossings. This finding positions Wikipedia as a near real-time indicator of emerging migration patterns during crises, and highlights Wikipedia’s potential role as an early-warning system in migration monitoring.


\section{Related work}
\label{sec:related}


\subsubsection{Refugees and online sources of information}
Information and communications technology, as well as online sources of information and social media platforms, have become increasingly important tools for refugees seeking information on which to base their migration decisions~\cite{merisalo2020digital,dekker2018smart,felton2015migrants}.
Studies about Ukrainian refugees report that 92\% of Ukrainian refugees in Poland have a mobile phone and 86\% have reliable internet access.\footnote{\url{https://www.socialprogress.org/thematic-webpages/ukraine-refugee-pulse}} Furthermore, the International Organization for Migration (IOM)\footnote{\url{https://dtm.iom.int/reports/germany-third-country-nationals-arriving-ukraine-germany-june-2022}} reports that Ukrainian refugees in Germany consider social media and the internet as their top sources of information.
Online sources of information help migrants and refugees with their decision-making processes~\cite{merisalo2020digital,dekker2018smart,felton2015migrants,zimmer2020age}. Additionally, the availability of digital trace data has paved the way for a growing literature that predicts and analyzes forced migration using innovative data sources~\cite{leasure2023nowcasting,gonzalez2024have,anastasiadou2024war,sanliturk2024search}.

Online search engines provide a useful tool to measure interest in migration-related topics and predict migration patterns~\cite{lin2019forecasting,bohme2020searching}. However, there are limitations associated with the use of online search engine data. For example, there are a variety of search engines available, but it is often difficult to compare search data across different platforms. While Google is the most popular search engine worldwide, Bing and Yandex also compete in various regions. Additionally, each search engine reports online search interest in its own way, using different parameters and algorithms. These algorithms may introduce biases. For instance, Google Trends provides only a normalized index for a given place and time and applies an unobservable threshold that prevents results from being produced when interest is sufficiently low.

Complementing online search engines, Wikipedia Pageviews data regarding the daily absolute number of views of Wikipedia articles are easily accessible through the API or via download from the website. Although Wikipedia use is often associated with deeper topical reading~\cite{kampf2015detection}, previous work has shown a high correlation between frequently searched keywords and views of Wikipedia articles. This suggests that Wikipedia Pageviews can be a valuable source for determining popular global web search trends~\cite{yoshida_wikipedia_2015}. Wikipedia has also been identified as a key source of information for asylum-seekers from the Middle East in Germany~\cite{zimmer2020age}.

\subsubsection{Wikipedia readership during crises}
Wikipedia is the most popular free online encyclopedia, enabling readers to engage with a variety of information content. Wikipedia readership is known to be influenced by real-world developments~\cite{miz2020trending}.
Numerous studies have used Wikipedia data to monitor trends in health information during outbreaks such as those of COVID-19 and the Zika virus~\cite{ribeiro_sudden_nodate,tizzoni2020impact}.
In addition, Wikipedia has been used to monitor, forecast, and assess information-seeking behavior with regard to diseases and natural disasters~\cite{jemielniak2021wikiproject,mciver2014wikipedia}. For instance,~\citet{mciver2014wikipedia} showed that Wikipedia-derived models have been effective in estimating influenza cases, outperforming Google Flu Trends in certain contexts.~\citet{jemielniak2021wikiproject} found that Wikipedia readership patterns are often correlated with external events, such as tropical cyclones, further supporting the utility of Wikipedia data in tracking responses to crises. These findings suggest that Wikipedia data can serve as a valuable complementary data source, especially when traditional surveillance systems are not available in real time in crisis contexts. 

In this study, we investigate the relationship between Wikipedia article views and forced migration flows resulting from the Russian invasion of Ukraine. Our aim is to assess how the Ukrainian refugee crisis affected information-seeking behavior on Wikipedia and to analyze the temporal relationship between information-seeking behavior on Wikipedia and Ukrainian refugee flows.
To the best of our knowledge, this is the first study to use Wikipedia Pageviews data to assess information-seeking behavior specifically within the context of forced migration.



\section{Data}
\label{sec:data}
We examine the relationship between changes in readership of Wikipedia articles about European cities and the situations of Ukrainian refugees across Europe. Our analysis integrates data collected from Wikipedia alongside supplementary data from various sources, enabling a comprehensive understanding of Ukrainian refugee flows across the continent, with a particular focus on Poland and Germany.

We compiled a list of European capitals to assess the changes in views of Wikipedia articles about European capitals. Then, we narrowed our focus to the Polish and German contexts. We specifically focused on the 19 most populous cities in Poland,\footnote{Białystok, Bydgoszcz, Częstochowa, Gdańsk, Gdynia, Gliwice, Katowice, Kielce, Kraków, Łódź, Lublin, Poznań, Radom, Rzeszów, Sosnowiec, Szczecin, Toruń, Warsaw, and Wrocław} which are also among the main destinations of Ukrainian refugees in Poland.
For the German context, we applied our methodology to the 40 most populous cities in Germany.\footnote{Cities with at least 200,000 inhabitants.}~\footnote{Berlin, Hamburg, Munich, Cologne, Frankfurt, Stuttgart, Düsseldorf, Leipzig, Dortmund, Essen, Bremen, Dresden, Hanover, Nuremberg, Duisburg, Bochum, Wuppertal, Bielefeld, Bonn, Münster, Mannheim, Karlsruhe, Augsburg, Wiesbaden, Mönchengladbach, Gelsenkirchen, Aachen, Braunschweig, Chemnitz, Kiel, Halle (Saale), Magdeburg, Freiburg im Breisgau, Krefeld, Mainz, Lübeck, Erfurt, Oberhausen, Rostock, Kassel}
To contextualize the magnitude and temporal dynamics of Wikipedia readership changes in these Polish cities, we also collected comparative data for five of the most populous capital cities globally that have been less directly impacted by Ukrainian refugee movements: Beijing, Tokyo, Kinshasa, Jakarta, and Lima.

\subsection{Official statistics}
\subsubsection{European data}
We collected data from Eurostat in order to compare these data with Wikipedia data. This allowed us to assess the association between the number of Ukrainian refugees across 31 European countries\footnote{Austria, Belgium, Bulgaria, Croatia, Cyprus, Czechia, Denmark, Estonia, Finland, France, Germany, Greece, Hungary, Ireland, Italy, Latvia, Lithuania, Luxembourg, Malta, the Netherlands, Poland, Portugal, Romania, Slovakia, Slovenia, Spain, and Sweden as members of the European Union, along with Iceland, Liechtenstein, Norway, and Switzerland.} and the proportion of views of Ukrainian-language Wikipedia articles about European capitals after the Russian invasion of Ukraine.
We used the Eurostat data on the number of Ukrainian beneficiaries of temporary protection at the end of each month by country.\footnote{\url{https://ec.europa.eu/eurostat/databrowser/view/migr_asytpsm__custom_17904294/default/table}} Some countries, such as Germany, started reporting the monthly stocks after August 2022. In addition, to ensure consistency across the other datasets used in our analysis, we selected the stocks as of December of each year (2022–2024) to construct annual stock measures.

\subsubsection{Polish data} 
To study the Polish context, we collected data provided by the Polish government on the daily number of Ukrainian refugees crossing the border from Ukraine to Poland from February 24, 2022, to March 7, 2023, from the UNHCR refugee data platform.\footnote{\url{https://data.unhcr.org/es/situations/ukraine/location/10781}}
After their arrival in Poland, Ukrainian refugees were offered transportation from the border to one of the 27 reception centers, where they received assistance related to transfers, accommodation, meals, and medical care.\footnote{\url{https://euaa.europa.eu/sites/default/files/2022-06/Booklet_Poland_EN.pdf}}

Additionally, we gathered data from Poland’s Data Portal (DANE)\footnote{\url{https://dane.gov.pl/en}} on the stocks of Ukrainian refugees who registered for temporary protection in Polish cities between April 2022 and August 2024. Ukrainian refugees registered for temporary protection in Poland, and were assigned a PESEL number (\textit{Powszechny Elektroniczny System Ewidencji Ludności}, in English: Universal Electronic System for Registration of the Population), 
which serves as an official identification number in Poland. While this dataset consists of a record of active applications of Ukrainians seeking protection, we used the latest release of each year (2022–2024)\footnote{The last reports for 2022, 2023, and 2024 are dated, respectively, from 12/26/2022, 12/12/2023, and 12/10/2024.} to construct the annual stock measures of Ukrainian refugees in Polish cities.

\subsubsection{German data} 
For the German context, we collected data from the German Federal Statistical Office (GENESIS)\footnote{\url{https://www-genesis.destatis.de/}} on the stocks of Ukrainians seeking protection by administrative district\footnote{For the 40 most populous cities in Germany, we collected the number of Ukrainians under temporary protection at the independent city level (\textit{kreisfreie Stadt}). For Hanover and Aachen, however, the data are available at the city-regional level (\textit{Städteregion}).} 
in Germany by reference date (December 31 of each year).\footnote{\url{https://www-genesis.destatis.de/datenbank/online/statistic/12531/table/12531-0041}} 
In terms of government support and coordination, as in Poland, border authorities in Germany directed refugees to the nearest reception center, where they were provided with accommodation, food, and other essential support services. At the reception center, Ukrainian refugees were given a place to sleep, food, and other support services until longer-term housing was found.\footnote{\url{https://euaa.europa.eu/sites/default/files/2022-06/Booklet_Germany_EN.pdf}}

\subsection{Wikipedia Pageviews}
We gathered data on Wikipedia article views using the library \texttt{WikiToolkit}\footnote{\url{https://github.com/pgilders/WikiToolkit/tree/main}} implemented in Python. The data contained daily counts of views (i.e., each time a page is loaded) of Wikipedia articles across different languages since July 2015. We collected data on user views, which included views by editors, anonymous editors, and readers. Views by search engine ``web crawlers'' or automated programs were not included. Additionally, the data used in this study included both direct pageviews
and those arriving via redirects.
By accounting for redirects, we captured views redirected through searches on Wikipedia using alternative spellings, abbreviations, misspellings, or variations in capitalization or spacing~\cite{hill2014consider}, thus ensuring more comprehensive coverage of pageviews.
For normalization purposes, we collected the total number of views across different languages from Wikimedia Statistics.\footnote{\label{note1}\url{https://stats.wikimedia.org/}}

Our analyses examined changes in the readership of Wikipedia articles about European capitals, Polish cities, and German cities across multiple languages.
English-language Wikipedia served as a proxy for international attention and as a baseline language, given that English is the most accessed language on Wikipedia.
Ukrainian and Russian are the most accessed languages in Ukraine. Readership of articles in Ukrainian, as the official language of Ukraine, was used as a strong proxy for Ukrainians accessing Wikipedia.
For articles about Polish and German cities, views in Polish and German, respectively, were included as proxies for domestic attention.

\subsubsection{Proportion of views}
To account for changes in the popularity of Wikipedia across different languages, such as the increased use of Ukrainian over Russian during the war~\cite{kulyk2024language}, we calculated the proportion of daily views for each specific Wikipedia article. Considering the number of views $WV$ on a page $p$ of a Wikipedia article in language $l$ at a time $t$ (e.g., day, week, month, or year), the proportion of views $PWV$ is given by:

{\small
\begin{equation}
    PWV_{p,l,t} = \frac{WV_{p,l,t}}{WV_{l,t}}
    \label{eq:PWV}
\end{equation}}

In the following sections, we present the methodology used to address each of our research questions, along with the key results that support our findings.
\section{RQ1: How did the Ukrainian refugee crisis affect information-seeking behavior on Wikipedia?}
\label{sec:rq1}

To address the first research question, we adopted a two-step approach. First, we examined the association between the proportion of views of Ukrainian-language Wikipedia articles and the stocks of Ukrainian refugees under temporary protection in EU countries and in the most populous cities in Poland and Germany. Second, we narrowed our focus to the Polish and German contexts. We assessed the percentage change in views of Ukrainian-language Wikipedia articles about Polish and German cities after the invasion. For comparison, we also analyzed the percentage change in views of Wikipedia articles about five of the most populous capital cities in the world that were less affected by the refugee flows, as well as of articles in language editions other than Ukrainian.

\subsubsection{Association between Wikipedia views and stocks of Ukrainian refugees}
First, we examined the association between the readership of Ukrainian-language Wikipedia articles and the numbers of Ukrainian refugees who applied for temporary protection across Europe and Polish and German cities. We compared the yearly proportion of views for each Wikipedia article in five languages: English, Ukrainian, Russian, Polish, and German. 

Most datasets on applications for temporary protection are reported on a yearly basis. To ensure consistency when comparing these data with Wikipedia views of Ukrainian refugee stocks across locations, we aggregated the Wikipedia pageviews data into yearly views. This aggregation was performed separately for each Wikipedia article and language, as defined in Equation~\ref{eq:PWV}. The resulting time series represents the proportion of yearly views of Wikipedia articles across different languages.

We hypothesized that the most viewed Ukrainian-language articles would show a stronger correlation with refugee presence, serving as a proxy for the information-seeking behavior of Ukrainian refugees. Building on this hypothesis, we further extended our analysis to explore the relationship between views of Wikipedia articles about EU capitals and the broader distribution of Ukrainian refugees across EU countries.

For the European context analysis, we created a ranking of EU countries sorted by the highest number of temporary protection applications from Ukrainian refugees, based on Eurostat data for 2022, 2023, and 2024. Similarly, we ranked EU capitals according to the proportion of views of their Wikipedia articles for the same years. We then calculated the Spearman’s rank correlation between the ranking of EU countries by number of Ukrainian refugees and the ranking of Wikipedia articles about EU capitals by number of views.

Table~\ref{tab:rank} summarizes the Spearman's rank correlation and shows that the proportion of views of Ukrainian-language Wikipedia articles about EU capitals was positively correlated with the number of Ukrainian refugees in EU countries for all three years. The correlation was lower when comparing the number of Ukrainian refugees in EU countries with the ranking of views of Wikipedia articles about EU capitals in Russian, English, Polish, and German. For a visual representation of these rankings, see Figure~\ref{fig:ranks-europe} in the Appendix.

\begin{table*}[t]
\centering
\footnotesize
\begin{tabular}{llllllllll}
\toprule
 & \multicolumn{3}{c}{\textbf{Europe}} & \multicolumn{3}{c}{\textbf{Poland}} & \multicolumn{3}{c}{\textbf{Germany}} \\
\cmidrule(lr){2-4} \cmidrule(lr){5-7} \cmidrule(lr){8-10}
Wikipedia language & 2022 & 2023 & 2024 & 2022 & 2023 & 2024 & 2022 & 2023 & 2024 \\
\midrule
Ukrainian & 0.69*** & 0.63*** & 0.63*** & 0.87*** & 0.87*** & 0.85*** & 0.77*** & 0.71*** & 0.71*** \\
Russian & 0.60*** & 0.53** & 0.51** & 0.82*** & 0.81*** & 0.76*** & 0.72*** & 0.68** & 0.72*** \\
English & 0.61*** & 0.56** & 0.59*** & 0.87*** & 0.84*** & 0.83*** & 0.73*** & 0.66** & 0.70*** \\
Polish & 0.57*** & 0.53** & 0.51** & 0.88*** & 0.88*** & 0.88*** & 0.73*** & 0.68** & 0.71*** \\
German & 0.45* & 0.39* & 0.43* & 0.85*** & 0.82*** & 0.83*** & 0.76*** & 0.75*** & 0.75*** \\
\bottomrule
\end{tabular}
\caption{Spearman's rank correlation between the yearly proportion of views of Wikipedia articles about EU capitals, Polish cities, and German cities and the stocks of Ukrainian refugees with temporarily recognized protection status by year. \\ \footnotesize{***$p<0.001$; **$p<0.01$; *$p<0.05$}}
\label{tab:rank}
\end{table*}

Poland and Germany emerged as key destination countries during the Ukrainian refugee crisis. On Ukrainian-language Wikipedia, the article on Poland’s capital city of Warsaw consistently ranked among the top articles in terms of the proportion of yearly views, while the article on Berlin appeared regularly within the top five articles. Building on these patterns, we narrowed our analysis to focus specifically on the Polish and German contexts.

For the analysis on Poland, we investigated the relationship between the number of Ukrainian refugees assigned a PESEL number in Polish cities and the proportion of yearly views of Wikipedia articles about those cities. 
We created a ranking of Polish cities based on the yearly stocks of Ukrainian refugees assigned PESEL numbers using data from Poland’s Data Portal (DANE) from 2022 to 2024. Similarly, we ranked Polish cities according to the proportion of yearly views of their Wikipedia articles. Following the approach used in the European context analysis, we calculated the Spearman’s rank correlation between the ranking of Polish cities by the number of Ukrainian refugees assigned PESEL numbers in those cities by year and the ranking of Polish cities by the proportion of yearly views of Wikipedia articles about those cities.

Table~\ref{tab:rank} also includes a summary of the Spearman’s rank correlations for the Polish context. We observed that views of Ukrainian-language Wikipedia articles about Polish cities were strongly aligned with the distribution of Ukrainian refugees across those cities, as measured by PESEL registrations. The correlation between Ukrainian-language views and PESEL numbers declined slightly after 2023, which may reflect a shift in refugees’ information needs over time as initial settlement challenges gave way to longer-term integration processes. By contrast, Polish-language Wikipedia views reflected more stable, long-term interest from the host population, and therefore produced rankings that remained consistent across years. Finally, the results for the Polish context also showed that Ukrainian-language correlations remained stronger than those for English-, Russian-, or German-language Wikipedia, underscoring the centrality of the Ukrainian language in the refugees' information-seeking behavior during displacement.

In the Appendix, Figure~\ref{fig:ranks-poland} shows a visual representation of these rankings. Despite the strong correlation between the proportion of the number of views of Ukrainian-language Wikipedia articles about Polish cities and the number of Ukrainian refugees registered in those cities, Rzeszów stands out. Although Rzeszów had fewer registered refugees than cities like Warsaw and Wrocław, the Wikipedia article about Rzeszów, and especially the Ukrainian-language article, had one of the highest proportions of views over the years 2022-2024. Rzeszów has emerged as a pivotal location in the Ukrainian crisis due to its proximity, of roughly 100 km, to the Ukrainian border. In the month and a half following Russia’s invasion of Ukraine in February 2022, Rzeszów hosted an estimated 100,000 refugees and served as a transit point for approximately 1.5 million more~\cite{pietka2025urban}. Wikipedia captured the information-seeking behavior focused on this city, which played a key role during the migration process.

For the analysis in Germany, we compared the ranking of German cities based on the yearly stocks of Ukrainian refugees under temporary protection from 2022 to 2024 with the ranking of German cities according to the proportion of yearly views of their Wikipedia articles across languages. Table~\ref{tab:rank} summarizes the Spearman’s rank correlations for the German context. Results show that views of Ukrainian-language Wikipedia articles about German cities were positively correlated with the official stocks of Ukrainian refugees in those cities, particularly in 2022, indicating that during the initial phase of displacement, refugees turned to Wikipedia in their native language to access information about German destinations as well. Similar to the pattern observed in the Polish context, the correlation between Ukrainian-language views and refugee stocks decreased after 2022, which may reflect changing information needs as the crisis evolved. By contrast, views of German-language Wikipedia articles remained more stable over time, reflecting host population interest, while views of English- and Russian-language articles showed weaker alignment with refugee distributions. In the Appendix, Figure~\ref{fig:ranks-germany} illustrates these rankings.


Next, we calculated the percentage change in Wikipedia article views about Polish and German cities to estimate the impact of the Russian invasion of Ukraine on readership across different languages.

\subsubsection{Relative change in Wikipedia views}
We quantified the percentage increase in views of Wikipedia articles about Polish cities across different language editions, using language as a proxy for the geographic origin of attention. In particular, we focused on Ukrainian-language Wikipedia to approximate the information-seeking behavior of Ukrainian users. To measure changes in the volume of views of articles after the Russian invasion of Ukraine, we computed the relative change in Wikipedia views across language editions. 

To reduce noise and provide a more stable representation of trends in Wikipedia readership over time, we aggregated the daily view data into weekly time series. This aggregation was performed separately for each Wikipedia article and language, as defined in Equation~\ref{eq:PWV}. The resulting time series represents the proportion of weekly views of Wikipedia articles across different languages. Then, we calculated the relative change in the proportion of weekly views of Wikipedia articles across language editions. 
This metric captured temporal fluctuations in attention, while accounting for seasonal patterns, by comparing each week to the corresponding week in the previous year.

The relative change ($RC_{p,l,t}$) in the proportion of weekly views for a given page $p$ of a Wikipedia article in language $l$ at a time $t$, is defined as follows:

{\small
\begin{equation}
RC_{p,l,t} = \frac{(PWV_{p,l,t} - PWV_{p,l,t-52})} {PWV_{p,l,t-52}} \times 100
\label{eq:RC}
\end{equation}}

where $PWV_{p,l,t}$ denotes the proportion of weekly views for a page $p$ in language $l$ during the week $t$, and $PWV_{p,l,t-52}$ is the proportion for the corresponding week one year earlier (i.e., 52 weeks prior). This year-over-year comparison helps isolate the impact of the refugee crisis by controlling for regular seasonal fluctuations in Wikipedia readership. 

\begin{figure*}[t]
\centering
\begin{overpic}[width=0.88\textwidth]{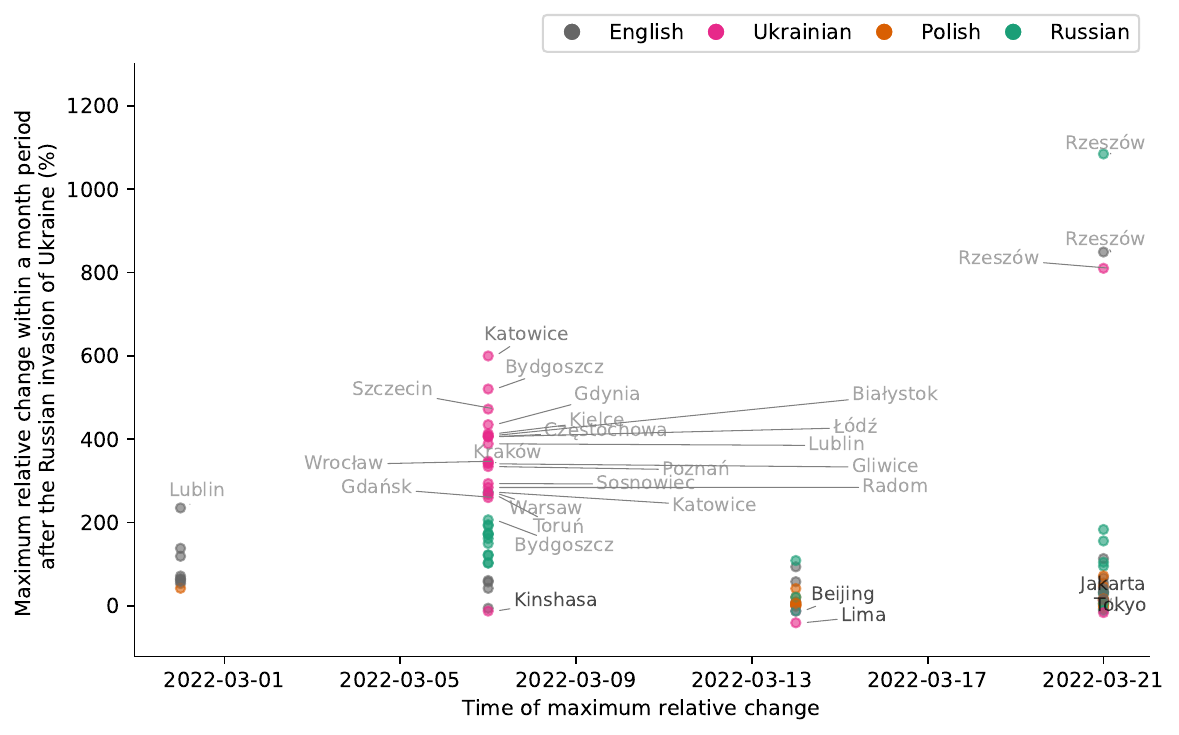}
    \put(11,31){ 
        \begin{tikzpicture}[baseline]
            \node[anchor=south west, inner sep=0] (inset) at (-0.02,-0.05) 
                {\includegraphics[scale=0.4]{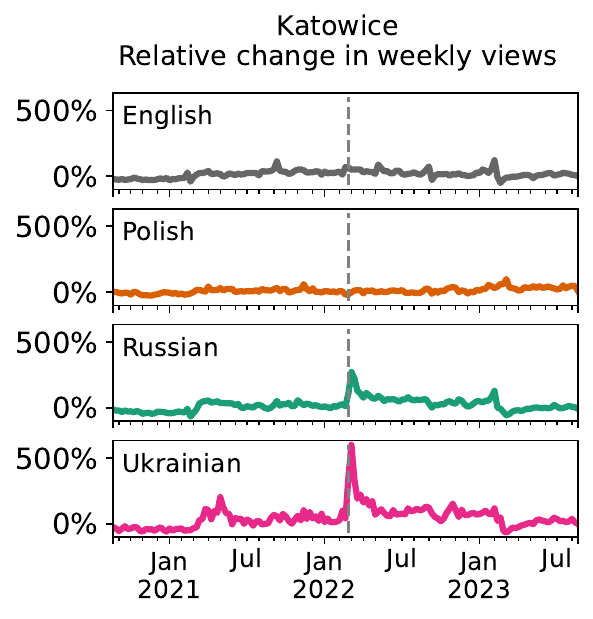}};
            \draw[black, thick] (0,0) rectangle (4.1,4.2); 
        \end{tikzpicture}
    }

    \put(26,32.2){ 
        \begin{tikzpicture}
            \draw[darkgray, thick, ->] (0,0) -- (-2.3,0.85); 
        \end{tikzpicture}
    }
\end{overpic}
\caption{Maximum relative change in the proportion of weekly views over the month following the Russian invasion of Ukraine, compared to the same period in the previous year. Results are shown for Wikipedia articles about the 19 most populous Polish cities and five of the most populous cities in the world (Beijing, Jakarta, Kinshasa, Lima, and Tokyo) across four languages (English, Polish, Russian, and Ukrainian). As an example, we also show the relative change in the proportion of weekly views compared to the previous year of the Wikipedia article about \textbf{Katowice} across four languages (English, Polish, Russian, and Ukrainian) from August 24, 2020, to August 24, 2023.}
\label{fig:rel_change}
\end{figure*}
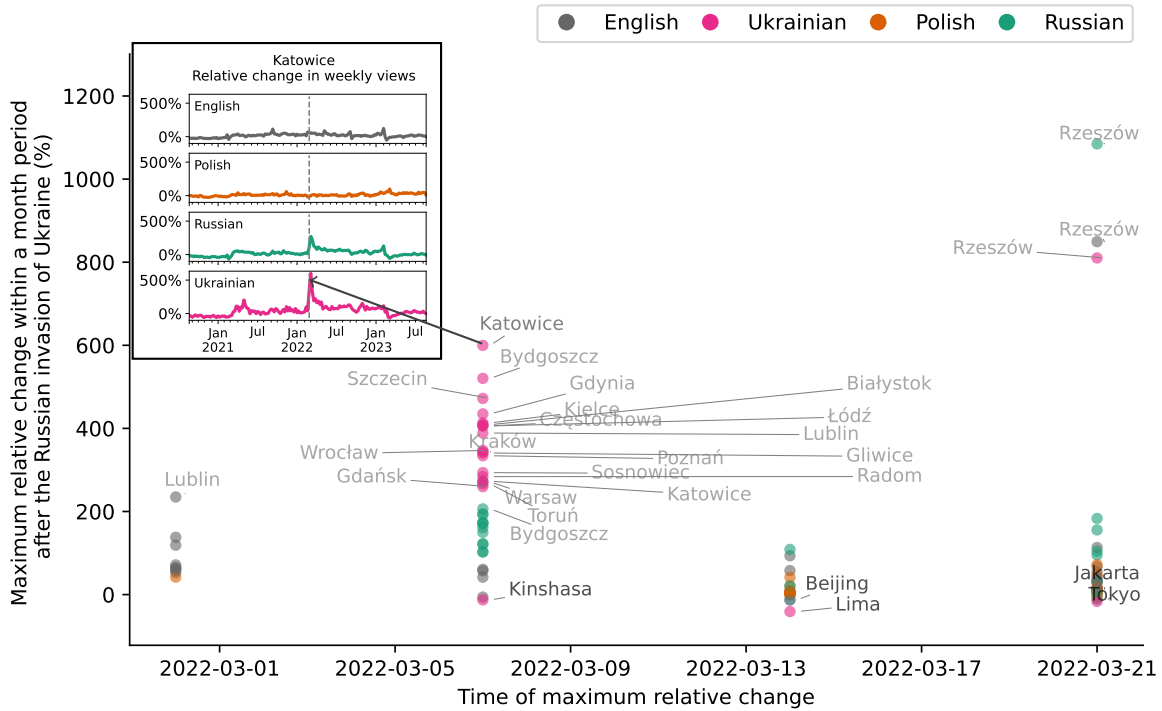

The inset in Figure \ref{fig:rel_change} illustrates the relative change in the proportion of weekly views, compared to the same week in the previous year, of the Wikipedia article about Katowice across four languages (English, Polish, Russian, and Ukrainian) from August 24, 2020, to August 24, 2023. 
While no noticeable increase in views of the English and Polish articles was observed after the Russian invasion of Ukraine, there was a marked relative change in views of both the Ukrainian and Russian articles. Most strikingly, a relative increase of over 500\% in views of the Ukrainian-language article for Katowice occurred immediately after the onset of the war on February 24, 2022. 
Moreover, during this period, the relative increase in views of Ukrainian-language Wikipedia articles for all other analyzed Polish cities exceeded 200\%. These findings underscore the potential impact of the Ukrainian refugee crisis on information-seeking behavior, as reflected in the increase in views of Ukrainian-language Wikipedia articles. For a complete overview of the time series showing relative changes for each one of the Polish cities, see Figure~\ref{fig:rel-changes-appendix} in the Appendix.

We contrasted the increases in views not only across languages, highlighting that the most significant increases were for views of Ukrainian-language articles, but also across cities that were affected differently by refugee flows in the early stages of the war. Specifically, we compared views of Wikipedia articles about Polish cities that received a large influx of Ukrainian refugees with views of articles about cities less directly impacted by Ukrainian refugee flows. 
Figure \ref{fig:rel_change} illustrates this contrast by showing the maximum relative change and the date of this peak in the month after the Russian invasion. The figure includes Wikipedia articles about the 19 most populous Polish cities and five of the most populous global capitals (Beijing, Jakarta, Kinshasa, Lima, and Tokyo) across the English, Polish, Russian, and Ukrainian editions. For clarity, the article names are shown only for the global capitals articles in Ukrainian (in black) and for the Polish city articles for which the relative change in views exceeded 200\% (in gray).

For the Wikipedia articles about five of the world's most populous capitals, we observed either marginal increases or overall decreases in views, suggesting that these cities were less affected by refugee-related information-seeking during this period. In contrast, we found relative increases of at least 200\% in views of all the Ukrainian-language articles about the most populous Polish cities in the week following the Russian invasion of Ukraine. This result points to a strong increase in information-seeking among Ukrainians about Polish cities immediately after the Russian invasion of Ukraine. The peak increase in views of Ukrainian-language Wikipedia articles about Polish cities was consistently much higher and more concentrated in the week following the invasion than the increase in views of Wikipedia articles about other cities or in other languages. 

We also observed clear spikes in views of Ukrainian-language Wikipedia articles about German cities shortly after the Russian invasion. 
The largest increases in views of the articles about cities such as Düsseldorf and Munich occurred within two weeks of the outbreak of the war, while views of the articles about Oberhausen, Münster, and Bielefeld peaked slightly later, at around three weeks after the invasion. These temporal dynamics suggest that different German destinations became the focus of the information-seeking behavior of refugees at varying stages of the initial displacement wave. Figure~\ref{fig:rel_change_de} in the Appendix illustrates these peaks by showing the maximum relative changes in views and the timing of these surges in the month immediately following the invasion.

To complement the findings regarding the timing of the changes in Wikipedia views, we employed an autoregressive model (AR(1)) and structural break tests. The aim was to identify break points in the time series of the proportion of daily views of Ukrainian-language Wikipedia articles about Polish cities. We used the \texttt{strucchange} package in R, which estimates the optimal number of breaks and their confidence intervals~\cite{strucchange}. 
The structural break analysis identified changes in the time series of views of Ukrainian-language Wikipedia articles about Polish cities around the time of the Russian invasion of Ukraine. Out of the 19 cities, 15 exhibited break points within one week of the invasion, indicating abrupt increases in information-seeking activity (Table \ref{tab:breaks_pol}). One city, Gliwice, showed a break point one week prior to the invasion (February 16, 2022), while Rzeszów showed a break point within three weeks after the invasion. Łódź and Lublin were the outliers, as for these cities a structural break in the time series was observed about three months after February 24, 2022. Similarly, the analysis of the time series of the proportion of daily views of Ukrainian-language Wikipedia articles about German cities showed that a structural break occurred within one week after the start of the invasion for 28 out of the 40 cities (Table \ref{tab:breaks_ger}). We observed a structural break within one month after the start of the invasion for all 40 cities except Krefeld, for which the break date was identified as one day before February 24, and Aachen, for which we did not observe a structural break until May 2022.
All observed breaks were followed by sharp rises in daily views of the respective Ukrainian-language articles. In contrast, we did not observe such a pattern for the five most populous world capitals (Table \ref{tab:breaks_w}). Out of the five most populous capitals, only Beijing showed a structural break in the time series in proximity to the start of the Russian invasion; however, this was followed by a decrease in the proportion of daily views of the Ukrainian-language Wikipedia article on Beijing.

As hypothesized, the increase in views of Ukrainian-language Wikipedia articles about European cities showed a strong association with the presence of refugees in those locations, particularly in Poland. Building on the consistently large increase in views of articles about Polish cities after the Russian invasion of Ukraine, we then investigated our second research question regarding the temporal relationship between Wikipedia article views and refugee flows.

\section{RQ2: What was the temporal relationship between information-seeking behavior on Wikipedia and Ukrainian refugee flows?}
\label{sec:rq2}

To answer the second research question, we investigated the relationship between the number of Ukrainian refugees in Poland and the increase in the number of views of Ukrainian-language Wikipedia articles about Polish cities. For the following analysis, we used the proportion of daily views of Wikipedia articles about Polish cities, along with official statistics provided by UNHCR on the number of refugees crossing the border from Ukraine into Poland from February 24, 2022 to March 7, 2023.

Figure~\ref{fig:series} presents the time series of UNHCR data on the daily number of Ukrainian refugees crossing into Poland, alongside data on Wikipedia views. Figure~\ref{fig:series-appendix} in the Appendix shows similar time series for each of the Polish cities analyzed. For the Wikipedia data, we reported the proportion of daily views of articles about Polish cities across four different languages: English, Polish, Russian, and Ukrainian. We observed that the proportion of daily views in Russian, and especially in Ukrainian, increased dramatically in 2022, following the start of the war.

We began this analysis by calculating the correlation between the time series. Figure~\ref{fig:corr-box} shows that the numbers of views of Ukrainian- and Russian-language Wikipedia articles about the 19 most populous cities in Poland were consistently positively correlated with the numbers of Ukrainian refugees who crossed the border into Poland.
In the figure, the whiskers represent the 5th and 95th percentiles, while the box spans from the first quartile to the third quartile of the data, with a horizontal line indicating the median.

\begin{figure}[t!]
    \centering
    \includegraphics[width=0.86\linewidth]{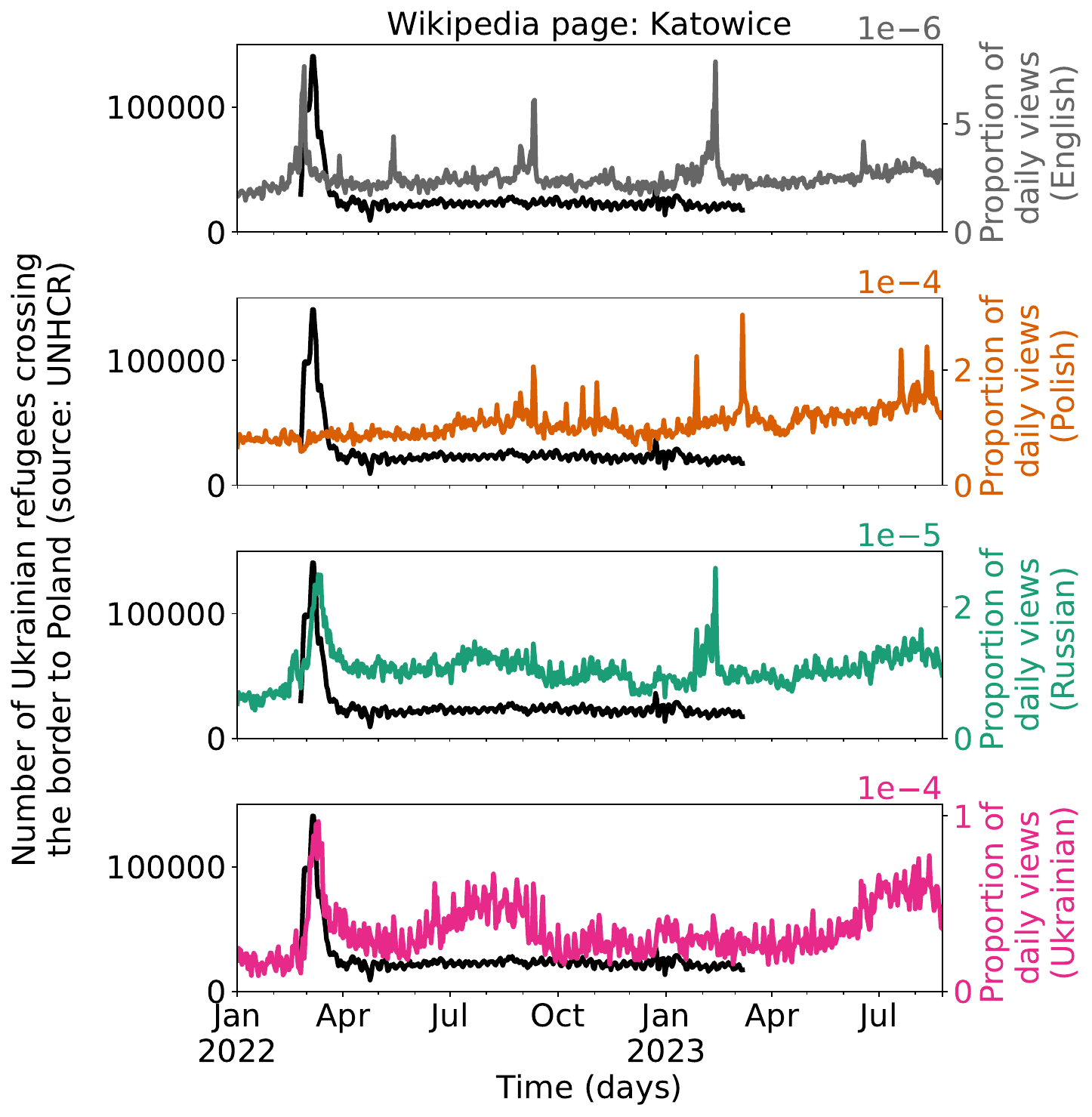}
    \caption{Time series representing, in black, the daily number of Ukrainian refugees crossing the border from Ukraine to Poland (from February 24, 2022 to March 7, 2023) and, in colors, the proportion of the daily number of views of Wikipedia articles about \textbf{Katowice} across four languages (English, Polish, Russian, and Ukrainian).}
    \label{fig:series}
\end{figure}

\begin{figure}[t]
    \centering
    \includegraphics[width=0.75\linewidth]{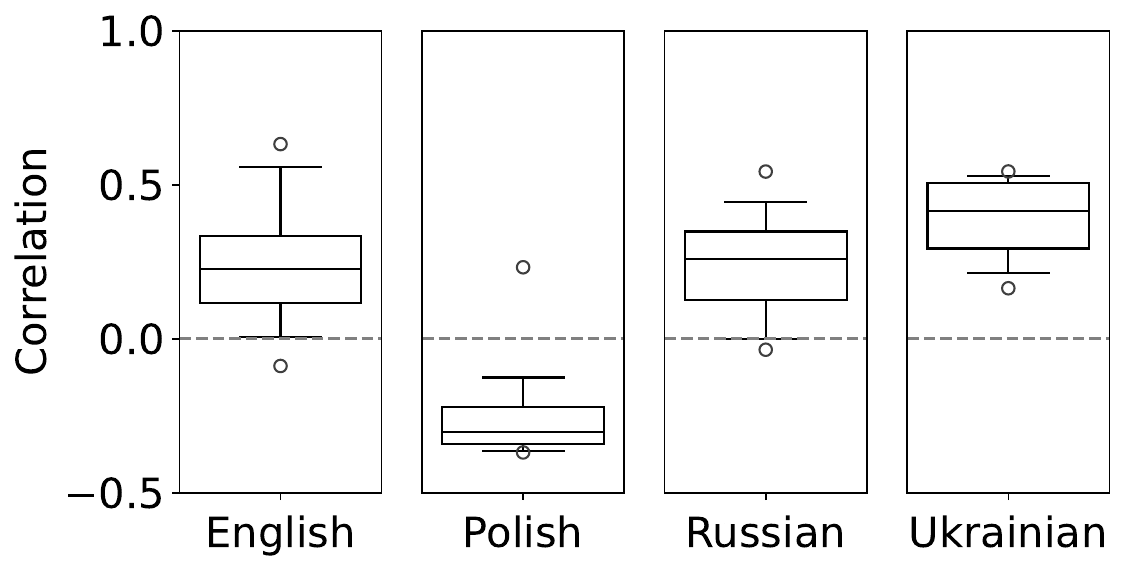}
    \caption{Correlation between the numbers of views of Wikipedia articles about the 19 most populous cities in Poland, across different languages, and the numbers of Ukrainian refugees crossing the border into Poland.} 
    \label{fig:corr-box}
\end{figure}

Next, to estimate the temporal relationship between information-seeking behavior as reflected in views of Wikipedia articles about Polish cities and Ukrainian refugee flows to Poland, we conducted Granger causality tests~\cite{granger2014forecasting}. Our aim was to assess whether Ukrainian refugee flows to Poland could predict subsequent increases in views of Wikipedia articles about Polish cities (or vice versa). Granger causality is a statistical test that evaluates temporal precedence, indicating whether past values of one time series improve the prediction of another. It is important to note that Granger causality assesses whether including past values of one variable enhances the predictive accuracy of another.

In this context, we tested whether the time series of the proportion of daily views of Wikipedia articles about Polish cities could help forecast the time series of the number of Ukrainian refugees crossing the border into Poland. The test involved regressing the dependent variable (e.g., daily number of Ukrainian refugees crossing the border) on both its own lagged values and the lagged values of the independent variable (e.g., daily views of Wikipedia articles about Polish cities). The null hypothesis stated that the independent variable would not improve the forecast of the dependent variable. If the associated p-value was below a chosen significance level (e.g., 0.05), we would reject the null hypothesis and infer evidence of Granger causality.
Thus, if the Granger causality test revealed a temporal relationship between daily views of Wikipedia articles about Polish cities and refugee flows, this would suggest that information-seeking on Wikipedia could serve as an indicator of migration patterns or decisions to cross into Poland.

Before conducting the Granger causality tests, we employed a Vector Autoregressive (VAR) modeling framework~\cite{lutkepohl2013vector}, following an approach similar to that in \citet{bailard2024keep}, using the implementation available in the Python library \texttt{statsmodels}.\footnote{\url{https://www.statsmodels.org/stable/index.html}} This framework models each variable as a function of its own lags and the lags of other variables in the analysis. We used this approach to identify statistically significant lag lengths for inclusion in the Granger causality tests. The Akaike information criterion (AIC)~\cite{bozdogan1987model} was employed to determine the optimal lag structure, which was found to be eight days for most Wikipedia articles about Polish cities. The only exceptions were for the articles about Radom and Wrocław, for which the optimal lag was, respectively, 15 and 17 days when modeling border crossings alongside the proportion of daily views in Ukrainian.
Next, we verified that all time series satisfied the stationarity requirement using the Augmented Dickey–Fuller (ADF) test~\cite{dickey1979distribution}, which confirmed stationarity at the identified lag length. We also tested for autocorrelation in the residuals with Lagrange Multiplier (LM) tests, confirming the absence of residual autocorrelation~\cite{johansen1995likelihood}. Additionally, the stability of the VAR models was assessed and confirmed using standard eigenvalue stability diagnostics.

After validating these assumptions and identifying the optimal lag for each pair of the proportion of daily Wikipedia article views and the border crossing data, we applied the Granger causality test. 
Figure~\ref{fig:boxplot-granger} presents the F-statistics from the Granger causality tests conducted for each relationship analyzed. Each box plot summarizes the distribution of F-statistics for one type of relationship across the Wikipedia articles about Polish cities. The whiskers extend from the 5th to the 95th percentile, and the box spans from the first quartile to the third quartile, with a vertical middle line indicating the median. Each dot represents the result for a specific Wikipedia article about a Polish city. Dots in blue and red indicate, respectively, statistically significant (p-value < 0.05) and non-significant relationships (p-value $\geq$ 0.05). 

Overall, we observed that the F-statistics were higher and significant (p-value < 0.05) in the direction where the number of Ukrainian refugees crossing the border to Poland Granger-caused the daily number of views of Wikipedia articles about Polish cities in Ukrainian and Russian. As Ukrainian and Russian are official or widely spoken languages in Ukraine, this finding is suggestive of information-seeking by Ukrainian refugees. This temporal ordering implies that spikes in information-seeking behavior tended to follow, rather than precede, refugee arrivals. 
For a detailed summary of the relationships tested, including the optimal lag, F-statistics, and p-values, see Table \ref{tab:granger} in the Appendix. This table includes only relationships with statistically significant p-values (p $<$ 0.05). 

Our findings align with the understanding that forced migration is often sudden and driven by urgent safety concerns, leaving little opportunity for extensive preparatory research before displacement. 
Unlike traditional migration processes, such as labor migration, in which individuals typically engage in preparatory information searches~\cite{bohme2020searching, wladyka2017queries} well in advance to inform their destination choices and logistical planning, forced migrants often make immediate decisions to flee and only subsequently seek out detailed information about their new surroundings, leading to an intensification of migration-related information searches after crises~\cite{anastasiadou2024war, sanliturk2024search}.


\begin{figure*}
    \centering
    \includegraphics[width=0.85\linewidth]{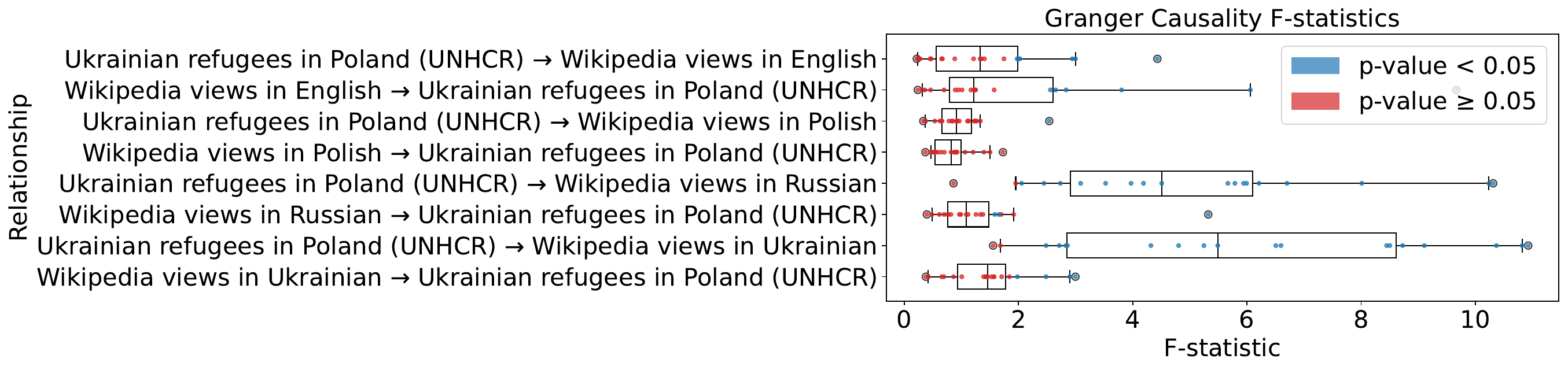}
    \caption{Distribution of F-statistics from Granger causality tests between time series of Wikipedia views of articles about Polish cities and Ukrainian refugees crossing the border to Poland. 
    Each colored dot represents the F-statistic for a Wikipedia article about one of the 19 most populous cities in Poland, with blue indicating statistically significant relationships (p < 0.05) and red indicating non-significant relationships (p $\geq$ 0.05).}
    \label{fig:boxplot-granger}
\end{figure*}

\section{Discussion}
\label{sec:discussion}

In this study, we provided empirical evidence that the Ukrainian refugee crisis significantly increased information-seeking behavior on Wikipedia. By using data from Wikipedia Pageviews, we identified large increases in views of Ukrainian-language Wikipedia articles about European cities after the Russian invasion of Ukraine. There were sharp increases in views of Wikipedia articles about cities that hosted large numbers of Ukrainian refugees, such as Warsaw and other major Polish urban centers, and particularly of articles in Ukrainian. This suggests a strong alignment between forced displacement and digital information-seeking behavior. Moreover, by comparing the Wikipedia views across different languages, we showed that while English- and Polish-language views remained relatively stable, Ukrainian-language views increased immediately after the onset of the war, highlighting Wikipedia’s role as a key information resource for displaced populations.

Our findings indicating that Wikipedia readership, especially in the Ukrainian language, increased not before but shortly after the Russian invasion of Ukraine, shed light on the temporal relationship between information-seeking behavior and refugee flows. This timing suggests that, unlike other types of migrants, such as labor migrants, who typically search for information during the pre-migration phase, forced migrants tend to seek information reactively, both during and after displacement. This pattern reflects the urgent and unplanned nature of forced migration. By providing evidence of these temporal dynamics between online information-seeking and forced migration, our study also contributes to the computational forced migration literature and paves the way for further research. Moreover, our findings highlight the potential of Wikipedia to serve as a real-time indicator of emerging information needs during humanitarian crises, and as a complementary tool for detecting shifts in forced migration flows and settlement patterns as they unfold.

\subsubsection{Limitations}
This study contributes to the growing literature on the use of digital trace data in migration research by introducing Wikipedia as a timely and scalable source for observing information-seeking behavior during forced migration. However, several limitations affect the generalizability of our findings. First, it is important to emphasize that the Wikipedia Pageviews data are available only from July 2015 onward. In our case, the Ukrainian refugee crisis is a recent event that fits within the period of data availability, but this would not be the case for some major refugee crises, such as those in Syria or Venezuela.

In addition, uneven internet penetration rates across populations may affect the use of Wikipedia as a potential source of information during a crisis. In the context of the Ukrainian refugee crisis and the flow of Ukrainian refugees to Poland, the internet penetration rate was not a major concern. Ukraine’s digital infrastructure is comparatively robust, with 80\% of individuals accessing the internet daily by early 2024. Poland and Germany have similarly high rates of digital connectivity, but this is not the case for many countries in the Global South.\footnote{\url{https://worldpopulationreview.com/country-rankings/internet-penetration-by-country}}

The most significant limitation is the use of language as a proxy for the users' origin. Information on the location of Wikipedia readers is not publicly available, and for our study, we used views of articles in the Ukrainian language as a proxy for views by Ukrainian refugees. While this methodology is less problematic in the case of Ukrainian, which is largely concentrated in Ukraine, it becomes a more serious limitation when it is applied to languages spoken widely across multiple countries, such as Arabic and Spanish. As a result, our findings may not extend to other major refugee crises, such as those in Syria or Venezuela, where digital access and language use patterns differ considerably. Consequently, the findings may not be generalizable to all forced displacement contexts, especially those with different digital infrastructures, official languages used in multiple countries, or crisis conditions predating 2015. Nevertheless, for the Ukrainian refugee crisis, we have provided empirical evidence of the impact of the Russian invasion of Ukraine on information-seeking behavior on Wikipedia, and established a temporal association between this behavior and Ukrainian refugee flows.

Finally, a further limitation concerns the extent to which our findings are specific to Wikipedia, or instead reflect broader patterns of online information-seeking mediated by search engine rankings. For instance, if much of Wikipedia’s traffic originates from Google, the relationship we documented may depend on Wikipedia’s current visibility in search results. Given the growing prominence of AI-driven search tools, the use of search engines or readership of online sources of information, such as Wikipedia, might change. These caveats underscore that our results should be interpreted within the present information-seeking ecosystem, in which Wikipedia continues to hold a central role.

Another limitation comes from the official data obtained from Eurostat and the Polish and German statistical offices about the number of temporary protection applications, which were used as ground-truth data in our analysis. These official data suffer from gaps or delays in reporting, particularly in tracking the number of Ukrainian refugees arriving in various cities in a structured and consistent manner. Additionally, once Ukrainian refugees have obtained a PESEL number in a Polish city, for instance, they are free to move within Poland, and we lack reliable data on their internal migration patterns.

\subsubsection{Implications}
To the best of our knowledge, this is the first study using Wikipedia Pageviews data to assess the relationship between information-seeking behavior and forced migration flows. We presented a timely, cost-effective, reproducible, and scalable methodology that leveraged Wikipedia as an innovative data source for studying migration. This approach has important implications.

First, from a policy perspective, the changes in Wikipedia readership in response to the Ukrainian refugee crisis were rapid, and having access to such timely information could provide significant benefits to decision-makers, especially in countries experiencing political instability. Governments can use such real-time data to respond more effectively to migration surges as they unfold, enabling better resource allocation and planning.

Second, the migration literature has previously shown evidence of a time gap between online information-seeking activity and migration flows, with the information search typically preceding the move by several months~\cite{bohme2020searching,wladyka2017queries}. In contrast, our study demonstrated how these temporal dynamics changed in the context of forced migration. We found evidence of a lag of about eight days between the onset of the Russian invasion of Ukraine, which led to Ukrainian refugees crossing the border into Poland, and the subsequent peak in views of Wikipedia articles about Polish cities.

Lastly, from a computational social science perspective, our work repurposed Wikipedia data to study a critical real-world phenomenon, demonstrating how digital trace data could be leveraged to provide insights into information-seeking behavior in the context of forced migration.

Beyond the methodological contributions, our work also has implications for civil society and humanitarian organizations. Looking at Wikipedia activity, as a near real-time complement to official statistics, can help organizations detect emerging migration flows during crises, when timely information is scarce. Because refugees often seek information in their native language shortly after displacement, monitoring Wikipedia pageviews could serve as a low-cost and scalable tool for anticipating where support services, such as housing, legal aid, or language assistance, are most urgently needed. While digital trace data are not a replacement for official records, integrating them into existing monitoring frameworks could enhance the responsiveness of humanitarian interventions and improve coordination across local and international actors.

Finally, our study has underscored the important role of Wikipedia, in particular the role of non-English-language editions such as those in Ukrainian and Russian, as an information resource for refugees. The recognition of Wikipedia as a valuable source of information for refugees can help motivate contributors and communities to strengthen public information goods that directly support vulnerable populations during crises. 

\subsubsection{Ethical considerations and reproducibility}
We collected only publicly available data from the UNHCR website, Poland’s official data portals, and the Wikimedia Pageviews platform, following established ethical guidelines~\cite{rivers2014ethical}. Our study used aggregated data on the number of refugees in European countries and in select Polish cities, as well as aggregated views of Wikipedia articles. In the Wikipedia data, the geographic location of readers is not available for privacy reasons. Finally, we did not attempt to identify or link any personal information to specific users.
All code and data necessary to reproduce our results are available at: \url{https://github.com/carolcoimbra/wikimig}.

\subsubsection{Future work}
In future work, the causal dimension of the relationship demonstrated in our study could be further investigated, and data collection could be expanded to cover more cities affected to varying degrees by Ukrainian refugee flows. Additionally, similar to changes in readership, Wikipedia edits are influenced by real-world events~\cite{ruprechter_volunteer_2021}. In this context, we aim to assess the impact of refugee crises on Wikipedia edits. Finally, the dedication, the size of the editors' community, and the quality of the information provided by Wikipedia editors may act as pull factors for certain destinations in refugees’ decision-making processes. We would like to explore to what extent online information sources, such as Wikipedia, compensate for the absence or weakness of migration networks for refugees during their movements, as highlighted in previous research~\cite{dekker2018smart,merisalo2020digital}.

\section{Conclusion}
In this study, we assessed the impact of forced migration on Wikipedia readership and provided evidence of a temporal relationship between information-seeking behavior on Wikipedia and refugee movements. Focusing on the Ukrainian refugee crisis, we showed that the number of views of Ukrainian-language Wikipedia articles about Polish cities increased by over 200\%, with these views serving as a proxy for information-seeking by Ukrainian users. In addition, we found a strong correlation between the number of views of Wikipedia articles about European cities and the number of Ukrainian refugees recorded across Europe. These findings demonstrated the potential of Wikipedia readership trends to reflect real-world migration patterns. 

Regarding the temporal dynamics, we identified a lag of approximately eight days between the onset of mass migration to Poland and the peak in Wikipedia views, which confirmed that spikes in information-seeking behavior tended to follow, rather than precede, refugee arrivals. These results highlight the reactive nature of information needs during forced migration. Despite this reactive pattern, Wikipedia activity increased almost immediately after border crossings, whereas official protection applications often lagged border crossings by weeks. This contrast underscores Wikipedia’s potential to serve as a near real-time indicator of migration during crises, or an early-warning system for monitoring it. Finally, our approach complements traditional data sources by offering faster and more dynamic insights into migration flows, thus opening up new avenues for research and policy development aimed at improving responses to large-scale displacement.

\small
\bibliography{refs}

@article{wladyka2017queries,
  title={Queries to Google Search as predictors of migration flows from Latin America to Spain},
  author={Wladyka, Dawid K},
  journal={Journal of Population and Social Studies},
  volume={25},
  number={4},
  pages={312},
  year={2017}
}

@book{granger2014forecasting,
  title={Forecasting economic time series},
  author={Granger, Clive William John and Newbold, Paul},
  year={2014},
  publisher={Academic press}
}

@article{anastasiadou2024war,
  title={War and mobility: Using Yandex web searches to characterize intentions to leave Russia after its invasion of Ukraine},
  author={Anastasiadou, Athina and Volgin, Artem and Leasure, Douglas R},
  journal={Demographic Research},
  volume={50},
  year={2024},
  publisher={Max Planck Institute for Demographic Research}
}

@article{gonzalez2024have,
  title={Where have Ukrainian refugees gone? Identifying potential settlement areas across European regions integrating digital and traditional geographic data},
  author={Gonz{\'a}lez-Leonardo, Miguel and Neville, Ruth and Gil-Clavel, Sof{\'\i}a and Rowe, Francisco},
  journal={Population, Space and Place},
  pages={e2790},
  year={2024},
  publisher={Wiley Online Library}
}

@article{leasure2023nowcasting,
  title={Nowcasting daily population displacement in Ukraine through social media advertising data},
  author={Leasure, Douglas R and Kashyap, Ridhi and Rampazzo, Francesco and Dooley, Claire A and Elbers, Benjamin and Bondarenko, Maksym and Verhagen, Mark and Frey, Arun and Yan, Jiani and Akimova, Evelina T and others},
  journal={Population and Development Review},
  volume={49},
  number={2},
  pages={231--254},
  year={2023},
  publisher={Wiley Online Library}
}

@article{felton2015migrants,
  title={Migrants, refugees, and mobility: How useful are information communication technologies in the first phase of resettlement},
  author={Felton, Emma},
  journal={Journal of Technologies in Society},
  volume={11},
  number={1},
  pages={1--13},
  year={2015},
  publisher={Common Ground Publishing}
}

@article{dekker2018smart,
  title={Smart refugees: How Syrian asylum migrants use social media information in migration decision-making},
  author={Dekker, Rianne and Engbersen, Godfried and Klaver, Jeanine and Vonk, Hanna},
  journal={Social Media+ Society},
  volume={4},
  number={1},
  pages={2056305118764439},
  year={2018},
  publisher={SAGE Publications Sage UK: London, England}
}

@article{robinson1998importance,
  title={The importance of information in the resettlement of refugees in the UK},
  author={Robinson, Vaughan},
  journal={Journal of refugee studies},
  volume={11},
  number={2},
  pages={146--160},
  year={1998},
  publisher={Oxford University Press}
}

@inproceedings{zimmer2020age,
  title={Age-and Gender-dependent Differences of Asylum Seekers' Information Behavior and Online Media Usage.},
  author={Zimmer, Franziska and Scheibe, Katrin},
  booktitle={HICSS},
  pages={1--10},
  year={2020}
}

@article{sanliturk2024search,
  title={Search for a New Home: Refugee Stock and Google Search},
  author={Sanliturk, Ebru and Billari, Francesco C},
  journal={International Migration Review},
  pages={01979183241275452},
  year={2024},
  publisher={SAGE Publications Sage CA: Los Angeles, CA}
}

@inproceedings{miz2020trending,
  title={What is trending on wikipedia? capturing trends and language biases across wikipedia editions},
  author={Miz, Volodymyr and Hanna, Jo{\"e}lle and Aspert, Nicolas and Ricaud, Benjamin and Vandergheynst, Pierre},
  booktitle={Companion proceedings of the Web conference 2020},
  pages={794--801},
  year={2020}
}

@inproceedings{ribeiro_sudden_nodate,
  title={Sudden attention shifts on wikipedia during the covid-19 crisis},
  author={Ribeiro, Manoel Horta and Gligori{\'c}, Kristina and Peyrard, Maxime and Lemmerich, Florian and Strohmaier, Markus and West, Robert},
  booktitle={Proceedings of the International AAAI Conference on Web and Social Media},
  volume={15},
  pages={208--219},
  year={2021}
}

@article{kampf2015detection,
  title={The detection of emerging trends using Wikipedia traffic data and context networks},
  author={K{\"a}mpf, Mirko and Tessenow, Eric and Kenett, Dror Y and Kantelhardt, Jan W},
  journal={PloS one},
  volume={10},
  number={12},
  pages={e0141892},
  year={2015},
  publisher={Public Library of Science San Francisco, CA USA}
}

@article{ruprechter_volunteer_2021,
	title = {Volunteer contributions to {Wikipedia} increased during {COVID}-19 mobility restrictions},
	volume = {11},
	copyright = {2021 The Author(s)},
	issn = {2045-2322},
	url = {https://www.nature.com/articles/s41598-021-00789-3},
	doi = {10.1038/s41598-021-00789-3},
	language = {en},
	number = {1},
	urldate = {2022-09-21},
	journal = {Scientific Reports},
	author = {Ruprechter, Thorsten and Horta Ribeiro, Manoel and Santos, Tiago and Lemmerich, Florian and Strohmaier, Markus and West, Robert and Helic, Denis},
	month = nov,
	year = {2021},
	pages = {21505},
}

@article{merisalo2020digital,
  title={Digital divides among asylum-related migrants: Comparing internet use and smartphone ownership},
  author={Merisalo, Maria and Jauhiainen, Jussi S},
  journal={Tijdschrift voor economische en sociale geografie},
  volume={111},
  number={5},
  pages={689--704},
  year={2020},
  publisher={Wiley Online Library}
}

@inproceedings{yoshida_wikipedia_2015,
	title = {Wikipedia {Page} {View} {Reflects} {Web} {Search} {Trend}},
	url = {https://dl.acm.org/doi/10.1145/2786451.2786495},
	doi = {10.1145/2786451.2786495},
	language = {en},
	urldate = {2022-11-15},
	booktitle = {Proceedings of the {ACM} {Web} {Science} {Conference}},
	author = {Yoshida, Mitsuo and Arase, Yuki and Tsunoda, Takaaki and Yamamoto, Mikio},
	month = jun,
	year = {2015},
	pages = {1--2},
}

@article{bohme2020searching,
  title={Searching for a better life: Predicting international migration with online search keywords},
  author={B{\"o}hme, Marcus H and Gr{\"o}ger, Andr{\'e} and St{\"o}hr, Tobias},
  journal={Journal of Development Economics},
  volume={142},
  pages={102347},
  year={2020},
  publisher={Elsevier}
}

@inproceedings{lin2019forecasting,
  title={Forecasting us domestic migration using internet search queries},
  author={Lin, Allen Yilun and Cranshaw, Justin and Counts, Scott},
  booktitle={The world wide web conference},
  pages={1061--1072},
  year={2019}
}

@article{rivers2014ethical,
  title={Ethical research standards in a world of big data},
  author={Rivers, Caitlin M and Lewis, Bryan L},
  journal={F1000Research},
  volume={3},
  number={38},
  pages={38},
  year={2014},
  publisher={F1000 Research Limited}
}

@article{mciver2014wikipedia,
  title={Wikipedia usage estimates prevalence of influenza-like illness in the United States in near real-time},
  author={McIver, David J and Brownstein, John S},
  journal={PLoS computational biology},
  volume={10},
  number={4},
  pages={e1003581},
  year={2014},
  publisher={Public Library of Science San Francisco, USA}
}

@article{tizzoni2020impact,
  title={The impact of news exposure on collective attention in the United States during the 2016 Zika epidemic},
  author={Tizzoni, Michele and Panisson, Andr{\'e} and Paolotti, Daniela and Cattuto, Ciro},
  journal={PLoS computational biology},
  volume={16},
  number={3},
  pages={e1007633},
  year={2020},
  publisher={Public Library of Science San Francisco, CA USA}
}

@article{jemielniak2021wikiproject,
  title={WikiProject Tropical Cyclones: the most successful crowd-sourced knowledge project with near real-time coverage of extreme weather phenomena},
  author={Jemielniak, Dariusz and Rychwalska, Agnieszka and Talaga, Szymon and Ziembowicz, Karolina},
  journal={Weather and Climate Extremes},
  volume={33},
  pages={100354},
  year={2021},
  publisher={Elsevier}
}

@article{duszczyk2022war,
  title={The war in Ukraine and migration to Poland: Outlook and challenges},
  author={Duszczyk, Maciej and Kaczmarczyk, Pawe{\l}},
  journal={Intereconomics},
  volume={57},
  number={3},
  pages={164--170},
  year={2022},
  publisher={Springer}
}

@article{kulyk2024language,
  title={Language shift in time of war: the abandonment of Russian in Ukraine},
  author={Kulyk, Volodymyr},
  journal={Post-Soviet Affairs},
  volume={40},
  number={3},
  pages={159--174},
  year={2024},
  publisher={Taylor \& Francis}
}

@book{johansen1995likelihood,
  title={Likelihood-based inference in cointegrated vector autoregressive models},
  author={Johansen, S{\o}ren},
  year={1995},
  publisher={OUP Oxford}
}

@article{pietka2025urban,
  title={The urban hierarchy: how the war in Ukraine has changed the status of the Polish city of Rzeszow},
  author={Pietka, Anna and Sielska, Agata},
  journal={The Annals of Regional Science},
  volume={74},
  number={2},
  pages={47},
  year={2025},
  publisher={Springer}
}

@article{dickey1979distribution,
  title={Distribution of the estimators for autoregressive time series with a unit root},
  author={Dickey, David A and Fuller, Wayne A},
  journal={Journal of the American statistical association},
  volume={74},
  number={366a},
  pages={427--431},
  year={1979},
  publisher={Taylor \& Francis}
}

@article{bozdogan1987model,
  title={Model selection and Akaike's information criterion (AIC): The general theory and its analytical extensions},
  author={Bozdogan, Hamparsum},
  journal={Psychometrika},
  volume={52},
  number={3},
  pages={345--370},
  year={1987},
  publisher={Springer-Verlag}
}

@incollection{lutkepohl2013vector,
  title={Vector autoregressive models},
  author={L{\"u}tkepohl, Helmut},
  booktitle={Handbook of research methods and applications in empirical macroeconomics},
  pages={139--164},
  year={2013},
  publisher={Edward Elgar Publishing}
}

@article{bailard2024keep,
  title={“Keep Your Heads Held High Boys!”: Examining the Relationship between the Proud Boys’ Online Discourse and Offline Activities},
  author={Bailard, Catie Snow and Tromble, Rebekah and Zhong, Wei and Bianchi, Federico and Hosseini, Pedram and Broniatowski, David},
  journal={American Political Science Review},
  volume={118},
  number={4},
  pages={2054--2071},
  year={2024},
  publisher={Cambridge University Press}
}

@Article{strucchange,
    title = {strucchange: An R Package for Testing for Structural
      Change in Linear Regression Models},
    author = {Achim Zeileis and Friedrich Leisch and Kurt Hornik and
      Christian Kleiber},
    journal = {Journal of Statistical Software},
    year = {2002},
    volume = {7},
    number = {2},
    pages = {1--38},
    doi = {10.18637/jss.v007.i02},
  }

@article{gebru2021datasheets,
  title={Datasheets for datasets},
  author={Gebru, Timnit and Morgenstern, Jamie and Vecchione, Briana and Vaughan, Jennifer Wortman and Wallach, Hanna and Iii, Hal Daum{\'e} and Crawford, Kate},
  journal={Communications of the ACM},
  volume={64},
  number={12},
  pages={86--92},
  year={2021},
  publisher={ACM New York, NY, USA}
}

@misc{fair,
    title="The FAIR Data principles",
year = 2020,
    author="{FORCE11}",
howpublished="\url{https://force11.org/info/the-fair-data-principles/}"
}

@inproceedings{hill2014consider,
  title={Consider the redirect: A missing dimension of Wikipedia research},
  author={Hill, Benjamin Mako and Shaw, Aaron},
  booktitle={Proceedings of the international symposium on open collaboration},
  pages={1--4},
  year={2014}
}

\clearpage
\normalsize
\subsection{Paper Checklist to be included in your paper}

\begin{enumerate}

\item For most authors...
\begin{enumerate}
    \item  Would answering this research question advance science without violating social contracts, such as violating privacy norms, perpetuating unfair profiling, exacerbating the socio-economic divide, or implying disrespect to societies or cultures?
    \answerYes{Yes.}
  \item Do your main claims in the abstract and introduction accurately reflect the paper's contributions and scope?
    \answerYes{Yes.}
   \item Do you clarify how the proposed methodological approach is appropriate for the claims made? 
    \answerYes{Yes.}
   \item Do you clarify what are possible artifacts in the data used, given population-specific distributions?
    \answerYes{Yes.}
  \item Did you describe the limitations of your work?
    \answerYes{Yes.}
  \item Did you discuss any potential negative societal impacts of your work?
    \answerYes{Yes.}
      \item Did you discuss any potential misuse of your work?
    \answerYes{Yes.}
    \item Did you describe steps taken to prevent or mitigate potential negative outcomes of the research, such as data and model documentation, data anonymization, responsible release, access control, and the reproducibility of findings?
    \answerYes{Yes.}
  \item Have you read the ethics review guidelines and ensured that your paper conforms to them?
    \answerYes{Yes.}
\end{enumerate}

\item Additionally, if your study involves hypotheses testing...
\begin{enumerate}
  \item Did you clearly state the assumptions underlying all theoretical results?
    \answerYes{Yes.}
  \item Have you provided justifications for all theoretical results?
    \answerYes{Yes.}
  \item Did you discuss competing hypotheses or theories that might challenge or complement your theoretical results?
    \answerYes{Yes.}
  \item Have you considered alternative mechanisms or explanations that might account for the same outcomes observed in your study?
    \answerYes{Yes.}
  \item Did you address potential biases or limitations in your theoretical framework?
    \answerYes{Yes.}
  \item Have you related your theoretical results to the existing literature in social science?
    \answerYes{Yes.}
  \item Did you discuss the implications of your theoretical results for policy, practice, or further research in the social science domain?
    \answerYes{Yes.}
\end{enumerate}

\item Additionally, if you are including theoretical proofs...
\begin{enumerate}
  \item Did you state the full set of assumptions of all theoretical results?
    \answerNA{NA}
	\item Did you include complete proofs of all theoretical results?
    \answerNA{NA}
\end{enumerate}

\item Additionally, if you ran machine learning experiments...
\begin{enumerate}
  \item Did you include the code, data, and instructions needed to reproduce the main experimental results (either in the supplemental material or as a URL)?
    \answerNA{NA}
  \item Did you specify all the training details (e.g., data splits, hyperparameters, how they were chosen)?
    \answerNA{NA}
     \item Did you report error bars (e.g., with respect to the random seed after running experiments multiple times)?
    \answerNA{NA}
	\item Did you include the total amount of compute and the type of resources used (e.g., type of GPUs, internal cluster, or cloud provider)?
    \answerNA{NA}
     \item Do you justify how the proposed evaluation is sufficient and appropriate to the claims made? 
    \answerNA{NA}
     \item Do you discuss what is ``the cost`` of misclassification and fault (in)tolerance?
    \answerNA{NA}
  
\end{enumerate}

\item Additionally, if you are using existing assets (e.g., code, data, models) or curating/releasing new assets, \textbf{without compromising anonymity}...
\begin{enumerate}
  \item If your work uses existing assets, did you cite the creators?
    \answerYes{Yes.}
  \item Did you mention the license of the assets?
    \answerYes{Yes.}
  \item Did you include any new assets in the supplemental material or as a URL?
    \answerYes{Yes.}
  \item Did you discuss whether and how consent was obtained from people whose data you're using/curating?
    \answerYes{Yes.}
  \item Did you discuss whether the data you are using/curating contains personally identifiable information or offensive content?
    \answerYes{Yes.}
\item If you are curating or releasing new datasets, did you discuss how you intend to make your datasets FAIR (see \citet{fair})?
\answerNA{NA}
\item If you are curating or releasing new datasets, did you create a Datasheet for the Dataset (see \citet{gebru2021datasheets})? 
\answerNA{NA}
\end{enumerate}

\item Additionally, if you used crowdsourcing or conducted research with human subjects, \textbf{without compromising anonymity}...
\begin{enumerate}
  \item Did you include the full text of instructions given to participants and screenshots?
    \answerNA{NA}
  \item Did you describe any potential participant risks, with mentions of Institutional Review Board (IRB) approvals?
    \answerNA{NA}
  \item Did you include the estimated hourly wage paid to participants and the total amount spent on participant compensation?
    \answerNA{NA}
   \item Did you discuss how data is stored, shared, and deidentified?
   \answerNA{NA}
\end{enumerate}

\end{enumerate}

\section{Appendix}

\begin{figure*}[ht!]
\caption*{\textbf{Rank comparison: Europe}}
    \centering
    \begin{subfigure}[h]{0.3\textwidth}
        \centering
        \includegraphics[trim=0cm 1.8cm 0cm 0cm, clip, width=\linewidth]{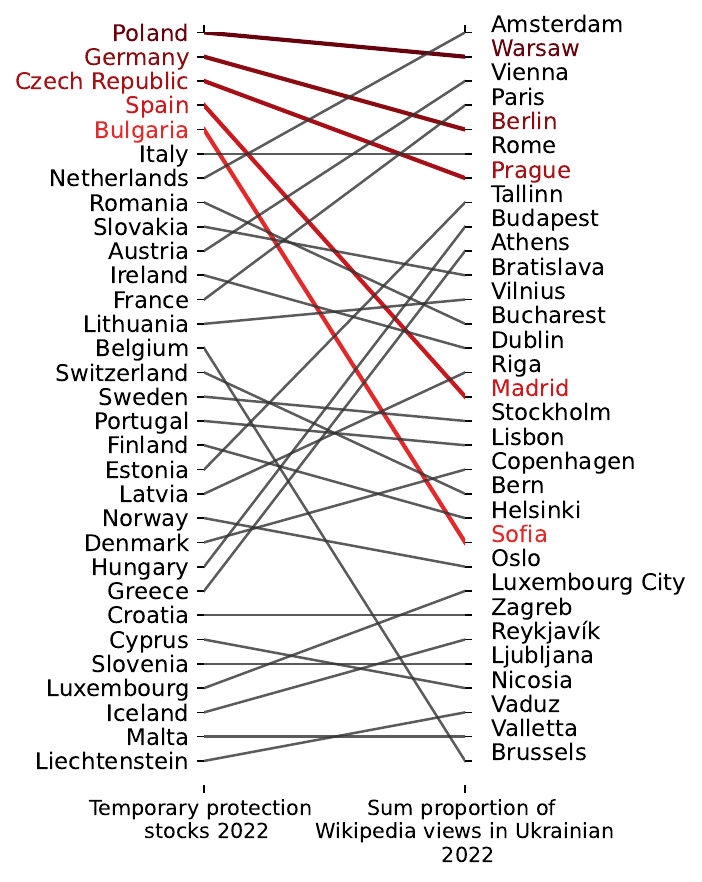}
        \caption{\textit{Ukrainian Wikipedia 2022}}
    \end{subfigure}
    \hfill
    \begin{subfigure}[h]{0.3\textwidth}
        \centering
        \includegraphics[trim=0cm 1.8cm 0cm 0cm, clip, width=\linewidth]{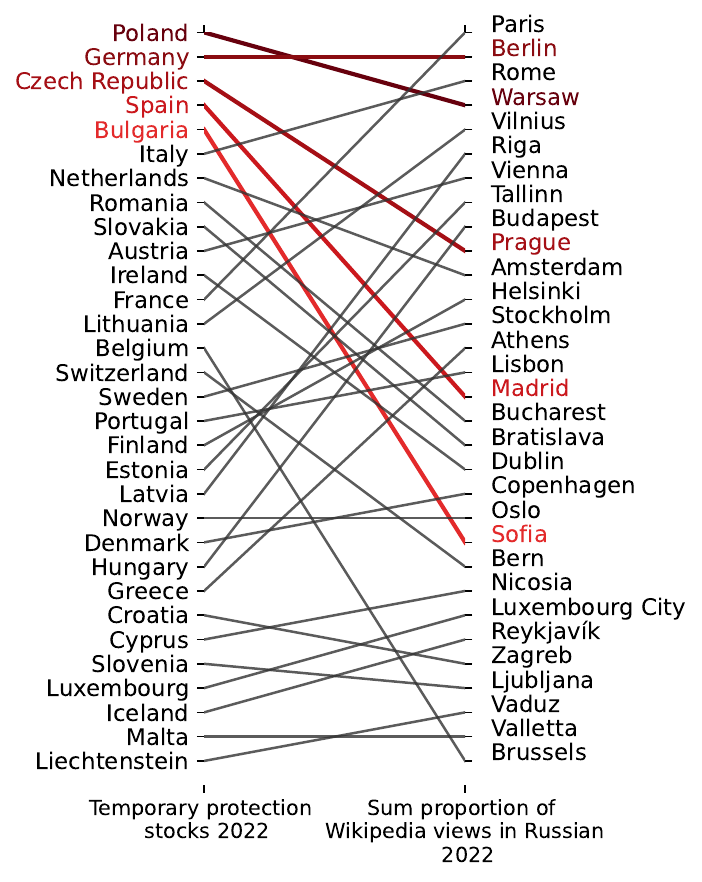}
        \caption{\textit{Russian Wikipedia 2022}}
    \end{subfigure}
    \hfill
    \begin{subfigure}[h]{0.3\textwidth}
        \centering
        \includegraphics[trim=0cm 1.8cm 0cm 0cm, clip, width=\linewidth]{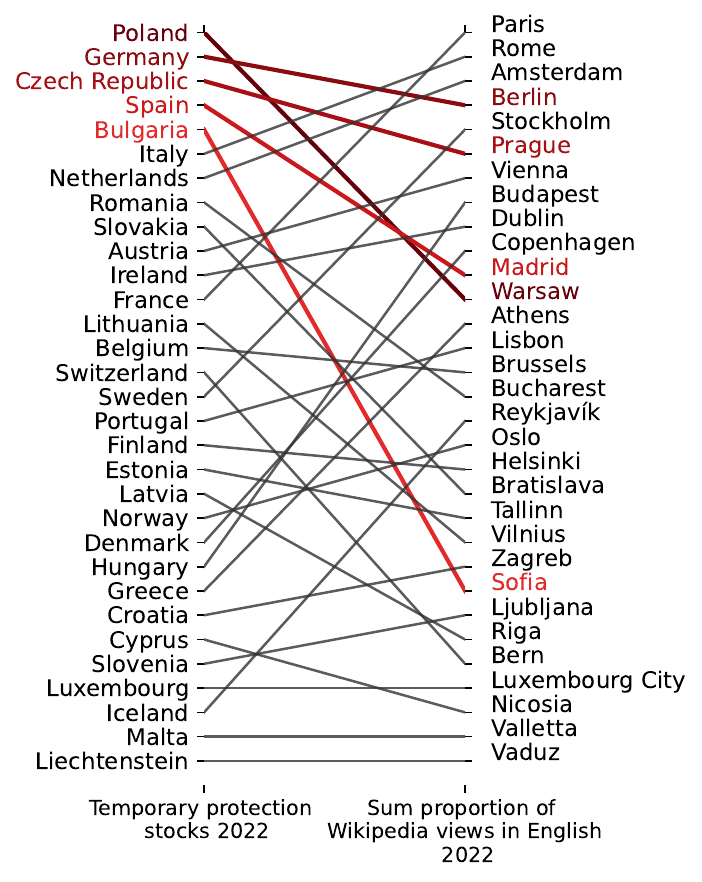}
        \caption{\textit{English Wikipedia 2022}}
    \end{subfigure}
    
    \begin{subfigure}[h]{0.3\textwidth}
        \centering
        \includegraphics[trim=0cm 1.8cm 0cm 0cm, clip, width=\linewidth]{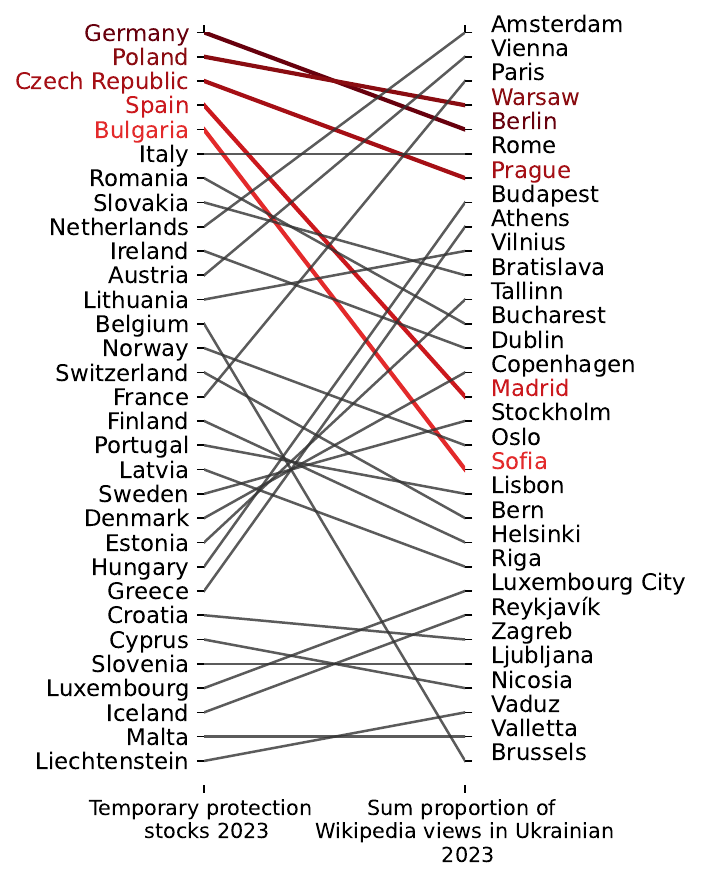}
        \caption{\textit{Ukrainian Wikipedia 2023}}
    \end{subfigure}
    \hfill
    \begin{subfigure}[h]{0.3\textwidth}
        \centering
        \includegraphics[trim=0cm 1.8cm 0cm 0cm, clip, width=\linewidth]{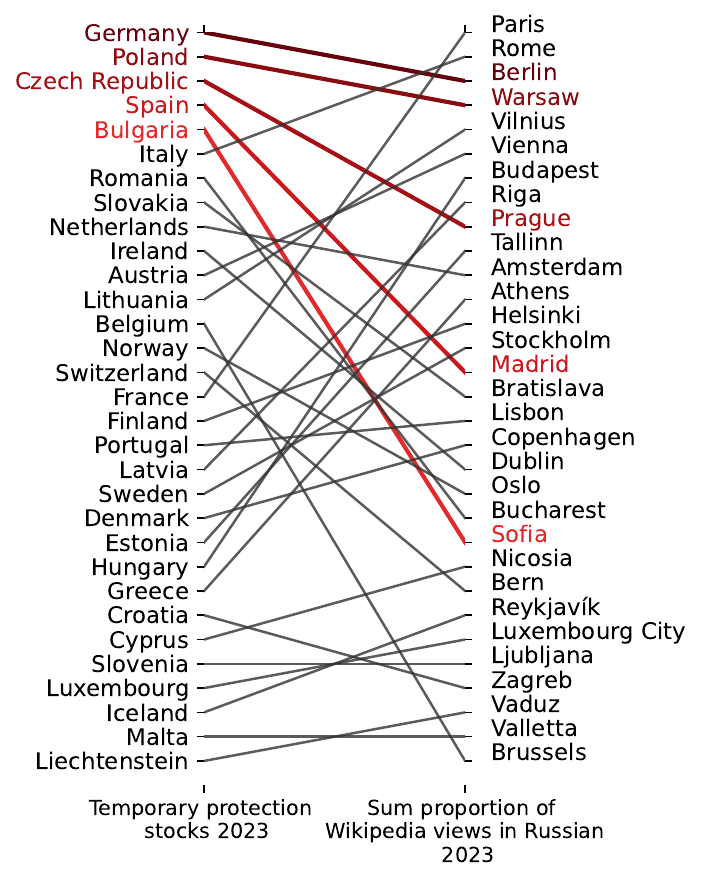}
        \caption{\textit{Russian Wikipedia 2023}}
    \end{subfigure}
    \hfill
    \begin{subfigure}[h]{0.3\textwidth}
        \centering
        \includegraphics[trim=0cm 1.8cm 0cm 0cm, clip, width=\linewidth]{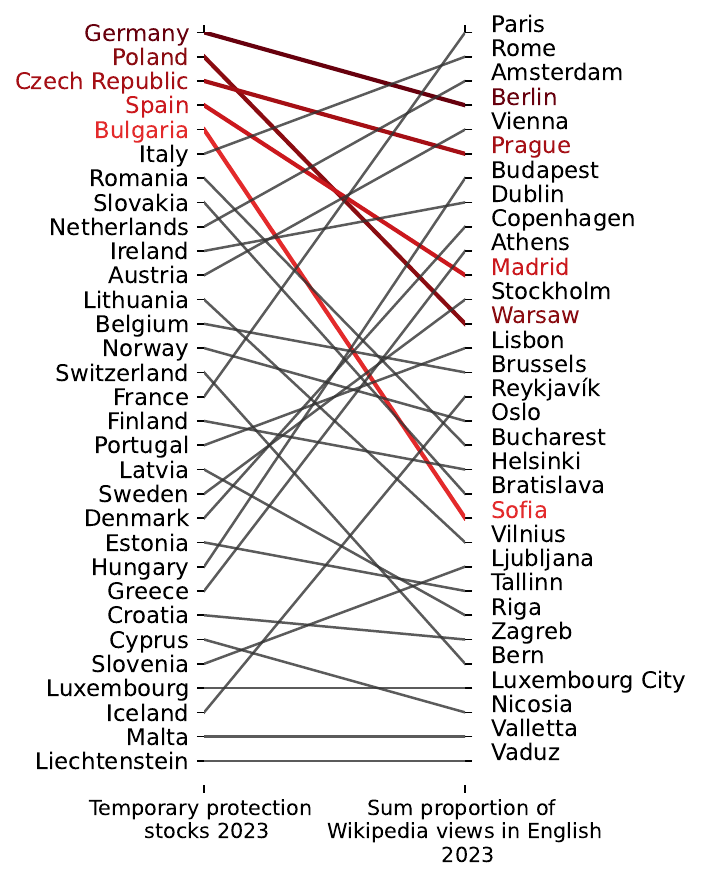}
        \caption{\textit{English Wikipedia 2023}}
    \end{subfigure}
    
    \begin{subfigure}[h]{0.3\textwidth}
        \centering
        \includegraphics[trim=0cm 1.8cm 0cm 0cm, clip, width=\linewidth]{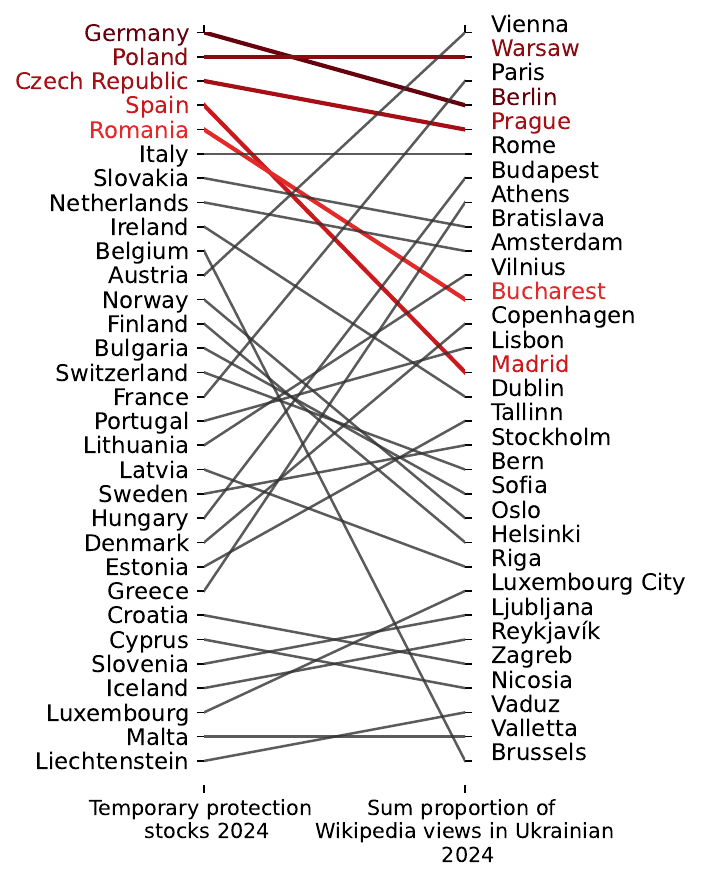}
        \caption{\textit{Ukrainian Wikipedia 2024}}
    \end{subfigure}
    \hfill
    \begin{subfigure}[h]{0.3\textwidth}
        \centering
        \includegraphics[trim=0cm 1.8cm 0cm 0cm, clip, width=\linewidth]{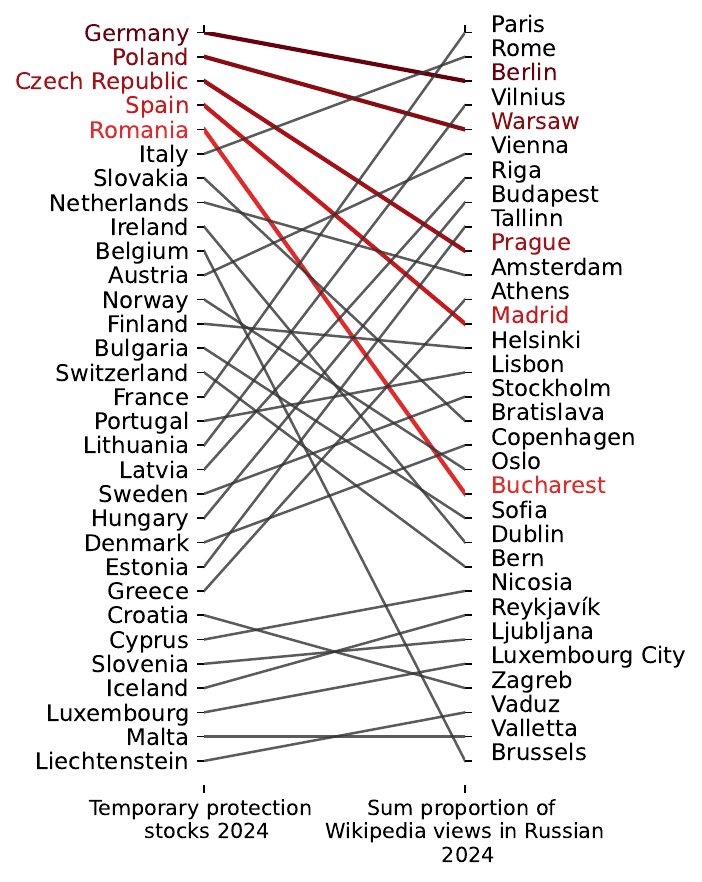}
        \caption{\textit{Russian Wikipedia 2024}}
    \end{subfigure}
    \hfill
    \begin{subfigure}[h]{0.3\textwidth}
        \centering
        \includegraphics[trim=0cm 1.8cm 0cm 0cm, clip, width=\linewidth]{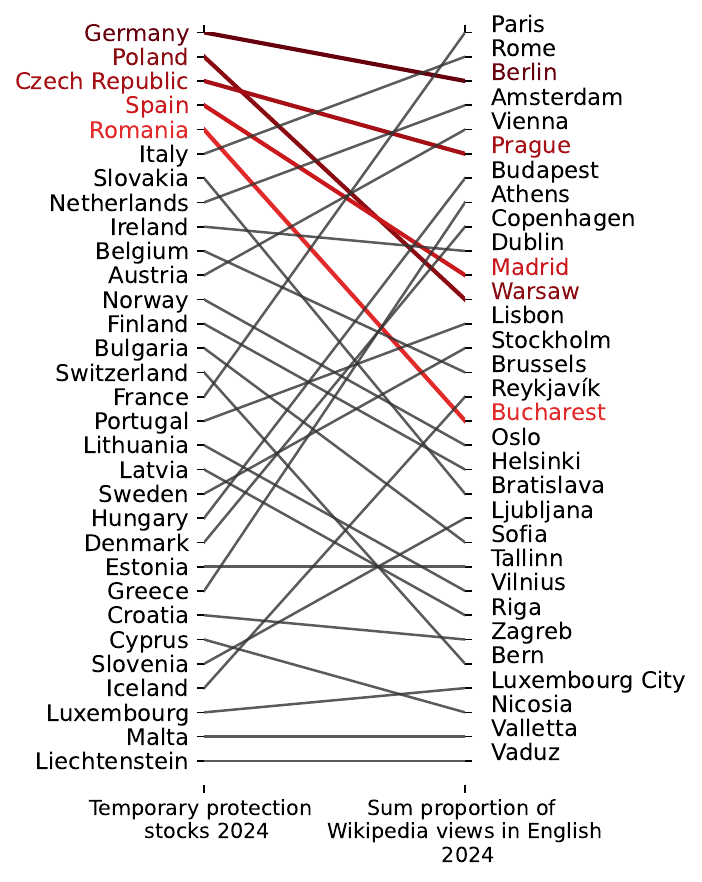}
        \caption{\textit{English Wikipedia 2024}}
    \end{subfigure}
    \caption{Correlation between rankings: stocks of Ukrainian refugees in EU countries (left) and the proportion of views of Wikipedia articles about EU capitals (right), by year. The five countries hosting the largest numbers of Ukrainian refugees are shown in color, while the remaining countries are shown in gray.}
    \label{fig:ranks-europe}
\end{figure*}

\begin{figure*}[ht!]
\caption*{\textbf{Rank comparison: Poland}}
    \centering
    \begin{subfigure}[h]{0.24\textwidth}
        \centering
        \includegraphics[trim=0cm 1.8cm 0cm 0cm, clip, width=\linewidth]{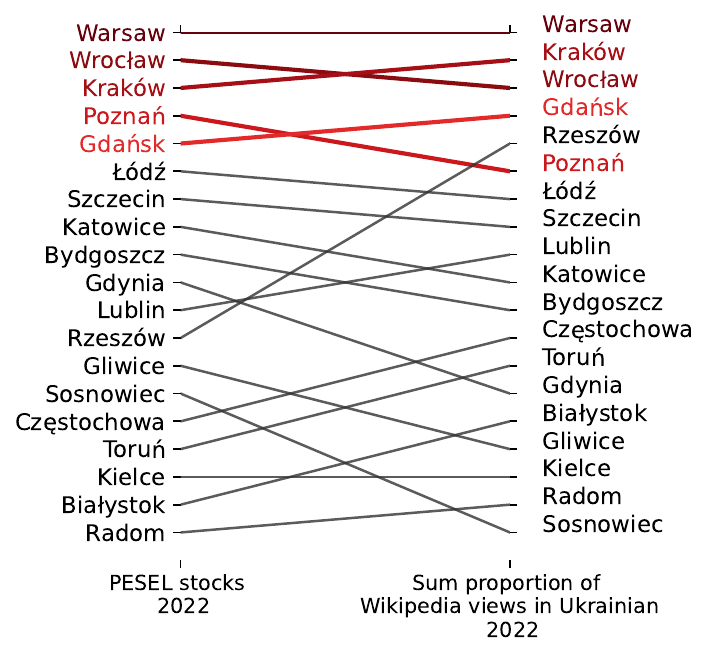}
        \caption{\textit{Ukrainian Wikipedia 2022}}
    \end{subfigure}
    \hfill
    \begin{subfigure}[h]{0.24\textwidth}
        \centering
        \includegraphics[trim=0cm 1.8cm 0cm 0cm, clip, width=\linewidth]{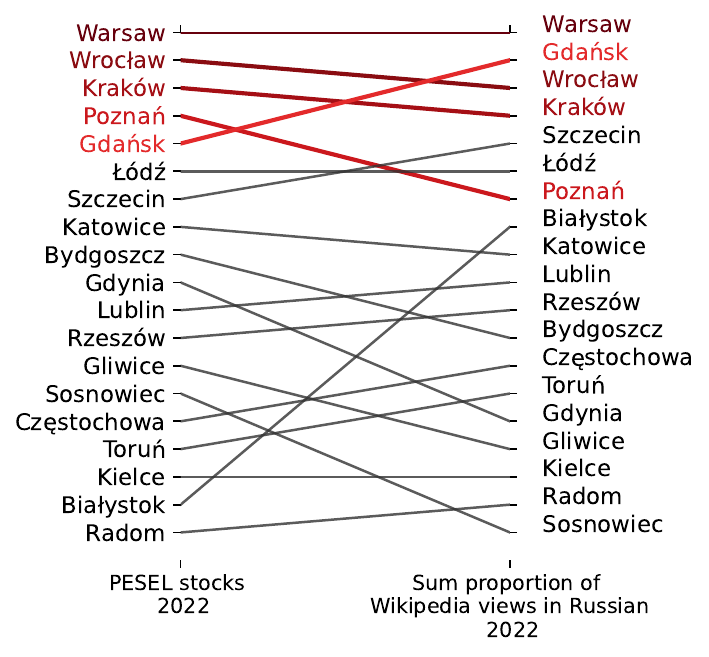}
        \caption{\textit{Russian Wikipedia 2022}}
    \end{subfigure}
    \hfill
    \begin{subfigure}[h]{0.24\textwidth}
        \centering
        \includegraphics[trim=0cm 1.8cm 0cm 0cm, clip, width=\linewidth]{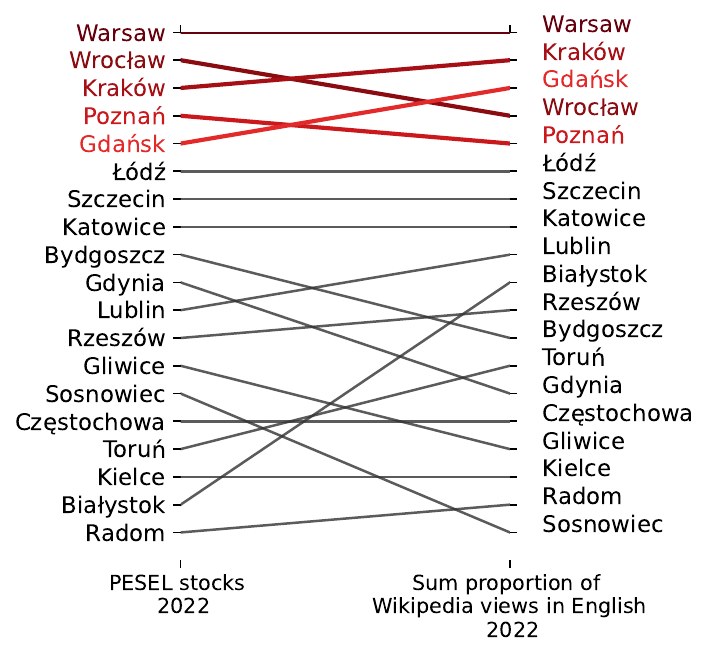}
        \caption{\textit{English Wikipedia 2022}}
    \end{subfigure}
    \hfill
    \begin{subfigure}[h]{0.24\textwidth}
        \centering
        \includegraphics[trim=0cm 1.8cm 0cm 0cm, clip, width=\linewidth]{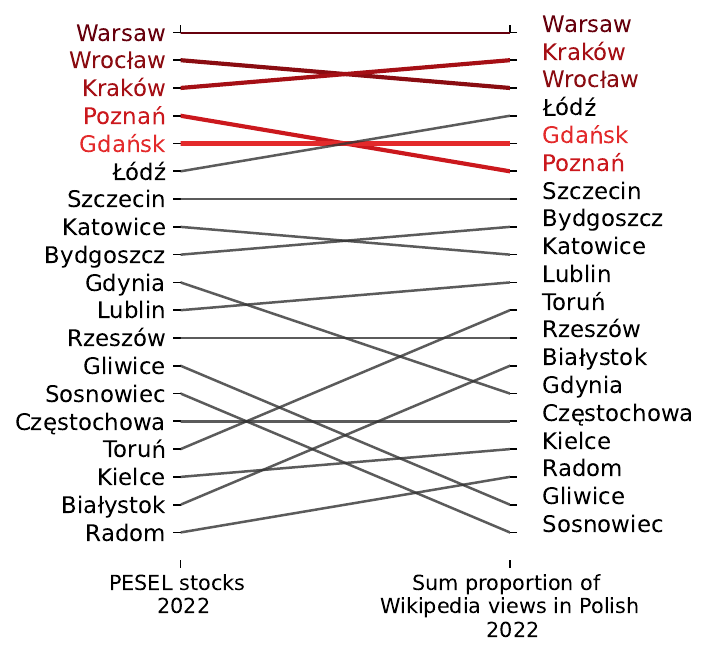}
        \caption{\textit{Polish Wikipedia 2022}}
    \end{subfigure}
    
    \begin{subfigure}[h]{0.24\textwidth}
        \centering
        \includegraphics[trim=0cm 1.8cm 0cm 0cm, clip, width=\linewidth]{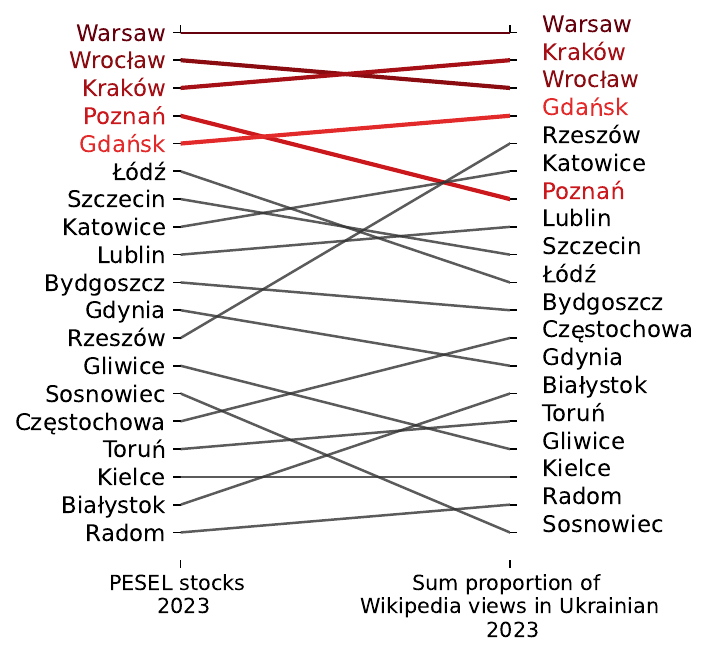}
        \caption{\textit{Ukrainian Wikipedia 2023}}
    \end{subfigure}
    \hfill
    \begin{subfigure}[h]{0.24\textwidth}
        \centering
        \includegraphics[trim=0cm 1.8cm 0cm 0cm, clip, width=\linewidth]{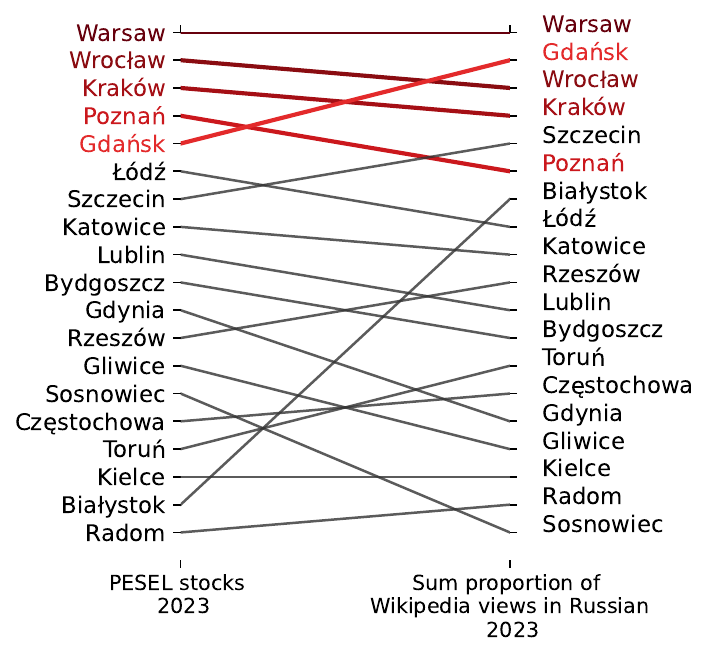}
        \caption{\textit{Russian Wikipedia 2023}}
    \end{subfigure}
    \hfill
    \begin{subfigure}[h]{0.24\textwidth}
        \centering
        \includegraphics[trim=0cm 1.8cm 0cm 0cm, clip, width=\linewidth]{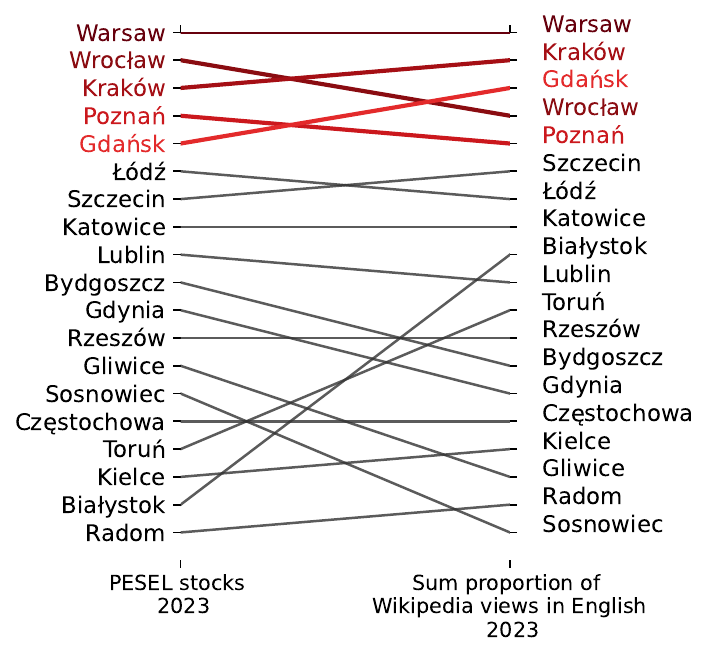}
        \caption{\textit{English Wikipedia 2023}}
    \end{subfigure}
    \hfill
    \begin{subfigure}[h]{0.24\textwidth}
        \centering
        \includegraphics[trim=0cm 1.8cm 0cm 0cm, clip, width=\linewidth]{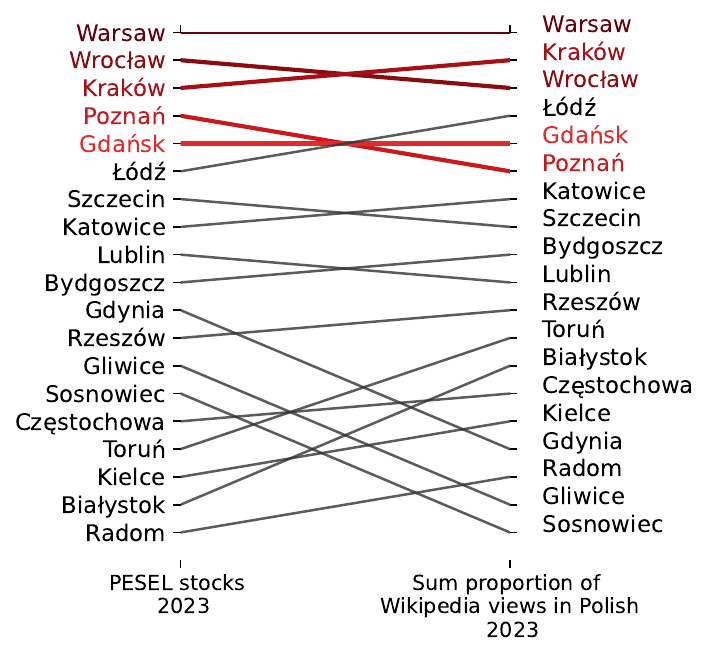}
        \caption{\textit{Polish Wikipedia 2023}}
    \end{subfigure}
    
    \begin{subfigure}[h]{0.24\textwidth}
        \centering
        \includegraphics[trim=0cm 1.8cm 0cm 0cm, clip, width=\linewidth]{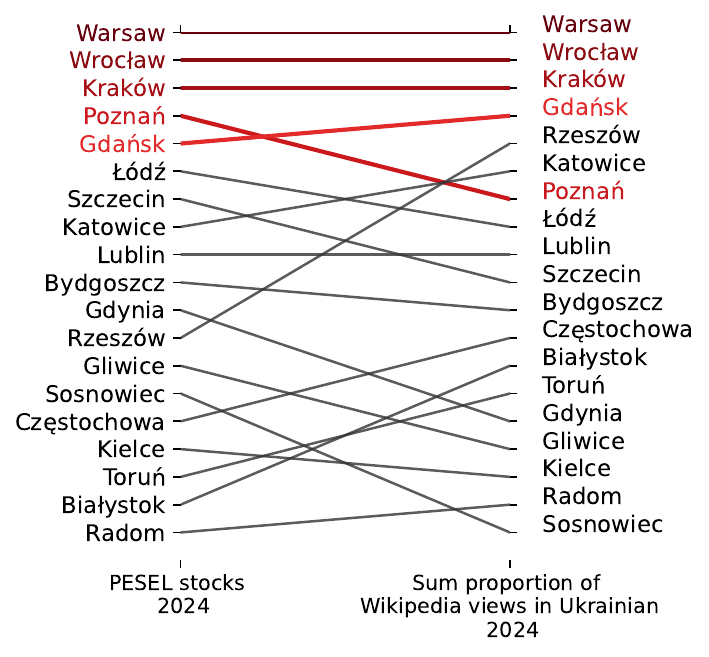}
        \caption{\textit{Ukrainian Wikipedia 2024}}
    \end{subfigure}
    \hfill
    \begin{subfigure}[h]{0.24\textwidth}
        \centering
        \includegraphics[trim=0cm 1.8cm 0cm 0cm, clip, width=\linewidth]{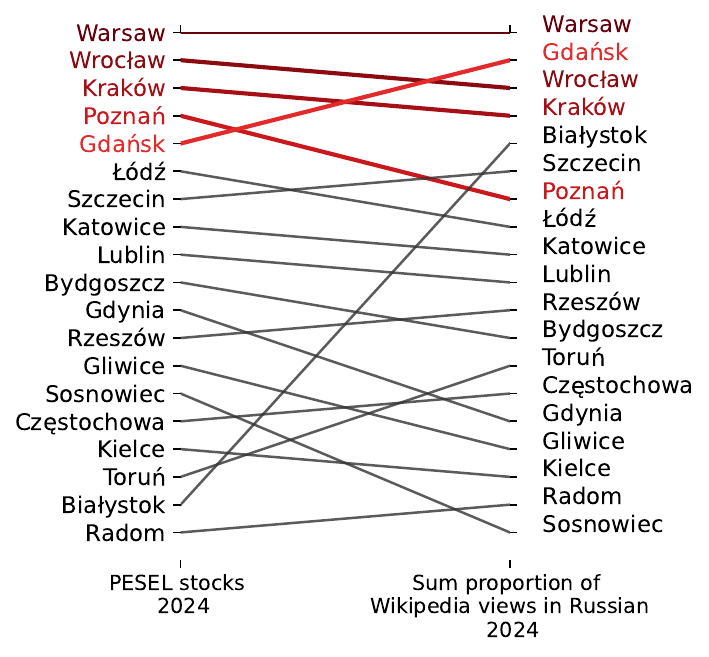}
        \caption{\textit{Russian Wikipedia 2024}}
    \end{subfigure}
    \hfill
    \begin{subfigure}[h]{0.24\textwidth}
        \centering
        \includegraphics[trim=0cm 1.8cm 0cm 0cm, clip, width=\linewidth]{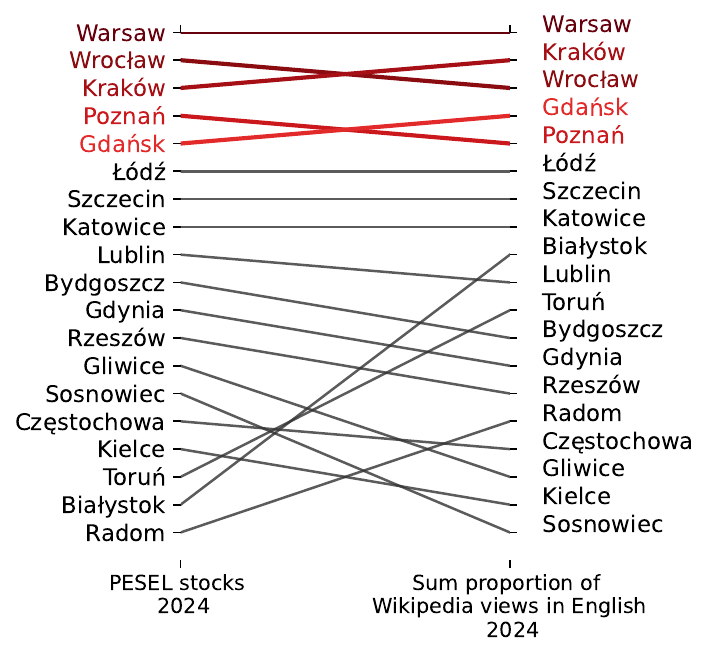}
        \caption{\textit{English Wikipedia 2024}}
    \end{subfigure}
    \hfill
    \begin{subfigure}[h]{0.24\textwidth}
        \centering
        \includegraphics[trim=0cm 1.8cm 0cm 0cm, clip, width=\linewidth]{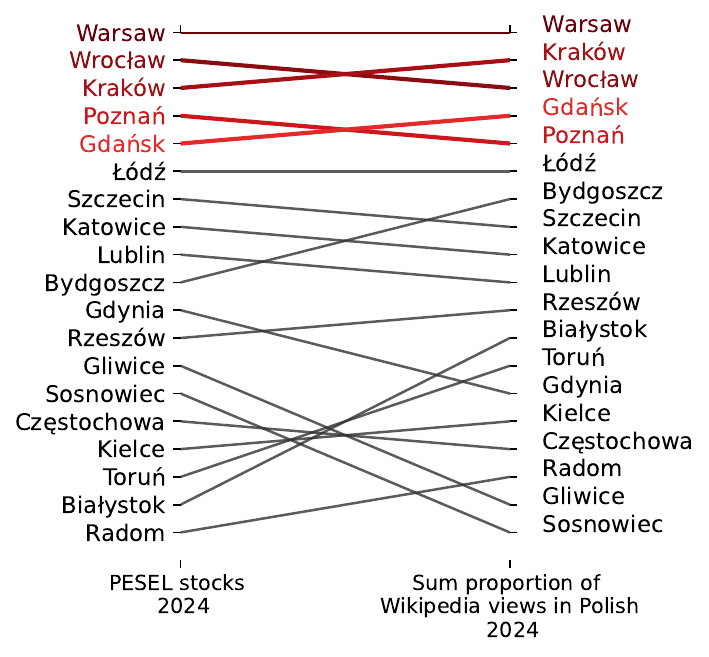}
        \caption{\textit{Polish Wikipedia 2024}}
    \end{subfigure}
    \caption{Correlation between rankings: stocks of Ukrainian refugees who have been assigned a PESEL number in Polish cities (left) and the proportion of views of Wikipedia articles about the 19 most populous cities in Poland (right), by year. The five cities hosting the largest numbers of PESEL-registered Ukrainian refugees are shown in color, while the remaining cities are shown in gray.}
    \label{fig:ranks-poland}
\end{figure*}

\begin{figure*}[ht!]
\caption*{\textbf{Rank comparison: Germany}}
\centering
    \begin{subfigure}[h]{0.24\textwidth}
        \centering
        \includegraphics[trim=0cm 1.8cm 0cm 0cm, clip, width=\linewidth]{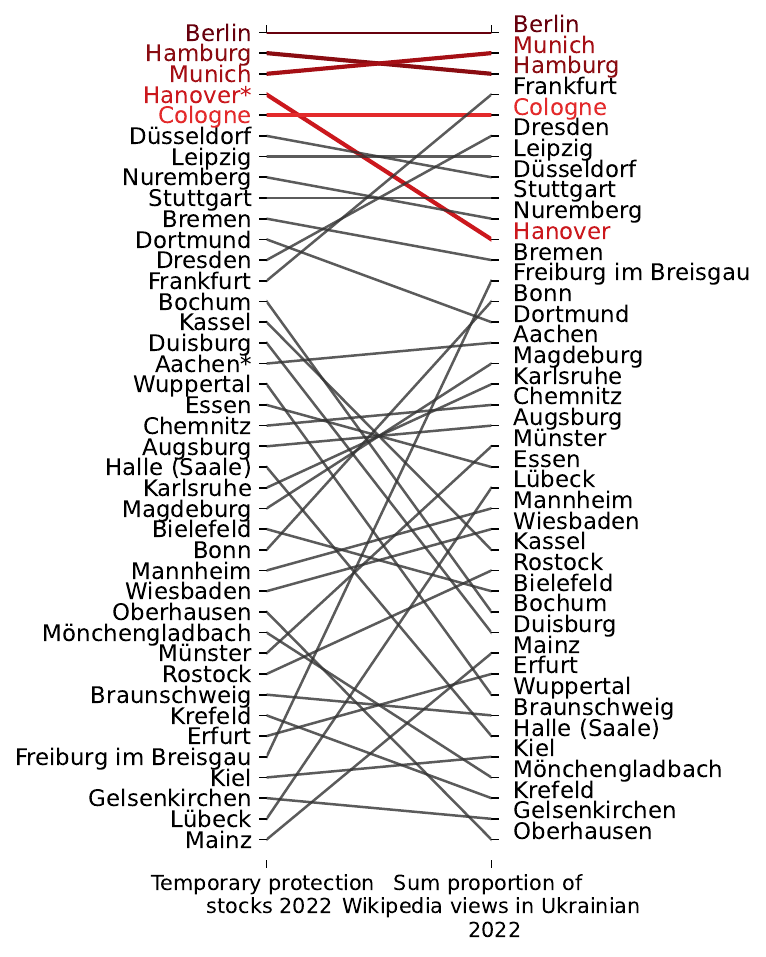}
        \caption{\textit{Ukrainian Wikipedia 2022}}
    \end{subfigure}
    \hfill
    \begin{subfigure}[h]{0.24\textwidth}
        \centering
        \includegraphics[trim=0cm 1.8cm 0cm 0cm, clip, width=\linewidth]{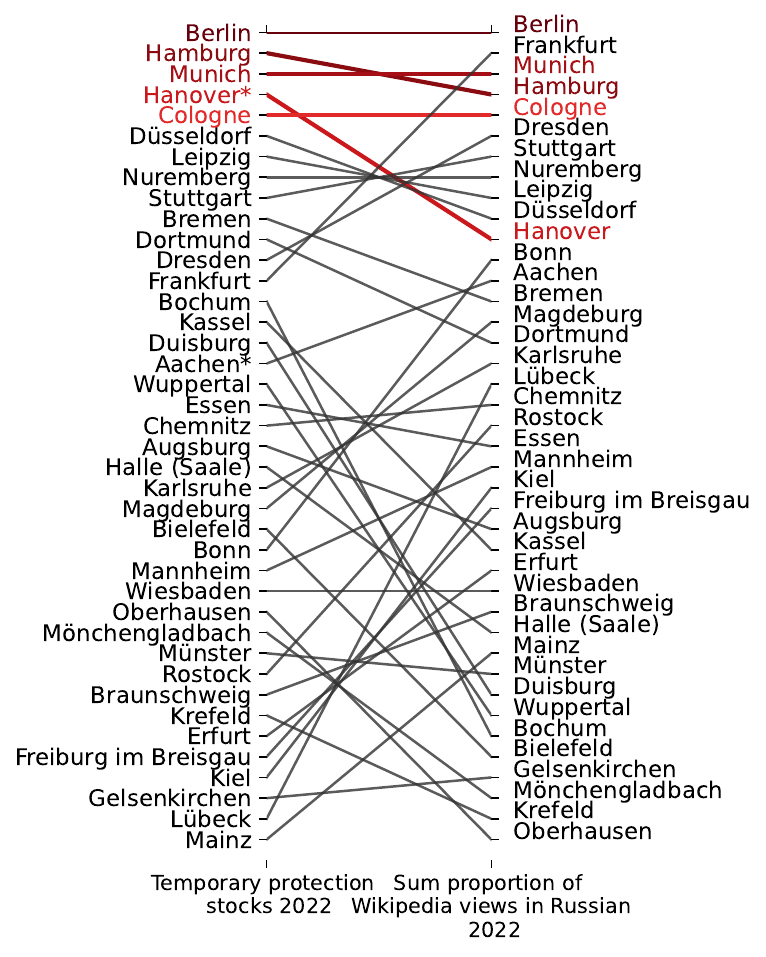}
        \caption{\textit{Russian Wikipedia 2022}}
    \end{subfigure}
    \hfill
    \begin{subfigure}[h]{0.24\textwidth}
        \centering
        \includegraphics[trim=0cm 1.8cm 0cm 0cm, clip, width=\linewidth]{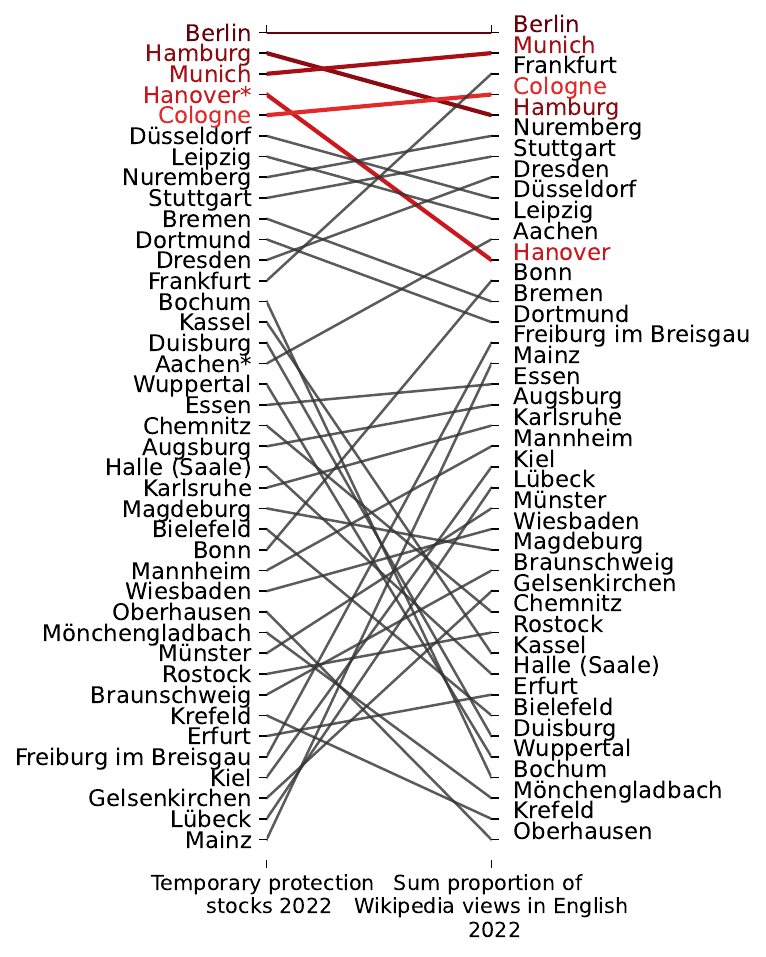}
        \caption{\textit{English Wikipedia 2022}}
    \end{subfigure}
    \hfill
    \begin{subfigure}[h]{0.24\textwidth}
        \centering
        \includegraphics[trim=0cm 1.8cm 0cm 0cm, clip, width=\linewidth]{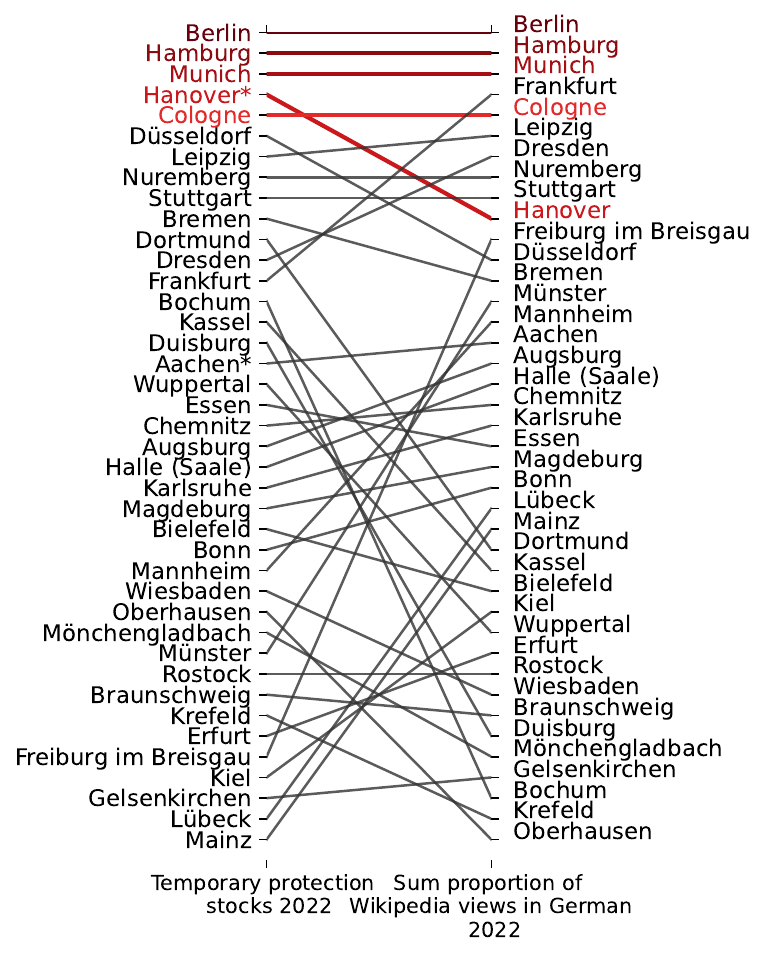}
        \caption{\textit{German Wikipedia 2022}}
    \end{subfigure}

    \begin{subfigure}[h]{0.24\textwidth}
        \centering
        \includegraphics[trim=0cm 1.8cm 0cm 0cm, clip, width=\linewidth]{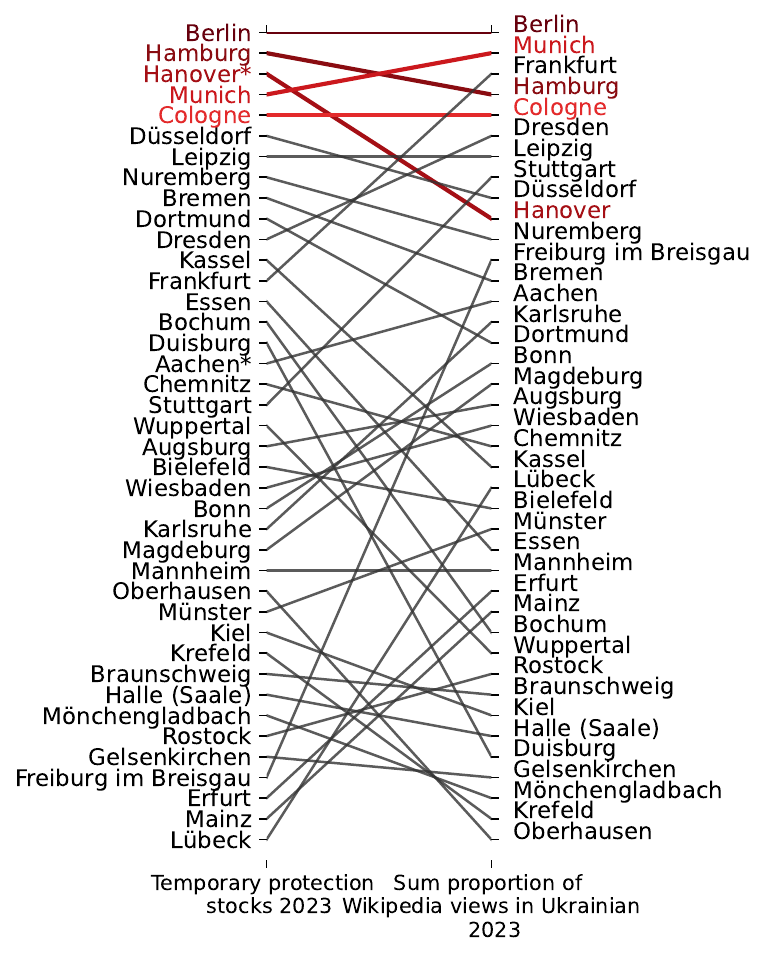}
        \caption{\textit{Ukrainian Wikipedia 2023}}
    \end{subfigure}
    \hfill
    \begin{subfigure}[h]{0.24\textwidth}
        \centering
        \includegraphics[trim=0cm 1.8cm 0cm 0cm, clip, width=\linewidth]{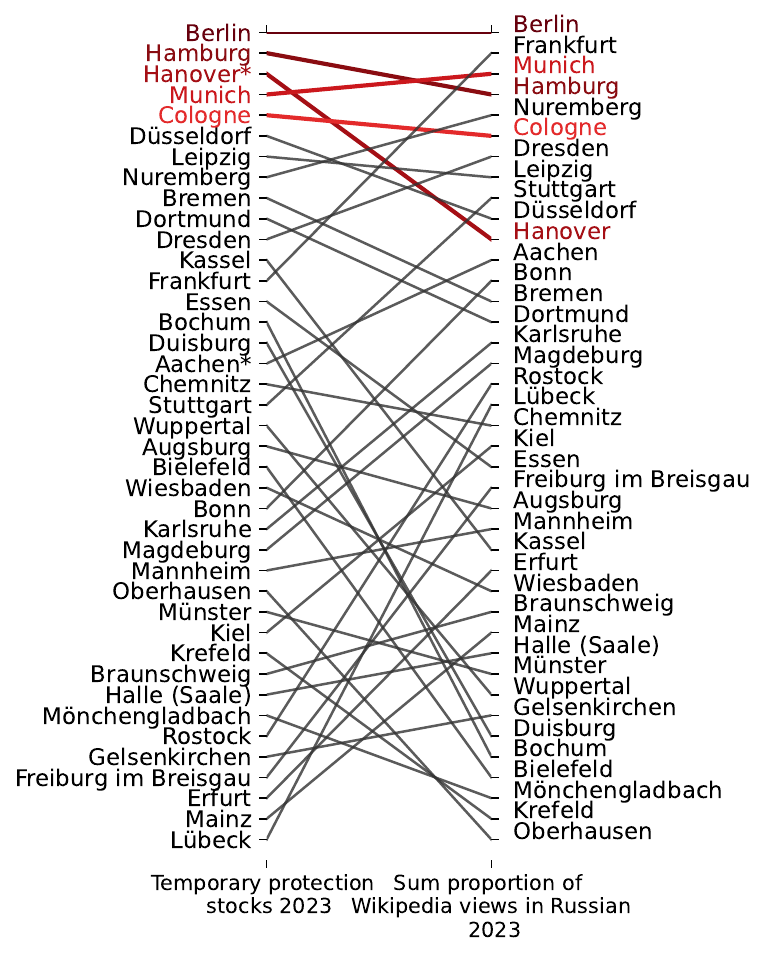}
        \caption{\textit{Russian Wikipedia 2023}}
    \end{subfigure}
    \hfill
    \begin{subfigure}[h]{0.24\textwidth}
        \centering
        \includegraphics[trim=0cm 1.8cm 0cm 0cm, clip, width=\linewidth]{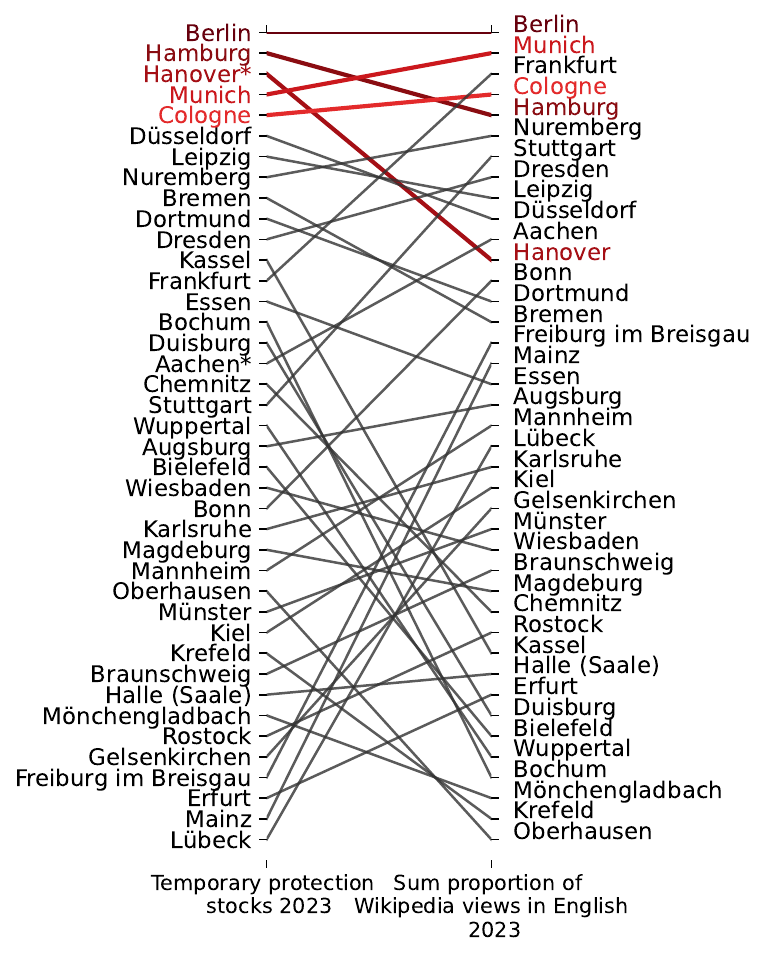}
        \caption{\textit{English Wikipedia 2023}}
    \end{subfigure}
    \hfill
    \begin{subfigure}[h]{0.24\textwidth}
        \centering
        \includegraphics[trim=0cm 1.8cm 0cm 0cm, clip, width=\linewidth]{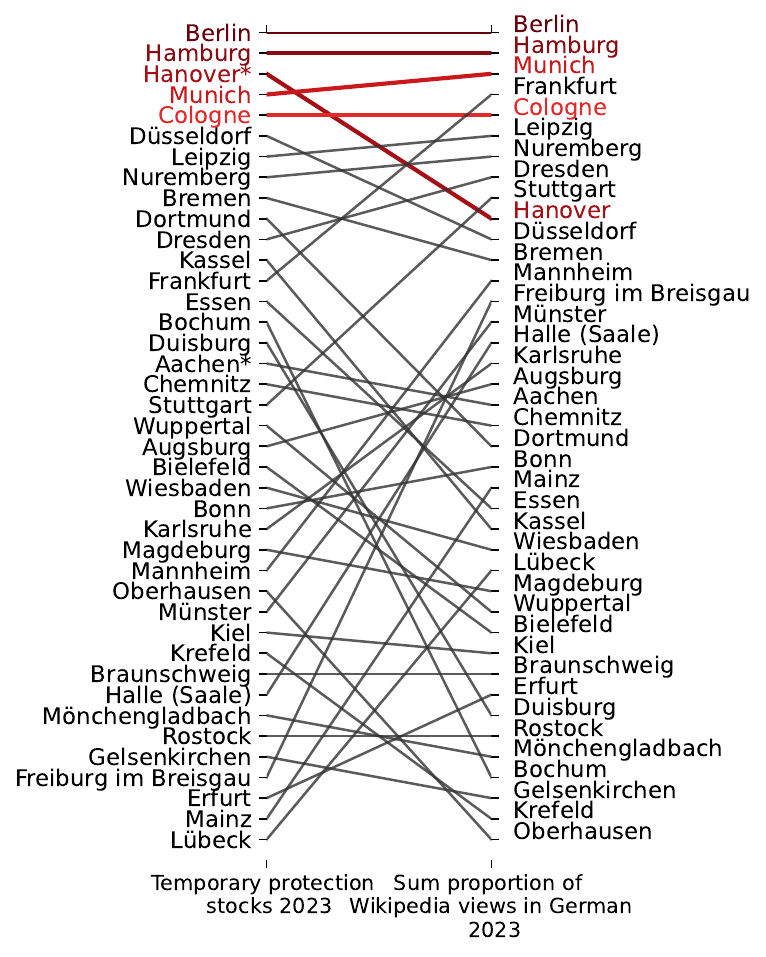}
        \caption{\textit{German Wikipedia 2023}}
    \end{subfigure}
    
    \begin{subfigure}[h]{0.24\textwidth}
        \centering
        \includegraphics[trim=0cm 1.8cm 0cm 0cm, clip, width=\linewidth]{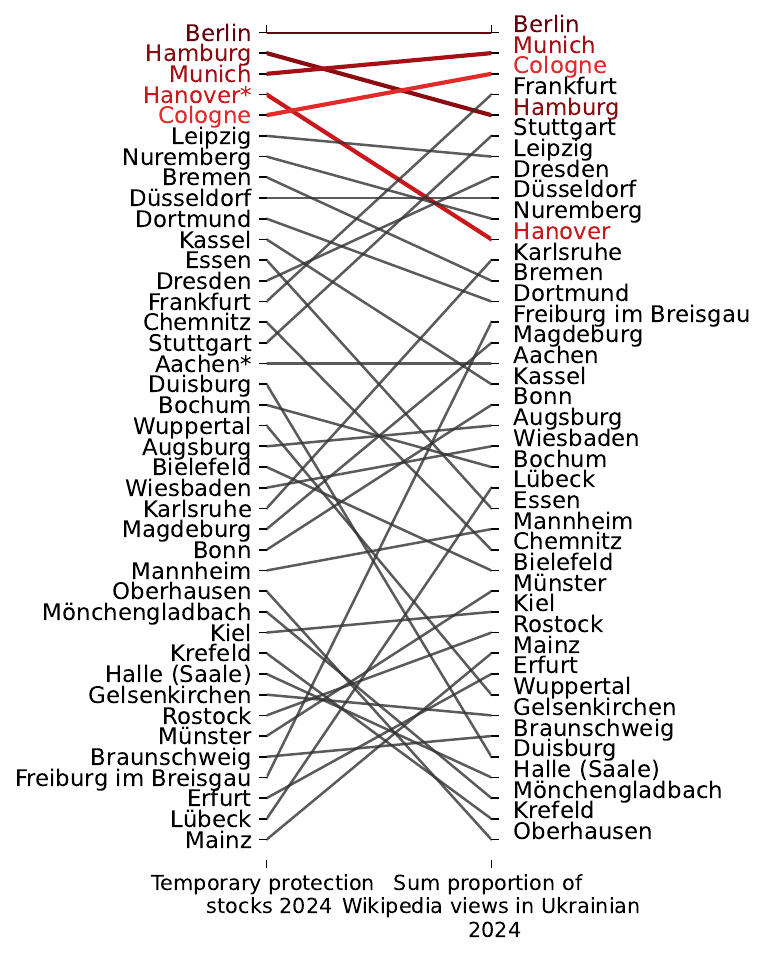}
        \caption{\textit{Ukrainian Wikipedia 2024}}
    \end{subfigure}
    \hfill
    \begin{subfigure}[h]{0.24\textwidth}
        \centering
        \includegraphics[trim=0cm 1.8cm 0cm 0cm, clip, width=\linewidth]{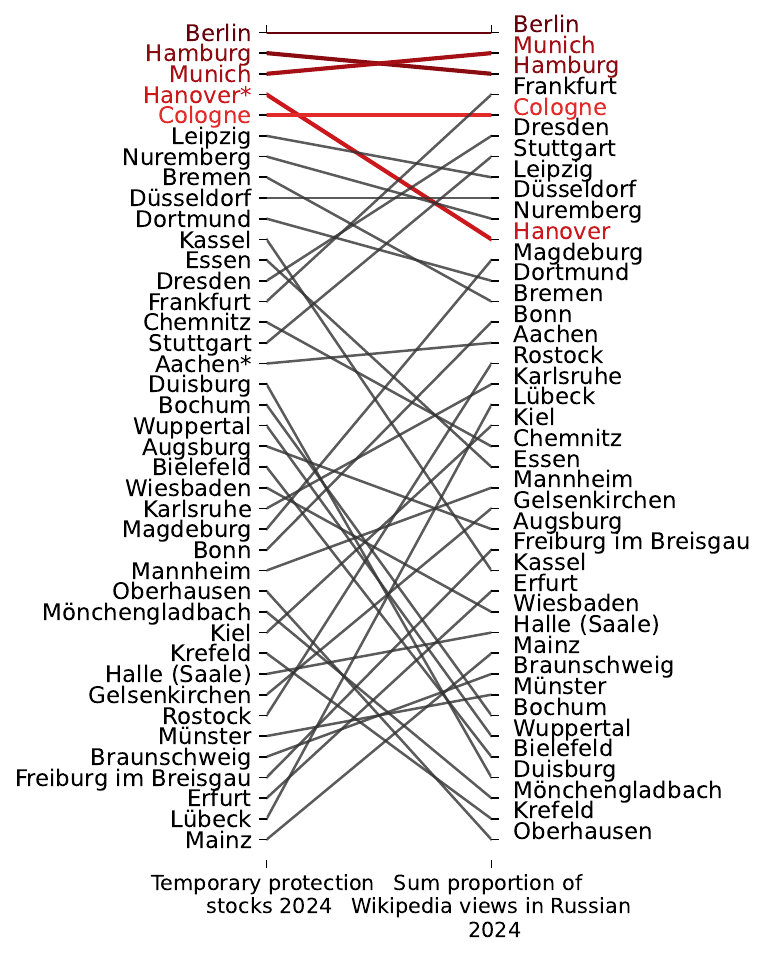}
        \caption{\textit{Russian Wikipedia 2024}}
    \end{subfigure}
    \hfill
    \begin{subfigure}[h]{0.24\textwidth}
        \centering
        \includegraphics[trim=0cm 1.8cm 0cm 0cm, clip, width=\linewidth]{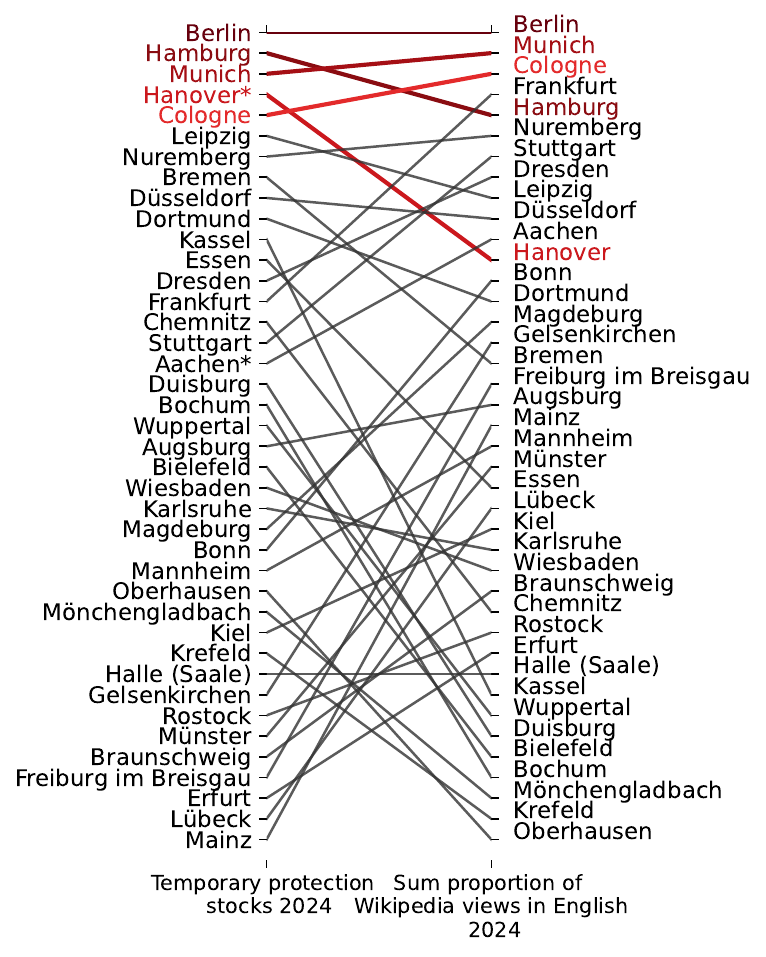}
        \caption{\textit{English Wikipedia 2024}}
    \end{subfigure}
    \hfill
    \begin{subfigure}[h]{0.24\textwidth}
        \centering
        \includegraphics[trim=0cm 1.8cm 0cm 0cm, clip, width=\linewidth]{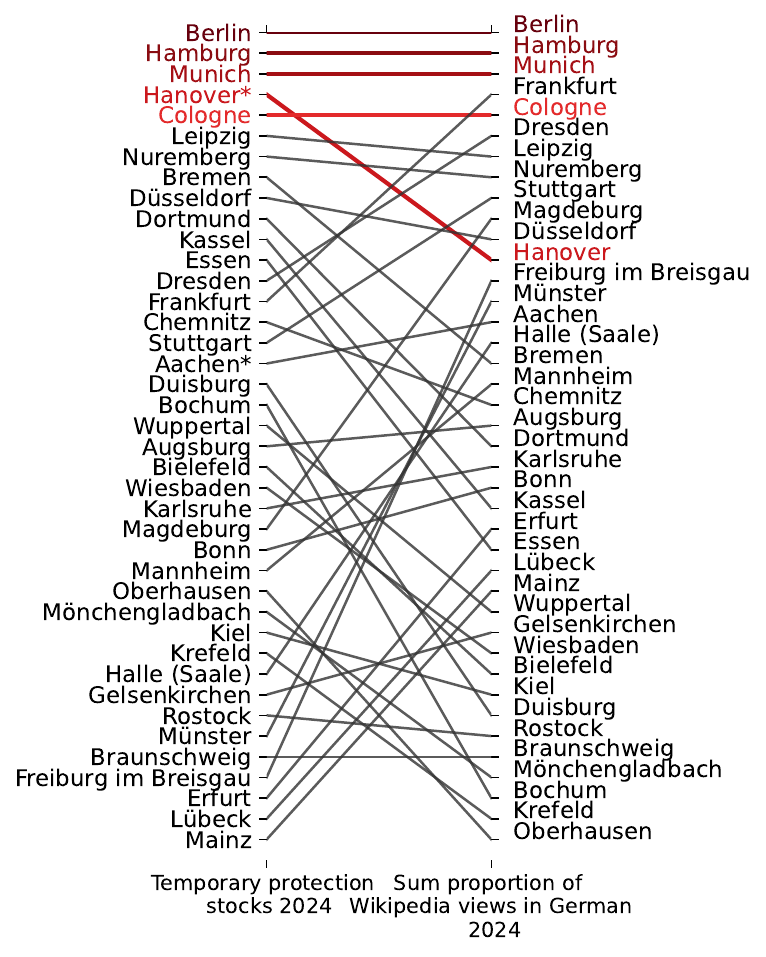}
        \caption{\textit{German Wikipedia 2024}}
    \end{subfigure}
    \caption{Correlation between rankings: stocks of Ukrainian refugees with temporary protection status in German cities (left) and the proportion of views of Wikipedia articles about the 40 most populous German cities (right), by year. The five cities hosting the largest numbers of Ukrainian refugees with temporary protection are shown in color, while the remaining cities are shown in gray. For Hanover and Aachen, data on Ukrainians under temporary protection are available only at the city-regional level (\textit{Städteregion}), rather than at the independent city level (\textit{kreisfreie Stadt}).}
    \label{fig:ranks-germany}
\end{figure*}

\begin{figure*}[ht!]
\caption*{\textbf{Relative change: Poland}}
    \centering
    \begin{subfigure}[h!]{0.19\textwidth}
        \centering
        \includegraphics[width=\linewidth]{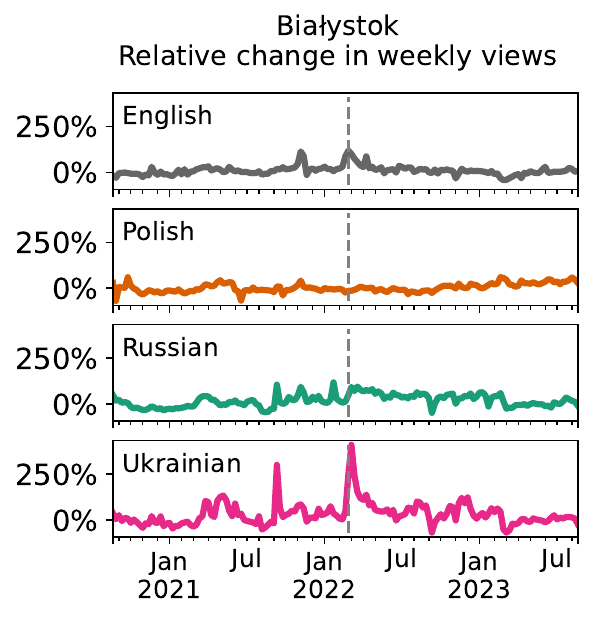}
        \caption{\textit{Białystok}}
    \end{subfigure}
    \begin{subfigure}[h!]{0.19\textwidth}
        \centering
        \includegraphics[width=\linewidth]{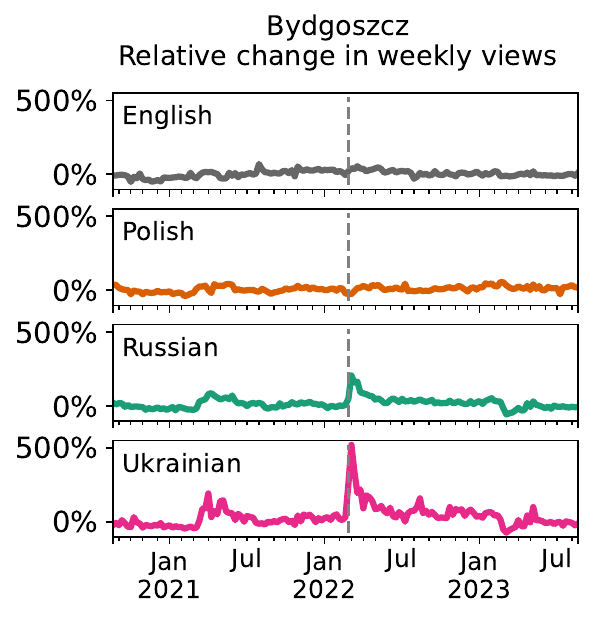}
        \caption{\textit{Bydgoszcz}}
    \end{subfigure}
    \begin{subfigure}[h!]{0.19\textwidth}
        \centering
        \includegraphics[width=\linewidth]{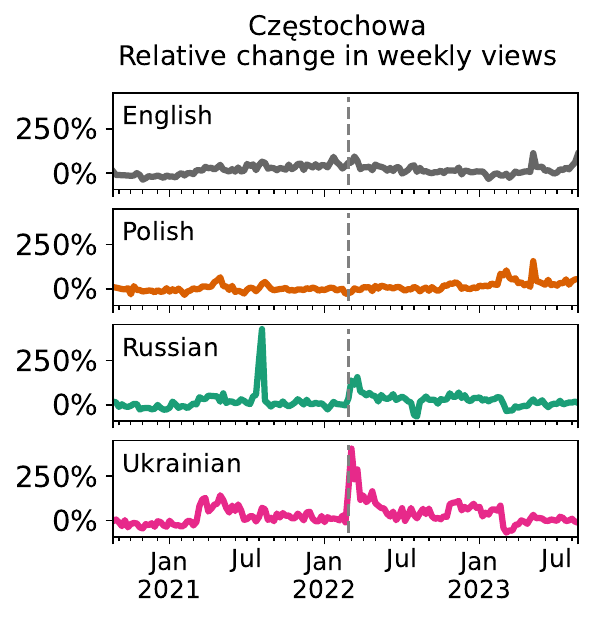}
        \caption{\textit{Częstochowa}}
    \end{subfigure}
    \begin{subfigure}[h!]{0.19\textwidth}
     \centering
     \includegraphics[width=\linewidth]{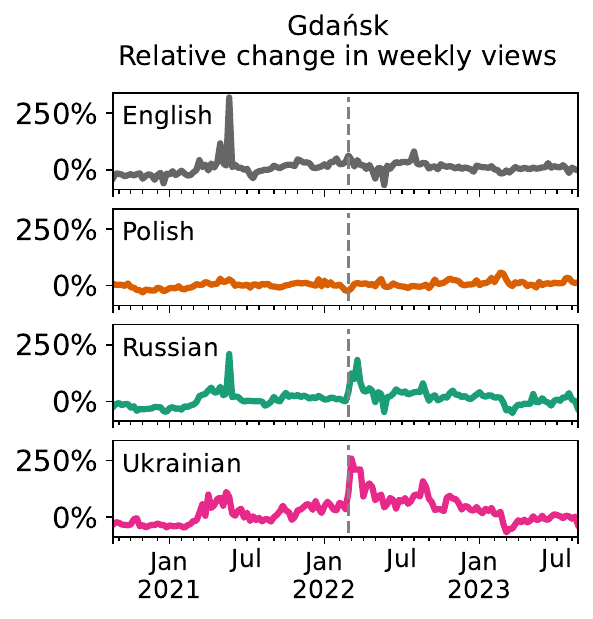}
     \caption{\textit{Gdańsk}}
    \end{subfigure}
    \begin{subfigure}[h!]{0.19\textwidth}
        \centering
        \includegraphics[width=\linewidth]{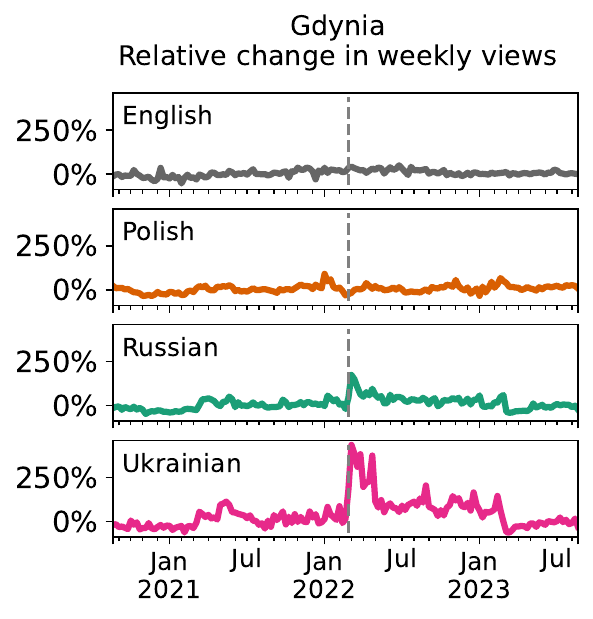}
        \caption{\textit{Gdynia}}
    \end{subfigure}
    \begin{subfigure}[h!]{0.19\textwidth}
        \centering
        \includegraphics[width=\linewidth]{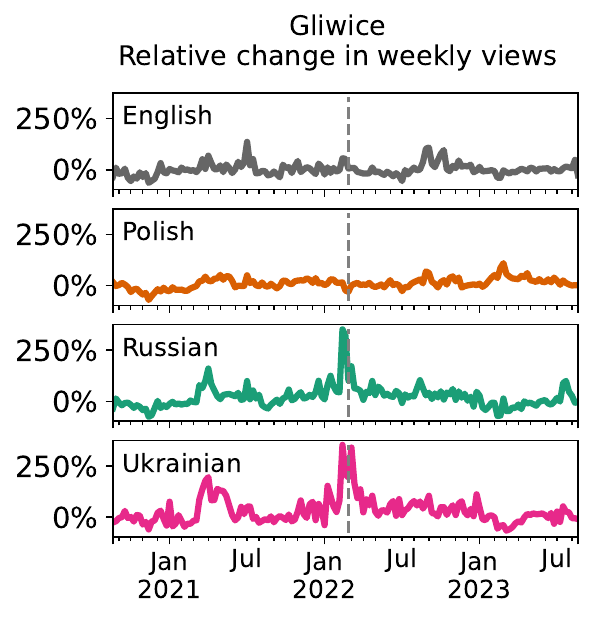}
        \caption{\textit{Gliwice}}
    \end{subfigure}
    \begin{subfigure}[h!]{0.19\textwidth}
        \centering
        \includegraphics[width=\linewidth]{figs/prop-relative-change-Katowice.pdf}
        \caption{\textit{Katowice}}
    \end{subfigure}
    \begin{subfigure}[h!]{0.19\textwidth}
        \centering
        \includegraphics[width=\linewidth]{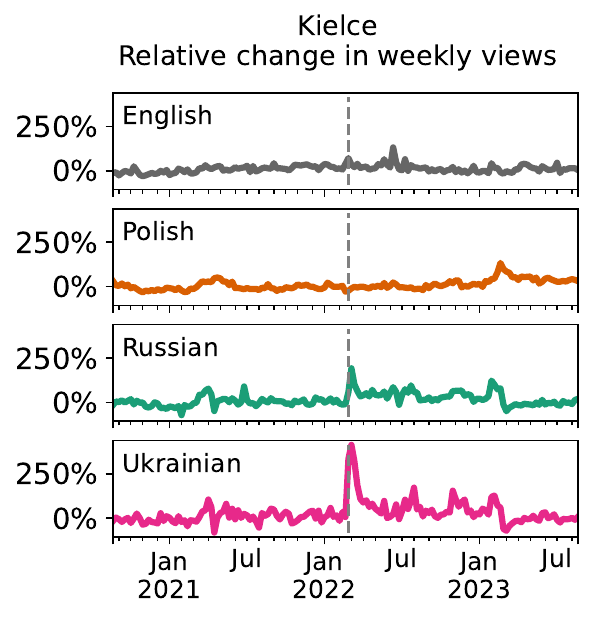}
        \caption{\textit{Kielce}}
    \end{subfigure}
    \begin{subfigure}[h!]{0.19\textwidth}
     \centering
     \includegraphics[width=\linewidth]{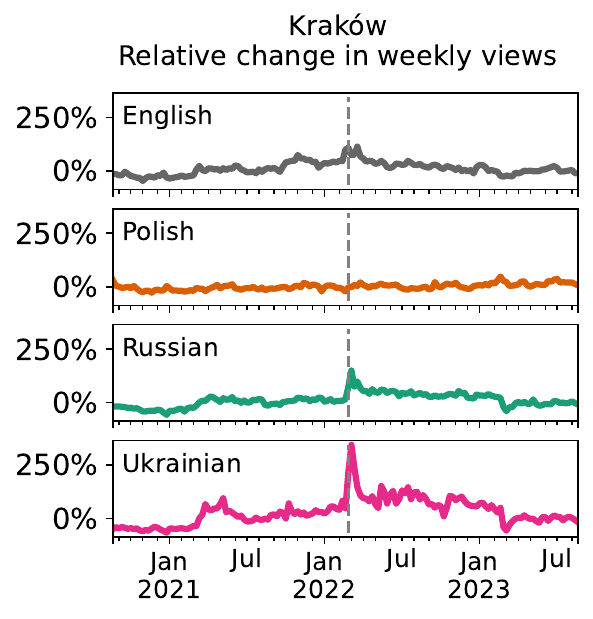}
     \caption{\textit{Kraków}}
    \end{subfigure}
    \begin{subfigure}[h!]{0.19\textwidth}
        \centering
        \includegraphics[width=\linewidth]{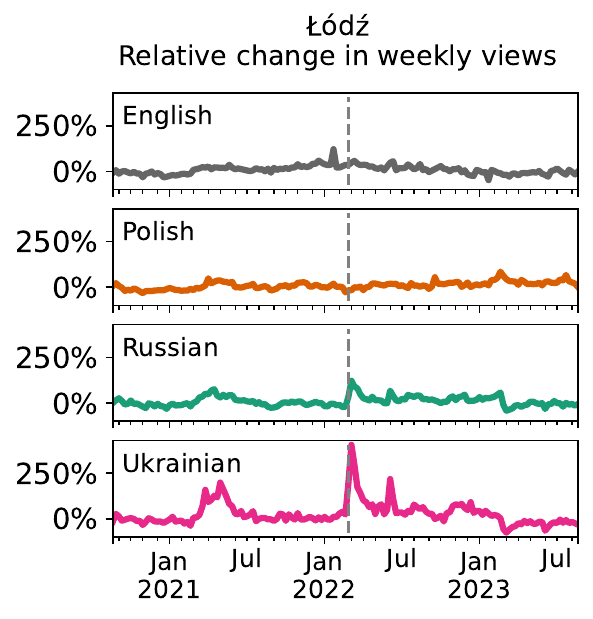}
        \caption{\textit{Łódź}}
    \end{subfigure}
    \begin{subfigure}[h!]{0.19\textwidth}
        \centering
        \includegraphics[width=\linewidth]{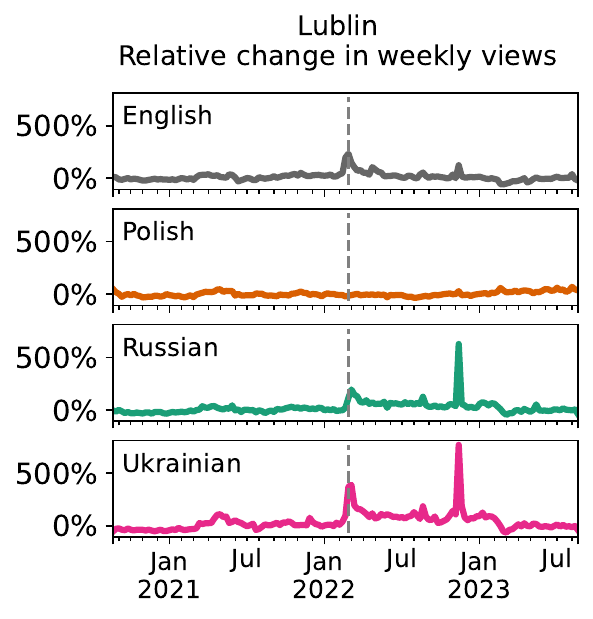}
        \caption{\textit{Lublin}}
    \end{subfigure}
    \begin{subfigure}[h!]{0.19\textwidth}
        \centering
        \includegraphics[width=\linewidth]{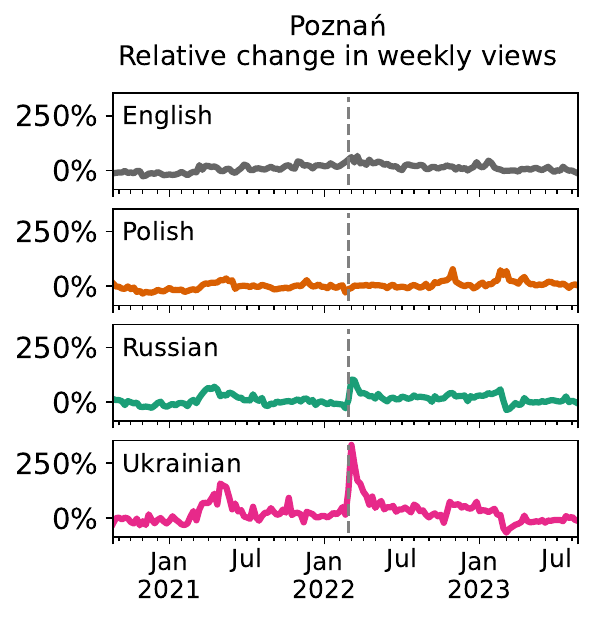}
        \caption{\textit{Poznań}}
    \end{subfigure}
    \begin{subfigure}[h!]{0.19\textwidth}
        \centering
        \includegraphics[width=\linewidth]{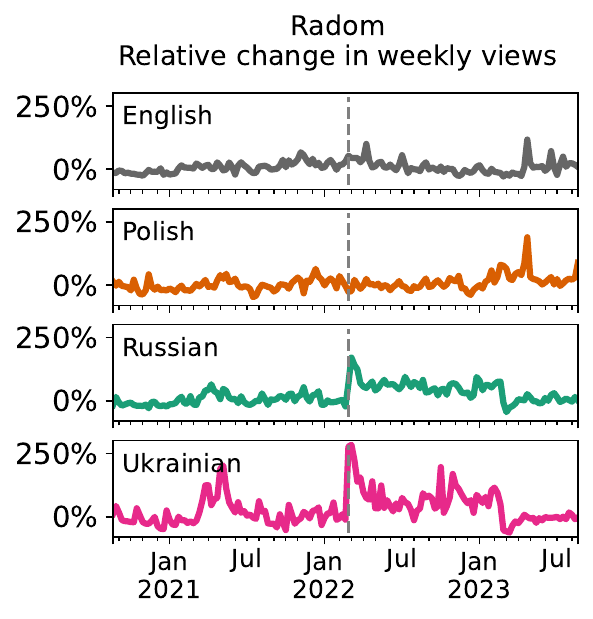}
        \caption{\textit{Radom}}
    \end{subfigure}
    \begin{subfigure}[h!]{0.19\textwidth}
        \centering
        \includegraphics[width=\linewidth]{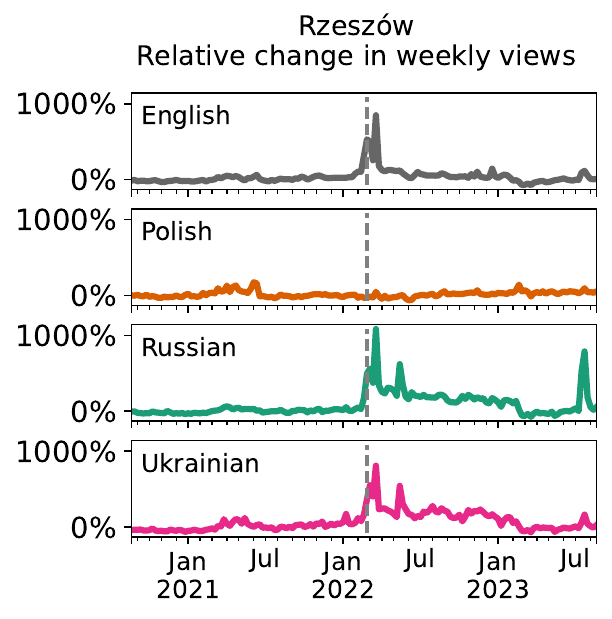}
        \caption{\textit{Rzeszów}}
    \end{subfigure}
    \begin{subfigure}[h!]{0.19\textwidth}
        \centering
        \includegraphics[width=\linewidth]{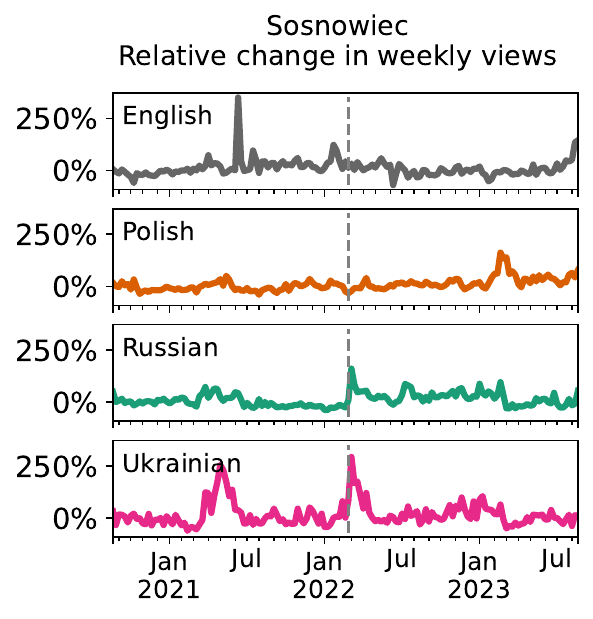}
        \caption{\textit{Sosnowiec}}
    \end{subfigure}
    \begin{subfigure}[h!]{0.19\textwidth}
        \centering
        \includegraphics[width=\linewidth]{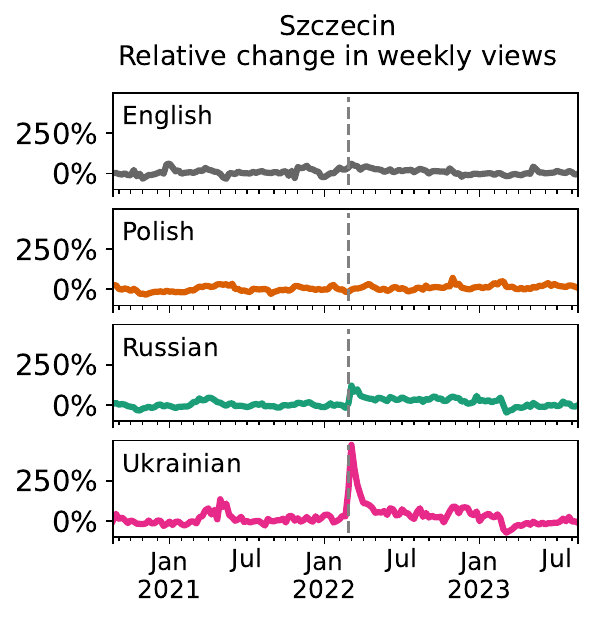}
        \caption{\textit{Szczecin}}
    \end{subfigure}
    \begin{subfigure}[h!]{0.19\textwidth}
        \centering
        \includegraphics[width=\linewidth]{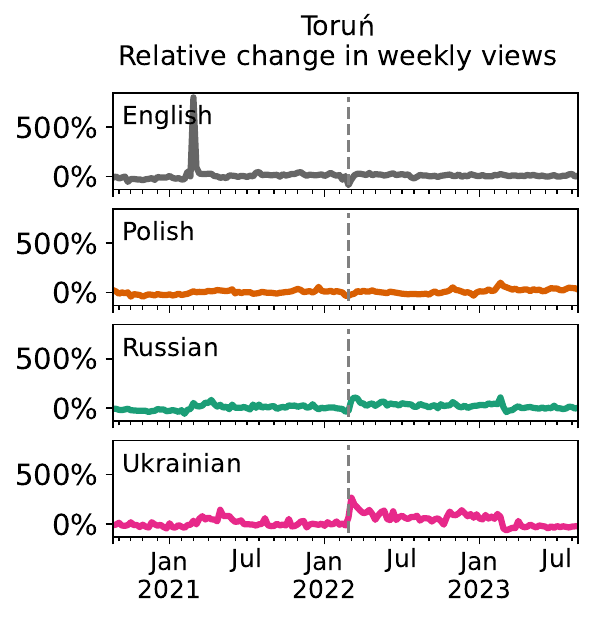}
        \caption{\textit{Toruń}}
    \end{subfigure}
    \begin{subfigure}[h!]{0.19\textwidth}
        \centering
        \includegraphics[width=\linewidth]{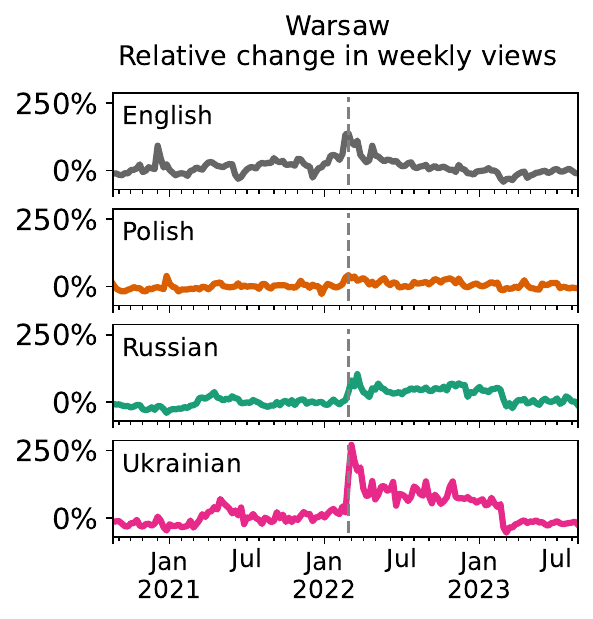}
        \caption{\textit{Warsaw}}
    \end{subfigure}
    \begin{subfigure}[h!]{0.19\textwidth}
        \centering
        \includegraphics[width=\linewidth]{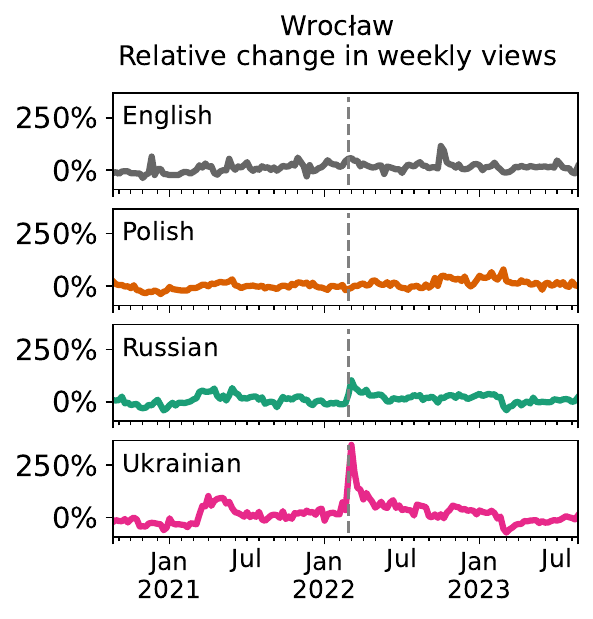}
        \caption{\textit{Wrocław}}
    \end{subfigure}
    \caption{Relative change in the proportion of weekly views, compared to the same period in the previous year, of Wikipedia articles about the 19 most populous Polish cities across four languages (English, Polish, Russian, and Ukrainian) from August 24, 2020, to August 24, 2023.}
    \label{fig:rel-changes-appendix}
\end{figure*}

\begin{figure*}[ht!]
\centering
    \includegraphics[width=\linewidth]{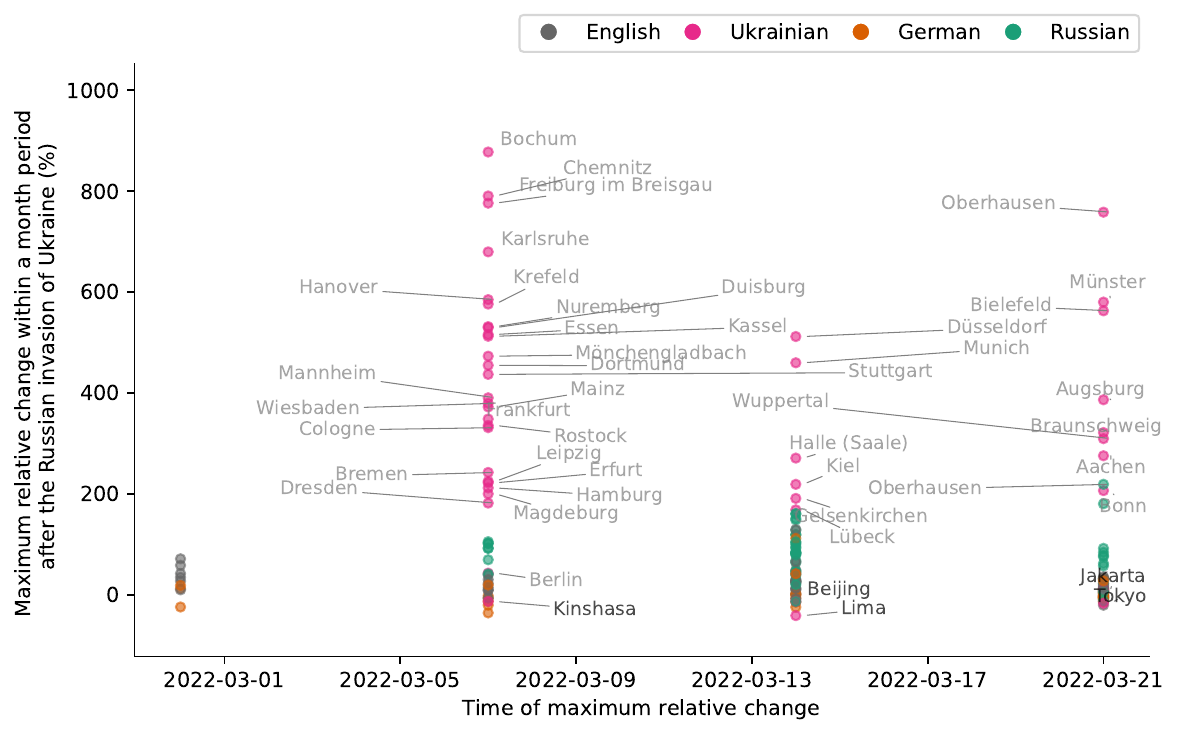}
    \caption{Maximum relative change in the proportion of weekly views over the month following the Russian invasion of Ukraine, compared to the same period in the previous year. Results are shown for Wikipedia articles about the 40 most populous German cities and five of the most populous cities in the world (Beijing, Jakarta, Kinshasa, Lima, and Tokyo) across four languages (English, German, Russian, and Ukrainian).}
    \label{fig:rel_change_de}
\end{figure*}

\newpage
\begin{table}[ht!]
\caption*{\textbf{Structural breaks: Poland}}
\footnotesize
\begin{tabular}{llll}
\hline
\textbf{City}  & \textbf{Break Date} & \textbf{CI lower} & \textbf{CI upper}  \\ \toprule
Bydgoszcz   & \textbf{2022-02-24} & 2022-02-14 & 2022-03-10\\
            & 2022-08-31 & 2022-08-16 & 2022-09-06\\ \hline
Częstochowa & \textbf{2022-03-01} & 2022-02-07 & 2022-03-08\\
            & 2022-08-28 & 2022-08-22 & 2022-09-16\\ \hline
Gdańsk      & \textbf{2022-03-01} & 2022-02-11 & 2022-03-06\\
            & 2022-08-23 & 2022-08-15 & 2022-09-03\\\hline
Gdynia      & \textbf{2022-03-01} & 2022-02-20 & 2022-03-07\\
            & 2022-09-04 & 2022-08-22 & 2022-09-08\\ \hline
Gliwice     & \textbf{2022-02-16} & 2022-02-07 & 2022-03-05\\
            & 2022-08-31 & 2022-08-18 & 2022-09-12\\ \hline
Katowice    & \textbf{2022-02-27} & 2022-02-22 & 2022-03-12\\
            & 2022-09-04 & 2022-08-07 & 2022-09-15\\ \hline
Kielce      & \textbf{2022-02-26} & 2022-02-20 & 2022-03-17\\
            & 2022-09-05 & 2022-08-17 & 2022-09-17\\ \hline
Kraków      & \textbf{2022-02-25} & 2022-02-20 & 2022-03-02\\
            & 2022-09-04 & 2022-08-10 & 2022-09-17\\ \hline
Łódź        & 2022-06-11 & 2022-06-01 & 2022-07-12\\ 
            & 2022-11-22 & 2022-11-02 & 2023-01-21\\ \hline
Lublin      & 2022-06-03 & 2022-04-26 & 2022-06-04\\
            & 2022-11-19 & 2022-11-18 & 2023-01-15\\ \hline
Poznań      & \textbf{2022-02-27} & 2022-02-16 & 2022-03-29\\
            & 2022-09-04 & 2022-07-09 & 2022-10-31\\ \hline
Radom       & \textbf{2022-02-25} & 2022-02-21 & 2022-02-28\\
            & 2022-08-31 & 2022-08-18 & 2022-09-11\\ \hline
Kraków      & \textbf{2022-02-25} & 2022-02-20 & 2022-03-02\\
            & 2022-09-04 & 2022-08-10 & 2022-09-17\\ \hline
Rzeszów     & \textbf{2022-03-14} & NA & NA\\
            & 2022-09-04 & 2022-08-22 & 2022-10-29\\ \hline
Sosnowiec   & \textbf{2022-03-02} & 2022-02-22 & 2022-03-09\\
            & 2022-09-18 & 2022-09-06 & 2022-10-02\\ \hline
Szczecin   & \textbf{2022-02-25} & 2022-02-12 & 2022-03-12\\
            & 2022-09-04 & 2022-08-21 & 2022-09-12\\ \hline
Toruń       & \textbf{2022-03-01} & 2022-02-21 & 2022-03-06\\
            & 2022-08-31 & 2022-08-20 & 2022-09-11\\ \hline
Warsaw      & \textbf{2022-02-28} & 2022-02-23 & 2022-03-03\\
            & 2022-09-04 & 2022-08-12 & 2022-10-01\\ \hline
Wrocław     & \textbf{2022-02-27} & 2022-02-09 & 2022-05-17\\ \hline
\bottomrule
\end{tabular}
\caption{Results of the structural break analysis using an autoregressive model (AR(1)) on the time series of the proportion of daily views of Ukrainian-language Wikipedia articles about the 19 most populous Polish cities. Only break points detected in 2022 are reported. For each city, the table reports the estimated break date in the second column, with the third and fourth columns indicating the lower and upper bounds of the corresponding confidence interval. Structural breaks identified within one month before or after the start of the Russian invasion of Ukraine (February 24, 2022) are shown in bold. Confidence intervals for break points are calculated by examining how the model fit improves when the relevant break point is shifted. For Rzeszów, the time series before the first estimated break point is highly monotonous, resulting in a distribution of the estimated break point with excessive probability at the boundary. When statistically meaningful confidence intervals cannot be computed, they are reported as NA.}
\label{tab:breaks_pol}
\end{table}

\begin{table}[ht!]
\caption*{\textbf{Structural breaks: World}}
\footnotesize
\begin{tabular}{llll}
\hline
\textbf{City}  & \textbf{Break Date} & \textbf{CI lower} & \textbf{CI upper}  \\ \toprule
Beijing & 2021-08-25 & 2021-08-05 & 2021-09-13\\
        & \textbf{2022-02-05} & 2022-01-15 & 2022-04-17\\
        & 2022-08-02 & 2022-05-03 & 2022-09-11\\ \hline
Jakarta & 2022-01-14 & 2021-10-04 & 2022-01-23\\
        & 2022-06-29 & 2022-06-27 & 2022-10-04\\ \hline
Kinshasa & 2022-11-04 & 2022-10-12 & 2022-12-04\\ \hline
Lima     & NA & NA & NA \\ \hline
Tokyo   & 2021-02-13 & 2021-01-24 & 2021-03-11\\
        & 2021-07-28 & 2021-07-14 & 2021-08-09\\
        & 2022-12-30 & 2022-12-09 & 2023-02-03\\ \hline
\bottomrule
\end{tabular}
\caption{Sensitivity checks in the structural break analysis. We conducted a structural break analysis using an autoregressive model (AR(1)) on the time series of the proportion of daily views of Ukrainian-language Wikipedia articles corresponding to five of the most populous capitals in the world. For each city, the table reports the estimated break date in the second column, with the third and fourth columns indicating the lower and upper bounds of the corresponding confidence interval. Structural breaks occurring within one month before or after the start of the Russian invasion of Ukraine (February 24, 2022) are shown in bold. A structural break was detected only for Beijing, about three weeks before the invasion (February 5, 2022). However, this break was followed by a decline rather than an increase in the proportion of Ukrainian-language views. For Lima, no statistically meaningful structural breaks were detected within the observed period, and the results are reported as NA.}
\label{tab:breaks_w}
\end{table}

\begin{table}[ht!]
\caption*{\textbf{Structural breaks: Germany}}
\footnotesize
\begin{tabular}{llll}
\toprule
\textbf{City}  & \textbf{Break Date} & \textbf{CI lower} & \textbf{CI upper}  \\ \midrule
Aachen      & 2022-05-20 & 2022-02-25 & 2022-05-21\\ \hline
Augsburg    & \textbf{2022-03-02} & 2022-02-25 & 2022-03-04\\
            & 2022-08-29 & 2022-07-13 & 2022-09-16\\ \hline
Berlin      & \textbf{2022-03-12} & 2022-01-29 & 2022-04-02\\ \hline
Bielefeld   & \textbf{2022-03-02} & 2022-02-26 & 2022-03-05\\
            & 2022-09-01 & 2022-08-23 & 2022-09-12\\ \hline
Bochum      & \textbf{2022-03-03} & 2022-02-25 & 2022-03-04\\
            & 2022-09-09 & 2022-08-06 & 2022-10-25\\ \hline
Bonn        & \textbf{2022-03-05} & 2022-02-28 & 2022-03-07\\
            & 2022-09-04 & 2022-08-31 & 2022-09-12\\ \hline
Braunschweig & \textbf{2022-03-03} & 2022-02-27 & 2022-03-06\\ 
            & 2022-09-05 & 2022-08-27 & 2022-09-13\\ \hline
Bremen      & \textbf{2022-03-02} & 2022-02-27 & 2022-03-03\\
            & 2022-08-31 & 2022-08-27 & 2022-09-06\\ \hline
Chemnitz    & \textbf{2022-03-04} & 2022-02-07 & 2022-03-14\\
            & 2022-09-04 & 2022-08-29 & 2022-09-16\\ \hline
Cologne     & \textbf{2022-03-02} & 2022-02-05 & 2022-03-12\\
            & 2022-09-04 & 2022-08-25 & 2022-09-16\\ \hline
Dortmund    & \textbf{2022-03-03} & 2022-02-22 & 2022-03-13\\
            & 2022-09-04 & 2022-08-23 & 2022-09-15\\ \hline
Dresden     & \textbf{2022-02-28} & 2022-02-24 & 2022-03-04\\
            & 2022-09-06 & 2022-08-14 & 2022-09-21\\ \hline
Duisburg    & \textbf{2022-03-05} & 2022-02-27 & 2022-03-07\\
            & 2022-10-30 & 2022-10-04 & 2022-11-11\\ \hline
Düsseldorf  & \textbf{2022-03-03} & 2022-02-13 & 2022-03-10\\
            & 2022-09-01 & 2022-08-15 & 2022-09-21\\ \hline
Erfurt      & \textbf{2022-03-02} & 2022-02-26 & 2022-03-07\\
            & 2022-09-01 & 2022-08-19 & 2022-09-11\\ \hline
Essen       & \textbf{2022-03-02} & 2022-02-26 & 2022-03-04\\
            & 2022-09-04 & 2022-08-25 & 2022-09-16\\ \hline
Frankfurt   & \textbf{2022-03-20} & 2022-02-25 & 2022-03-21\\
            & 2022-09-01 & 2022-08-30 & 2022-10-11\\ \hline
Freiburg & \textbf{2022-02-28} & 2022-02-23 & 2022-03-01\\
im Breisgau & 2022-09-04 & 2022-08-25 & 2022-09-25\\ \hline
Gelsenkirchen & \textbf{2022-03-08} & 2022-02-20 & 2022-03-15\\ \hline
Halle       & \textbf{2022-03-05} & 2022-03-01 & 2022-03-08\\
            & 2022-09-01 & 2022-08-23 & 2022-09-30\\ \hline
Hamburg     & \textbf{2022-03-02} & 2022-02-26 & 2022-03-04\\
            & 2022-09-04 & 2022-08-25 & 2022-09-21\\ \hline
Hanover     & \textbf{2022-03-01} & 2022-02-25 & 2022-03-06\\
            & 2022-08-29 & 2022-08-03 & 2022-09-26\\ \hline
Karlsruhe   & \textbf{2022-03-02} & 2022-02-25 & 2022-03-06\\
            & 2022-09-04 & 2022-08-24 & 2022-09-11\\ \hline
Kassel      & \textbf{2022-03-03} & 2022-02-25 & 2022-03-05\\
            & 2022-09-04 & 2022-08-31 & 2022-09-13\\ \hline
Kiel        & \textbf{2022-03-03} & 2022-02-22 & 2022-03-10\\
            & 2022-09-04 & 2022-08-31 & 2022-09-17\\ \hline
Krefeld     & \textbf{2022-02-23} & 2022-02-13 & 2022-02-26\\
            & 2022-09-18 & 2022-09-04 & 2022-10-17\\ \hline
Leipzig     & \textbf{2022-03-01} & 2022-02-03 & 2022-03-05\\
            & 2022-09-07 & 2022-08-31 & 2022-09-24\\ \hline
Lübeck      & \textbf{2022-03-09} & 2022-03-02 & 2022-03-13\\
            & 2022-09-06 & 2022-09-01 & 2022-09-15\\ \hline
\vdots      & \vdots     & \vdots     & \vdots      \\ \hline
\end{tabular}
\end{table}

\begin{table}[ht!]
\footnotesize
\begin{tabular}{llll}
\toprule
\textbf{City}  & \textbf{Break Date} & \textbf{CI lower} & \textbf{CI upper}  \\ \midrule
\vdots      & \vdots     & \vdots     & \vdots      \\ \hline
Magdeburg   & \textbf{2022-03-03} & 2022-02-25 & 2022-03-05\\
            & 2022-09-04 & 2022-08-28 & 2022-09-20\\ \hline
Mainz       & \textbf{2022-03-03} & 2022-02-27 & 2022-03-06\\
            & 2022-09-03 & 2022-08-26 & 2022-09-18\\ \hline
Mannheim    & \textbf{2022-03-05} & 2022-02-27 & 2022-03-07\\
            & 2022-09-04 & 2022-08-30 & 2022-09-18\\ \hline
Mönchengladbach & \textbf{2022-03-04} & 2022-01-22 & 2022-09-18\\ \hline
Munich      & \textbf{2022-02-28} & 2022-02-18 & 2022-03-15\\
            & 2022-08-28 & 2022-08-07 & 2022-09-08\\ \hline
Münster     & \textbf{2022-03-02} & 2022-02-25 & 2022-03-04\\
            & 2022-08-31 & 2022-08-17 & 2022-09-16\\ \hline
Nuremberg   & \textbf{2022-02-28} & 2022-02-24 & 2022-03-06\\
            & 2022-09-04 & 2022-08-27 & 2022-09-13\\ \hline
Oberhausen  & \textbf{2022-03-03} & 2022-02-25 & 2022-03-06\\
            & 2022-09-07 & 2022-05-28 & 2022-12-09\\ \hline
Rostock     & \textbf{2022-03-02} & 2022-02-21 & 2022-03-14\\
            & 2022-09-03 & 2022-08-20 & 2022-09-22\\ \hline
Stuttgart   & \textbf{2022-03-02} & 2022-02-27 & 2022-03-04\\
            & 2022-09-03 & 2022-08-28 & 2022-09-10\\ \hline
Wiesbaden   & \textbf{2022-03-06} & 2022-03-02 & 2022-03-09\\
            & 2022-09-04 & 2022-08-21 & 2022-09-20\\ \hline
Wuppertal   & \textbf{2022-03-01} & 2022-02-21 & 2022-03-03\\
            & 2022-09-11 & 2022-08-20 & 2022-10-06\\ \hline
\bottomrule
\end{tabular}
\caption{Results of the structural break analysis using an autoregressive model (AR(1)) on the time series of the proportion of daily views of Ukrainian-language Wikipedia articles about the 40 most populous German cities. Only break points detected in 2022 are reported. For each city, the table reports the estimated break date in the second column, with the third and fourth columns indicating the lower and upper bounds of the corresponding confidence interval. When a structural break occurred within one month before or after the start of the Russian invasion of Ukraine (February 24, 2022), the date is shown in bold.}
\label{tab:breaks_ger}
\end{table}

\newpage
\begin{figure*}[ht!]
    \caption*{\textbf{Wikipedia proportion of daily views and UNHCR data on Ukrainian refugees crossing the border to Poland}}
    \centering
    \begin{subfigure}[h]{0.19\textwidth}
        \centering
        \includegraphics[width=\linewidth]{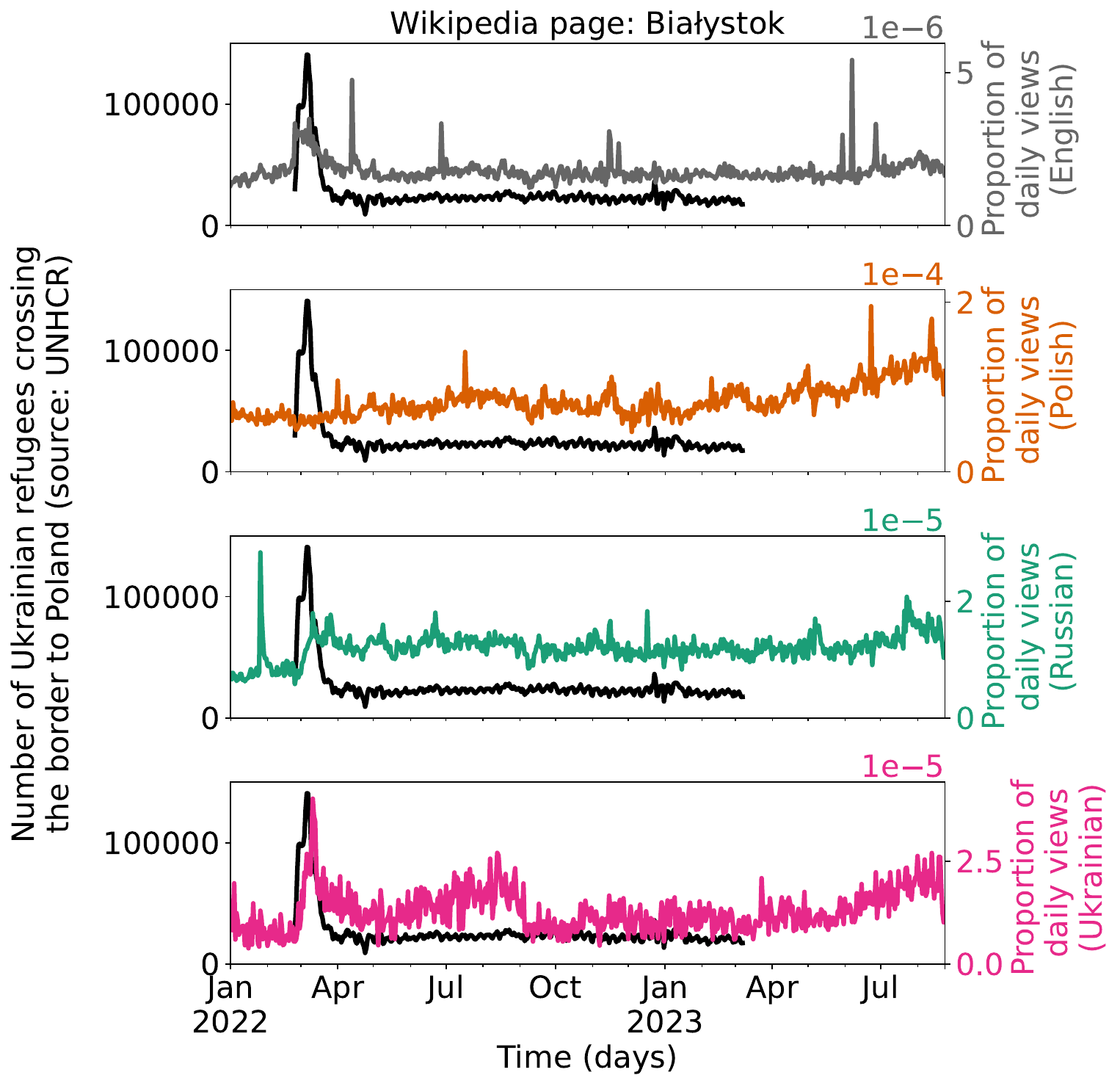}
        \caption{\textit{Białystok}}
    \end{subfigure}
    \begin{subfigure}[h]{0.19\textwidth}
        \centering
        \includegraphics[width=\linewidth]{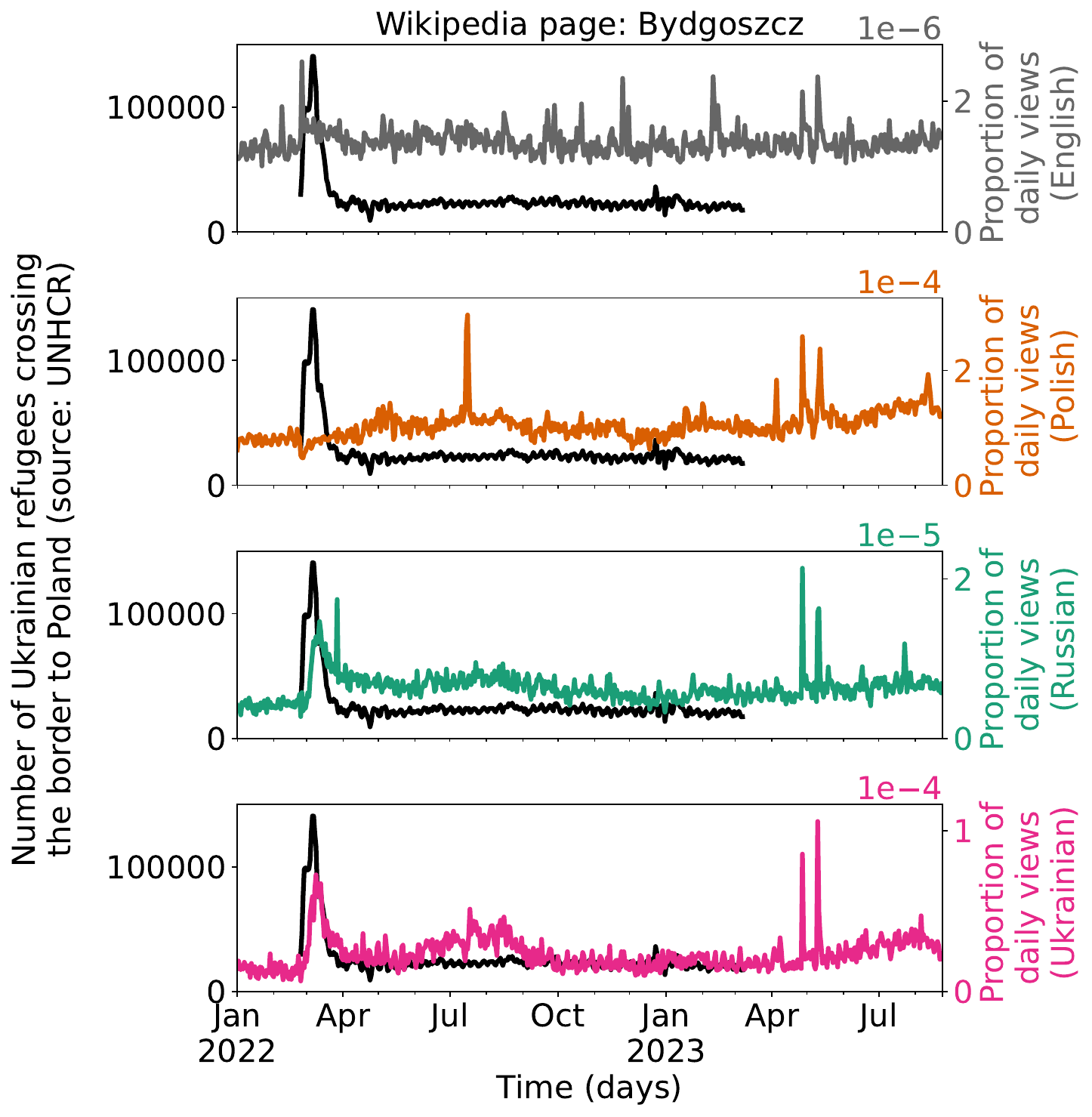}
        \caption{\textit{Bydgoszcz}}
    \end{subfigure}
    \begin{subfigure}[h]{0.19\textwidth}
        \centering
        \includegraphics[width=\linewidth]{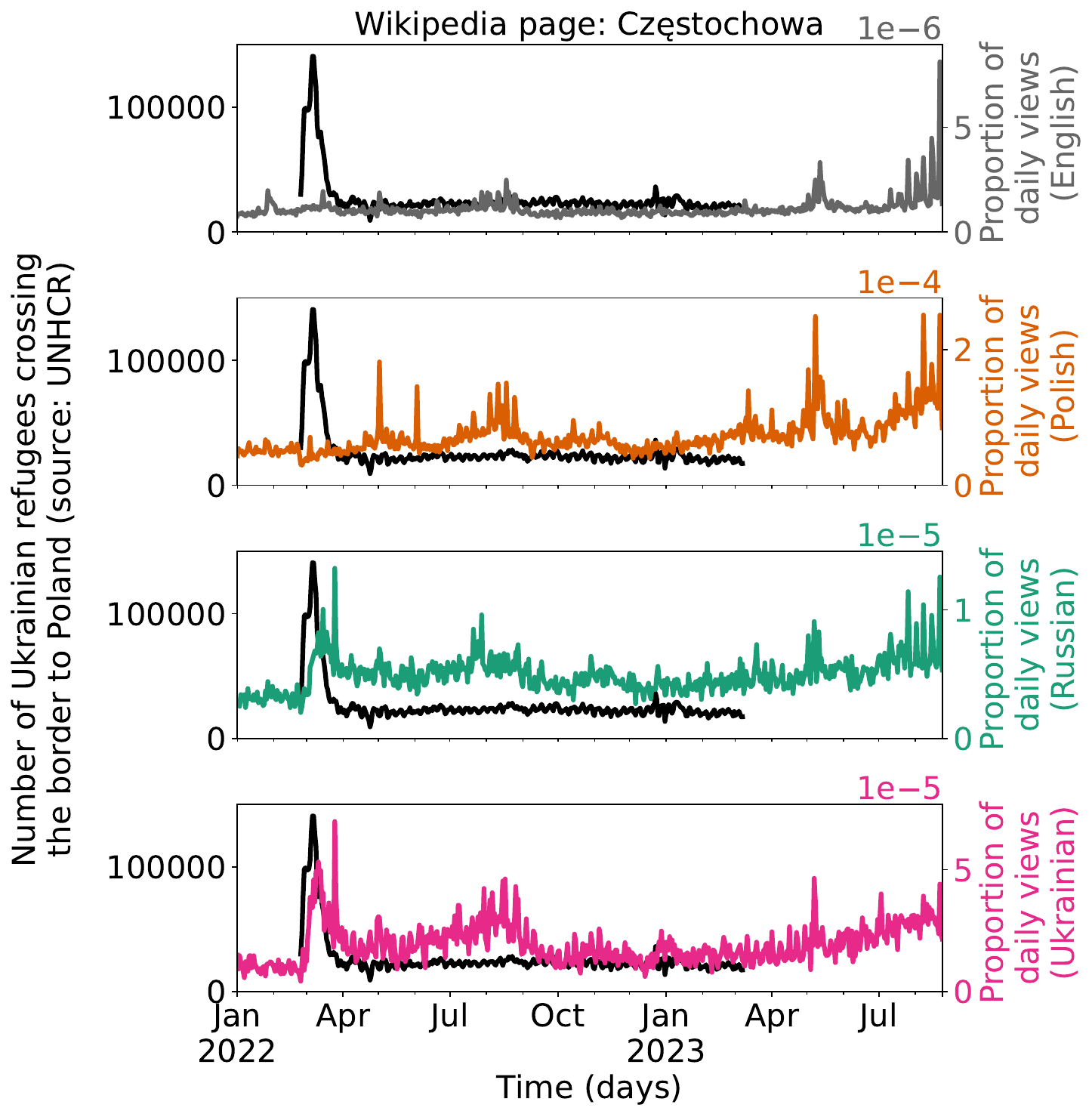}
        \caption{\textit{Częstochowa}}
    \end{subfigure}
    \begin{subfigure}[h]{0.19\textwidth}
     \centering
     \includegraphics[width=\linewidth]{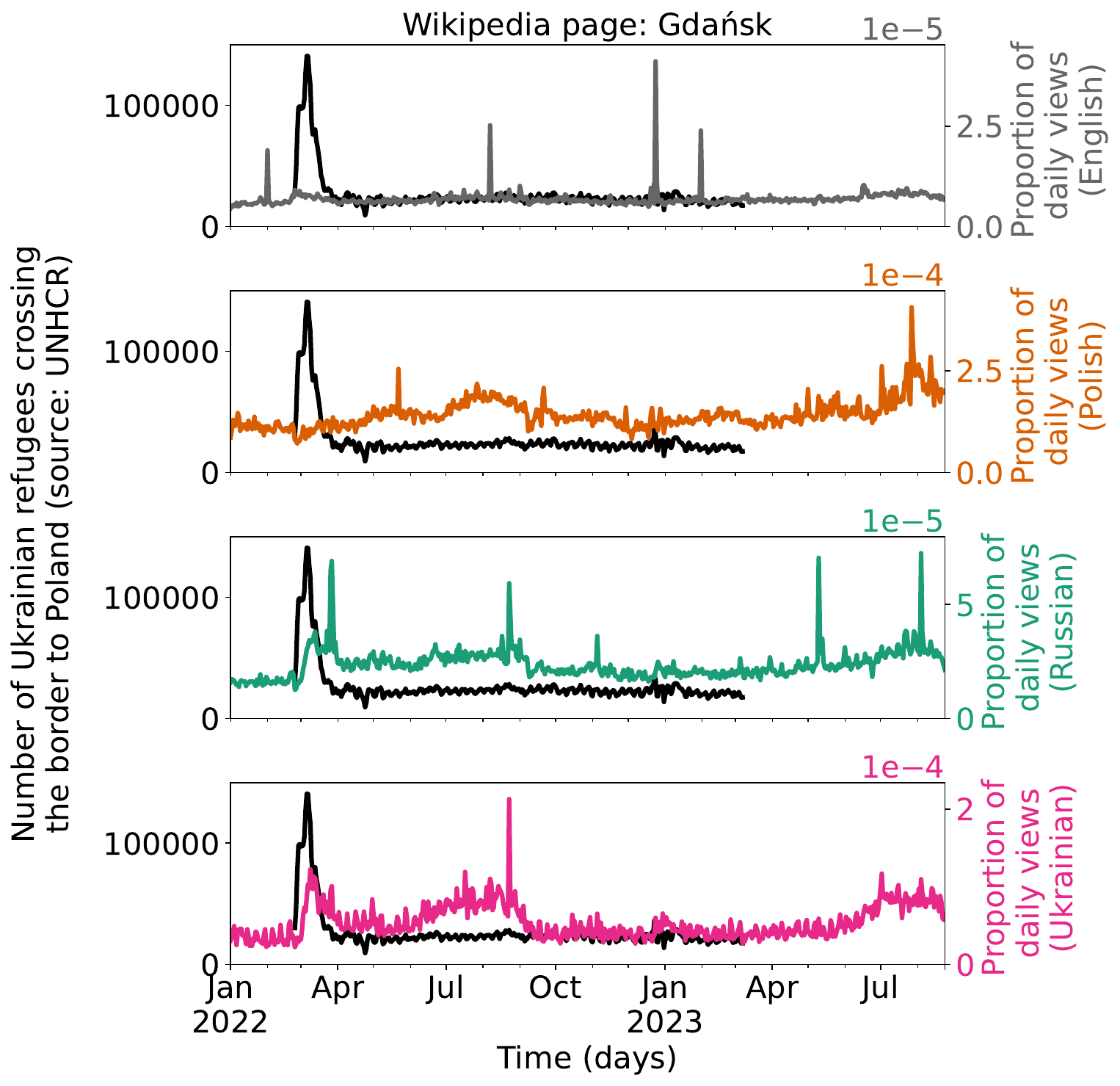}
     \caption{\textit{Gdańsk}}
    \end{subfigure}
    \begin{subfigure}[h]{0.19\textwidth}
        \centering
        \includegraphics[width=\linewidth]{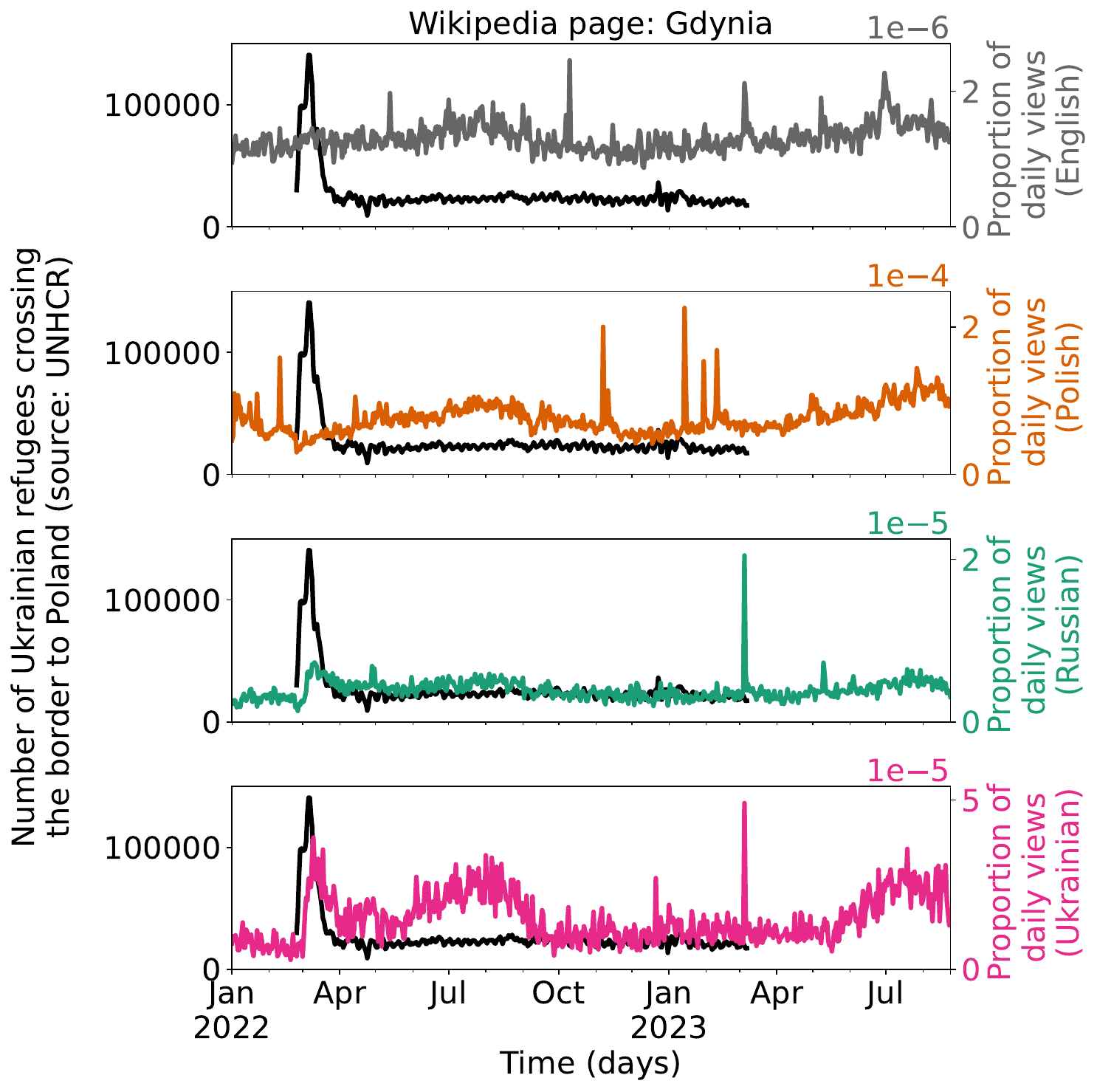}
        \caption{\textit{Gdynia}}
    \end{subfigure}
    \begin{subfigure}[h]{0.19\textwidth}
        \centering
        \includegraphics[width=\linewidth]{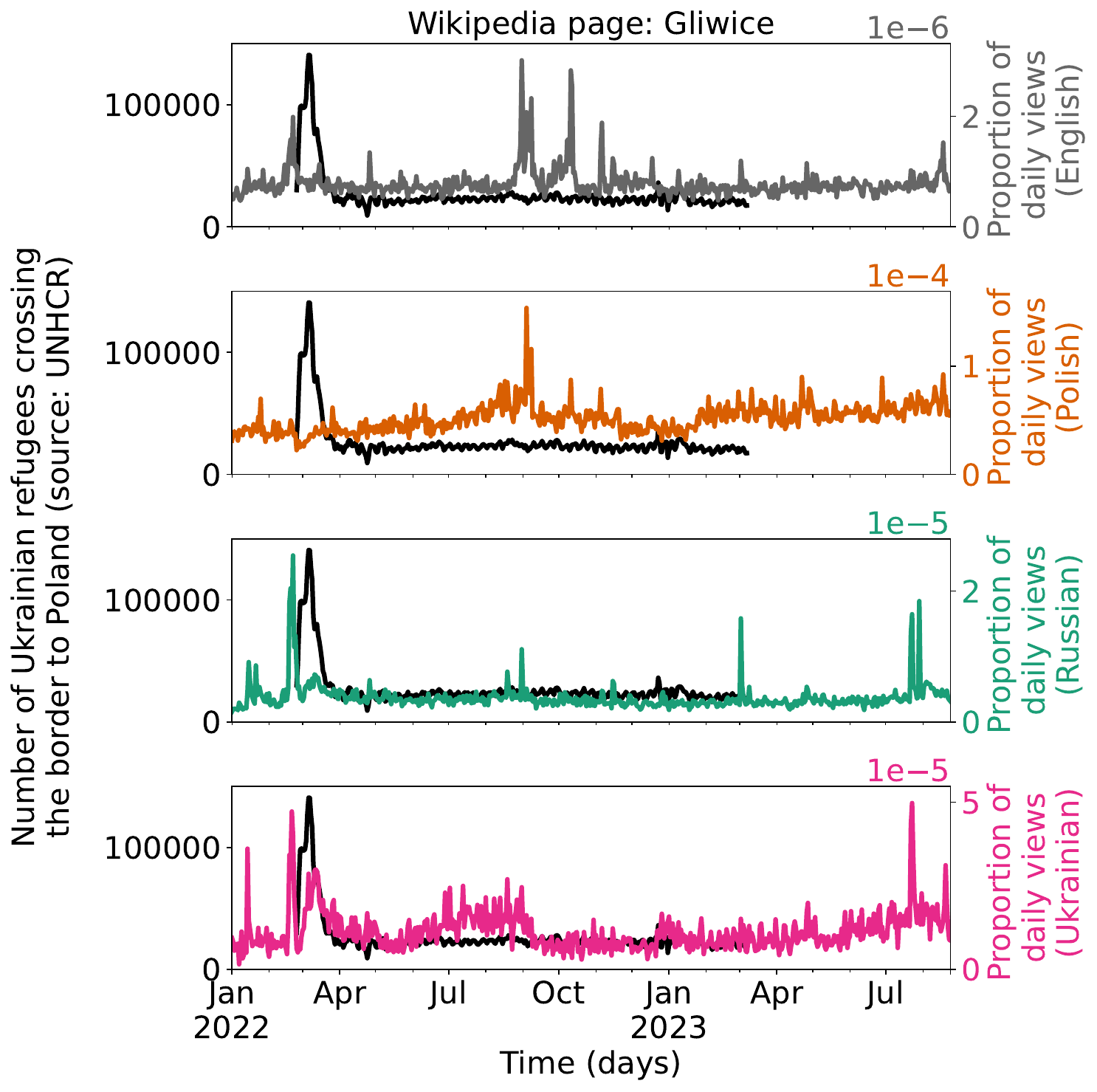}
        \caption{\textit{Gliwice}}
    \end{subfigure}
    \begin{subfigure}[h]{0.19\textwidth}
        \centering
        \includegraphics[width=\linewidth]{figs/plot-Katowice.pdf}
        \caption{\textit{Katowice}}
    \end{subfigure}
    \begin{subfigure}[h]{0.19\textwidth}
        \centering
        \includegraphics[width=\linewidth]{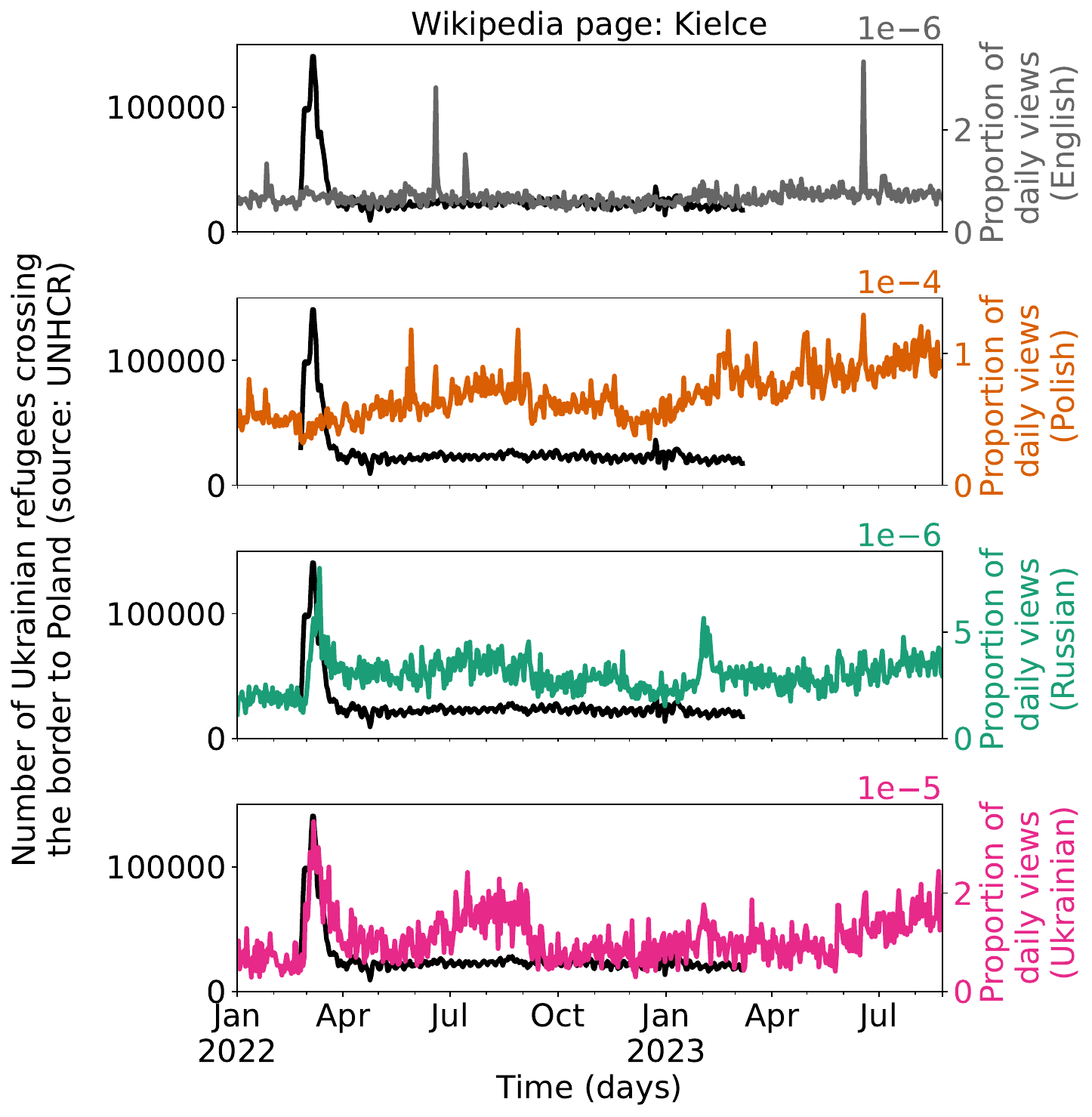}
        \caption{\textit{Kielce}}
    \end{subfigure}
    \begin{subfigure}[h]{0.19\textwidth}
     \centering
     \includegraphics[width=\linewidth]{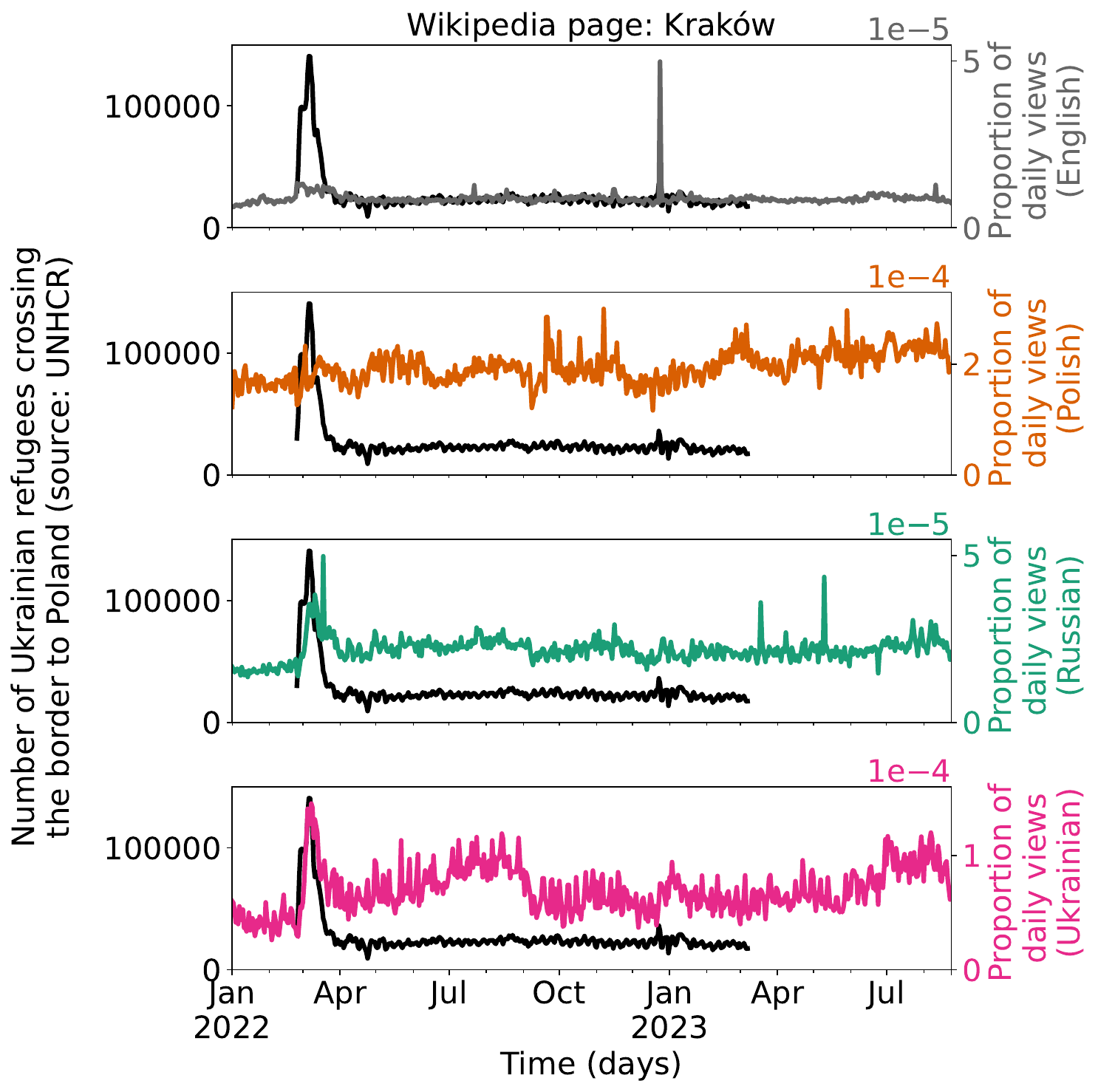}
     \caption{\textit{Kraków}}
    \end{subfigure}
    \begin{subfigure}[h]{0.19\textwidth}
        \centering
        \includegraphics[width=\linewidth]{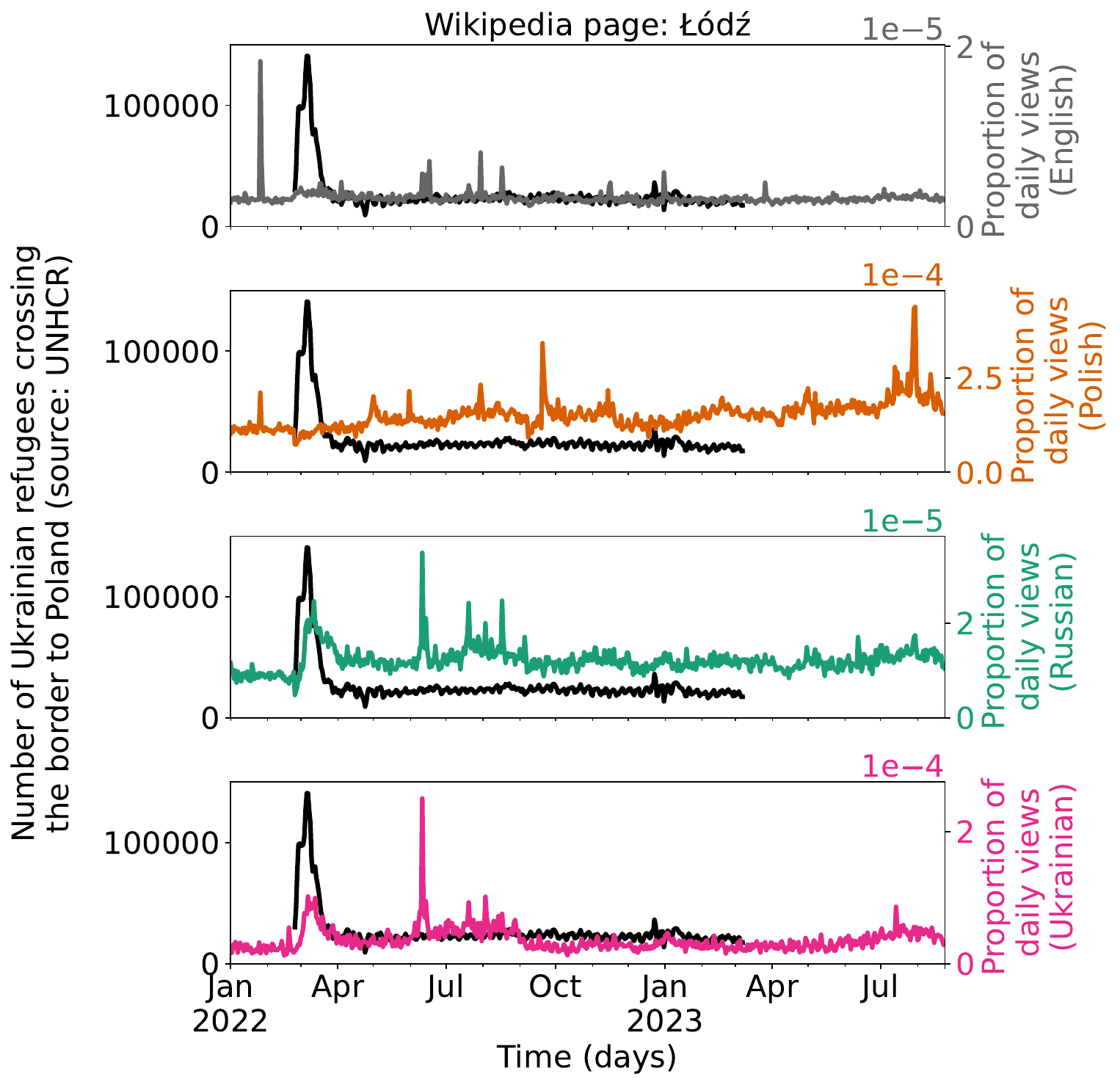}
        \caption{\textit{Łódź}}
    \end{subfigure}
    \begin{subfigure}[h]{0.19\textwidth}
        \centering
        \includegraphics[width=\linewidth]{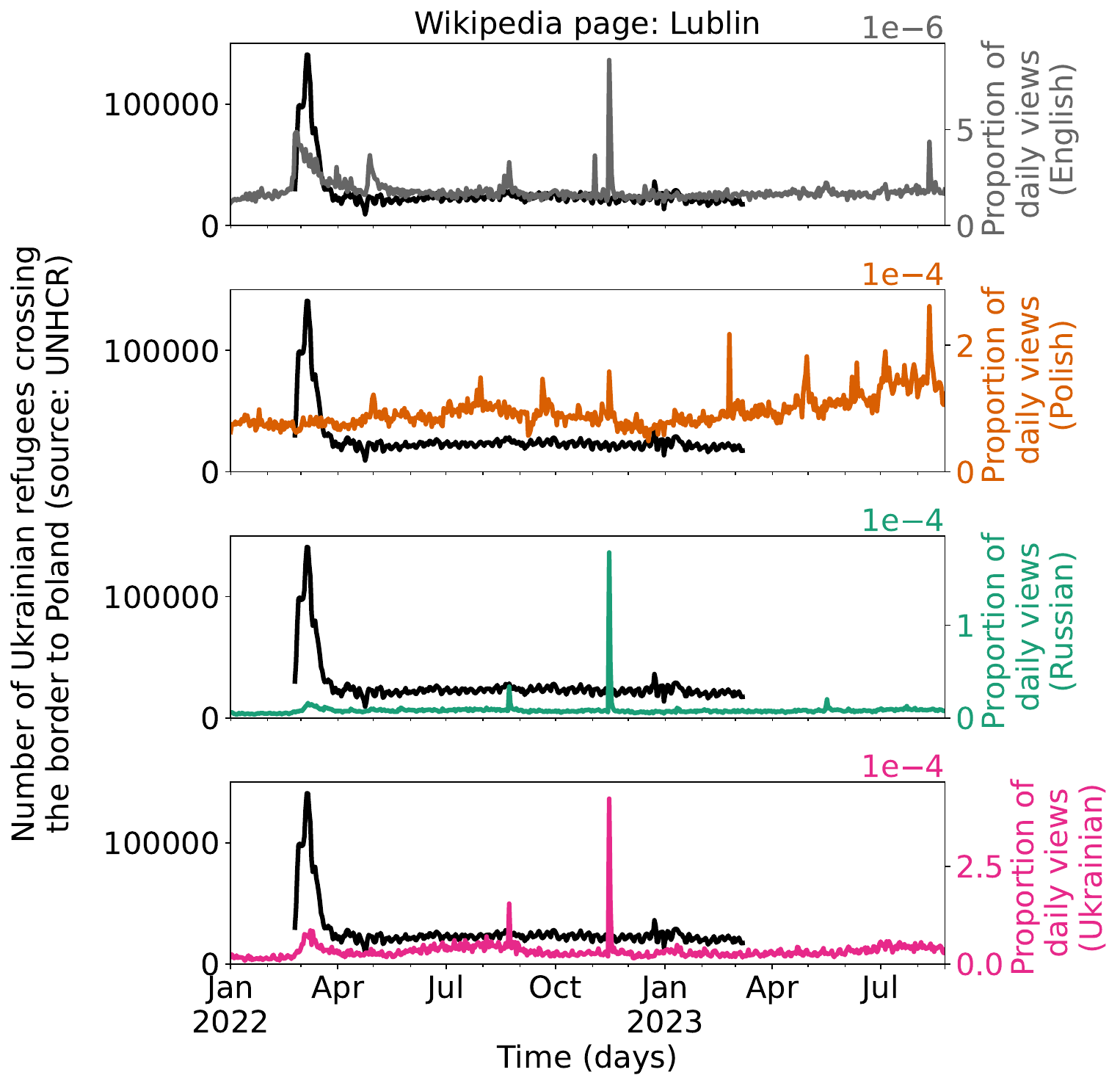}
        \caption{\textit{Lublin}}
    \end{subfigure}
    \begin{subfigure}[h]{0.19\textwidth}
        \centering
        \includegraphics[width=\linewidth]{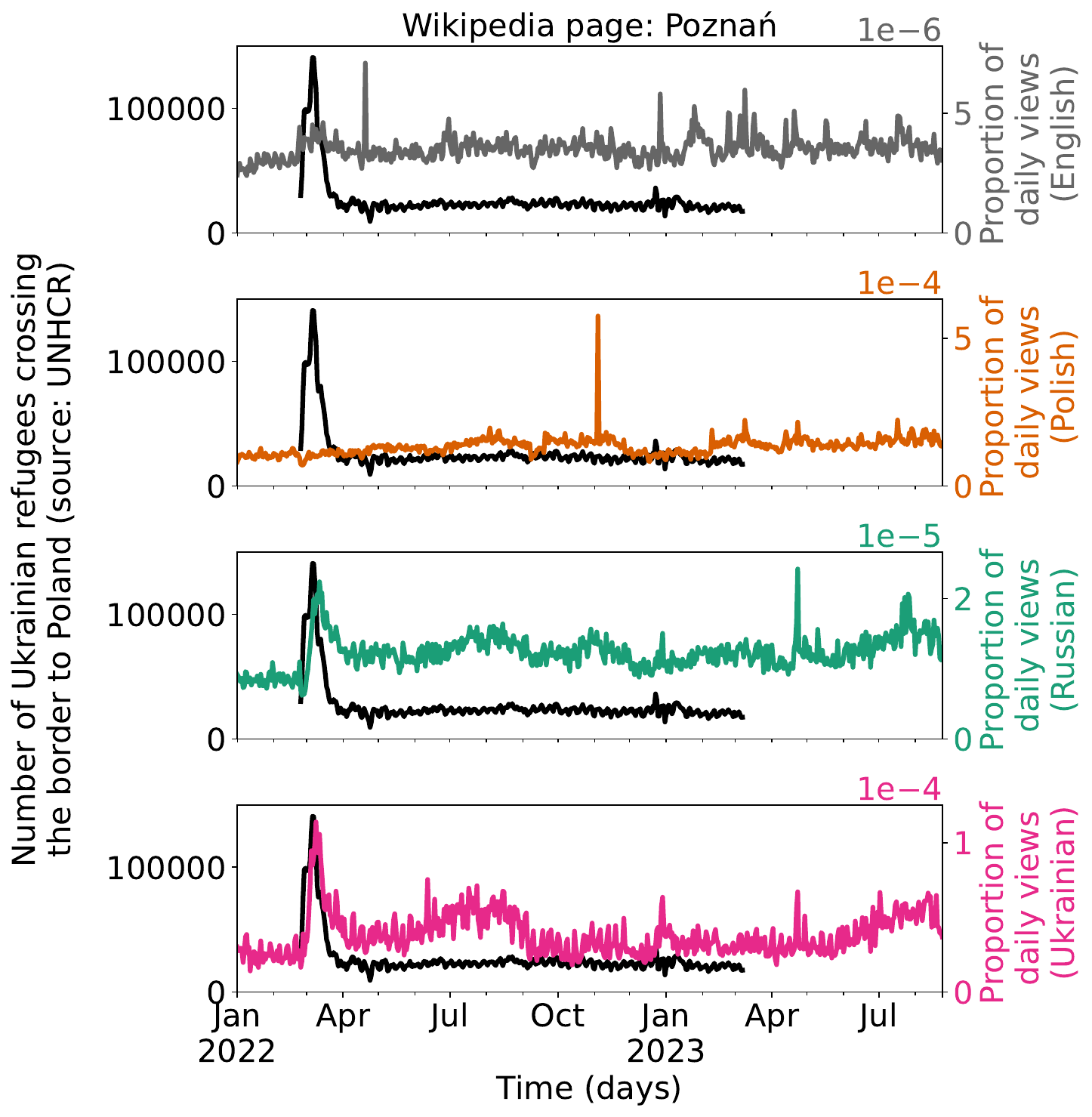}
        \caption{\textit{Poznań}}
    \end{subfigure}
    \begin{subfigure}[h]{0.19\textwidth}
        \centering
        \includegraphics[width=\linewidth]{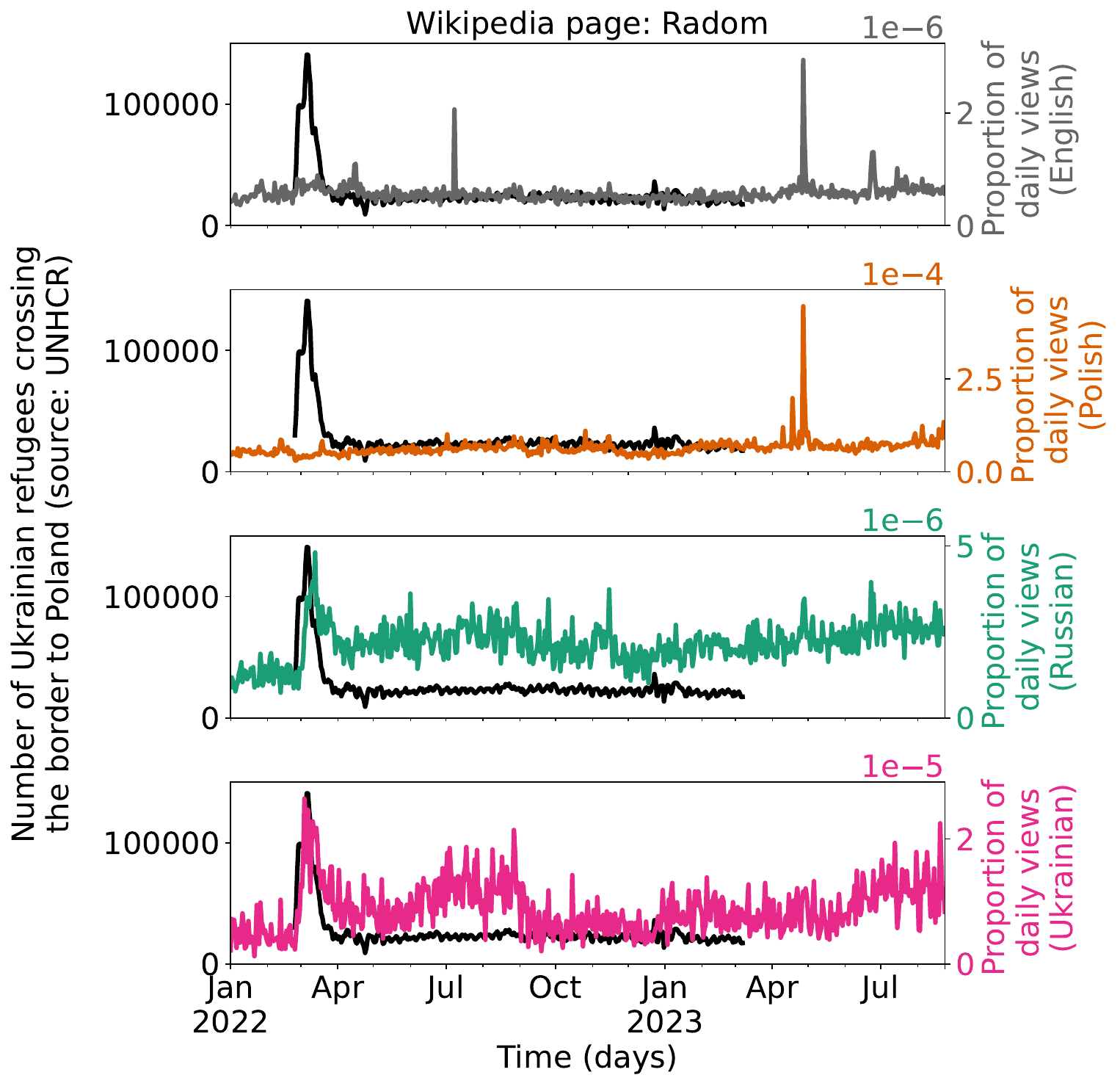}
        \caption{\textit{Radom}}
    \end{subfigure}
    \begin{subfigure}[h]{0.19\textwidth}
        \centering
        \includegraphics[width=\linewidth]{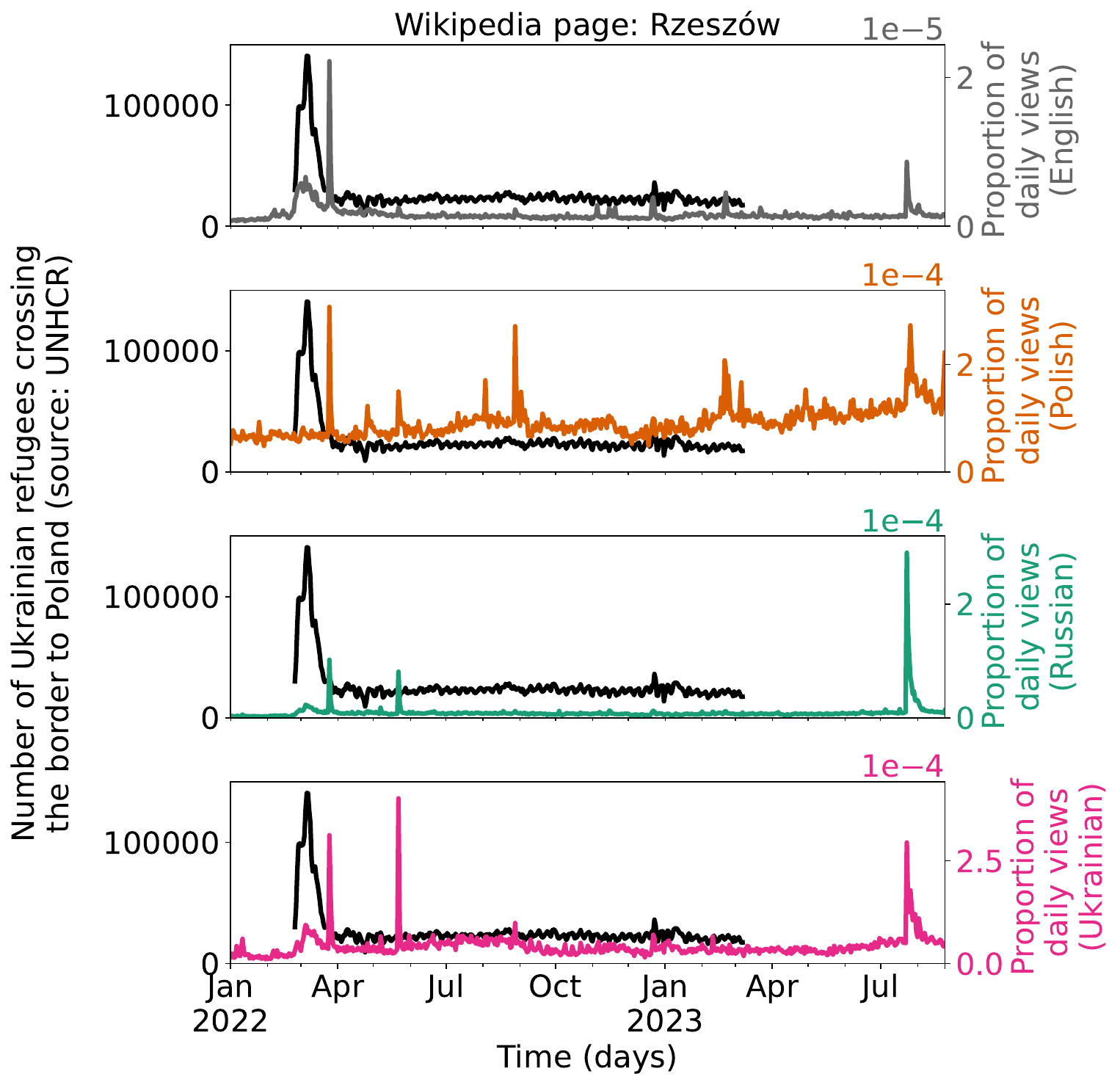}
        \caption{\textit{Rzeszów}}
    \end{subfigure}
    \begin{subfigure}[h]{0.19\textwidth}
        \centering
        \includegraphics[width=\linewidth]{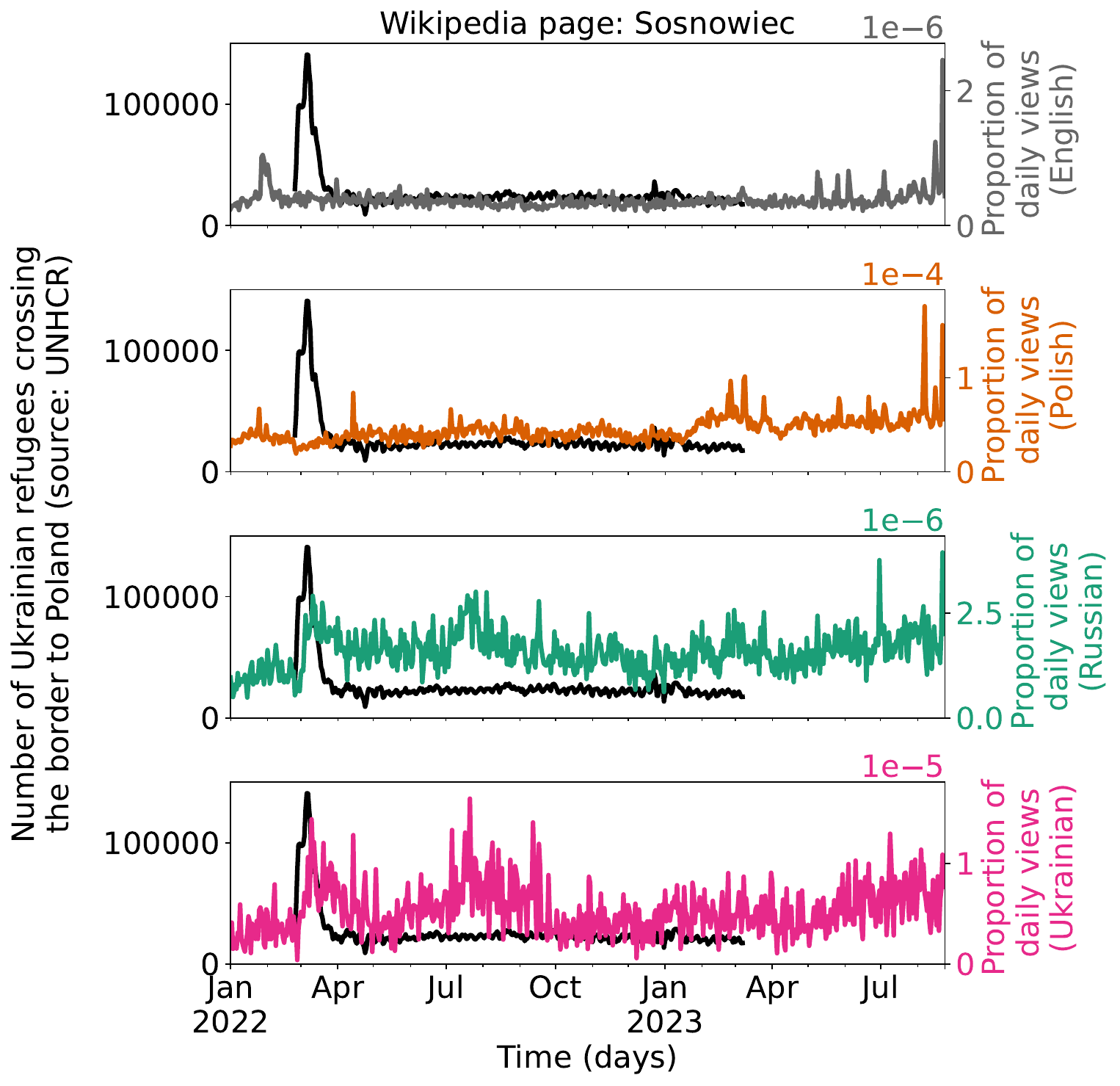}
        \caption{\textit{Sosnowiec}}
    \end{subfigure}
    \begin{subfigure}[h]{0.19\textwidth}
        \centering
        \includegraphics[width=\linewidth]{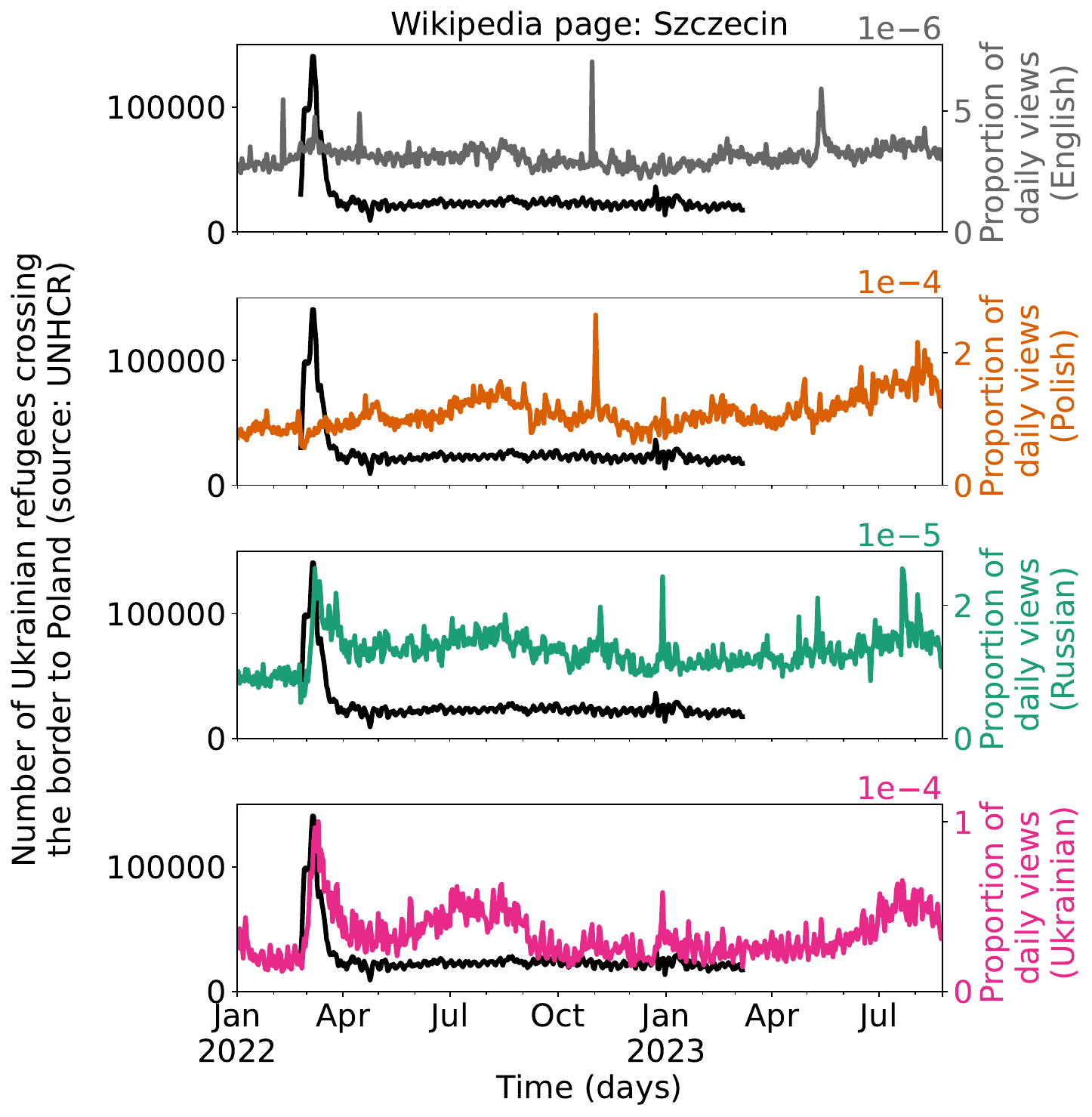}
        \caption{\textit{Szczecin}}
    \end{subfigure}
    \begin{subfigure}[h]{0.19\textwidth}
        \centering
        \includegraphics[width=\linewidth]{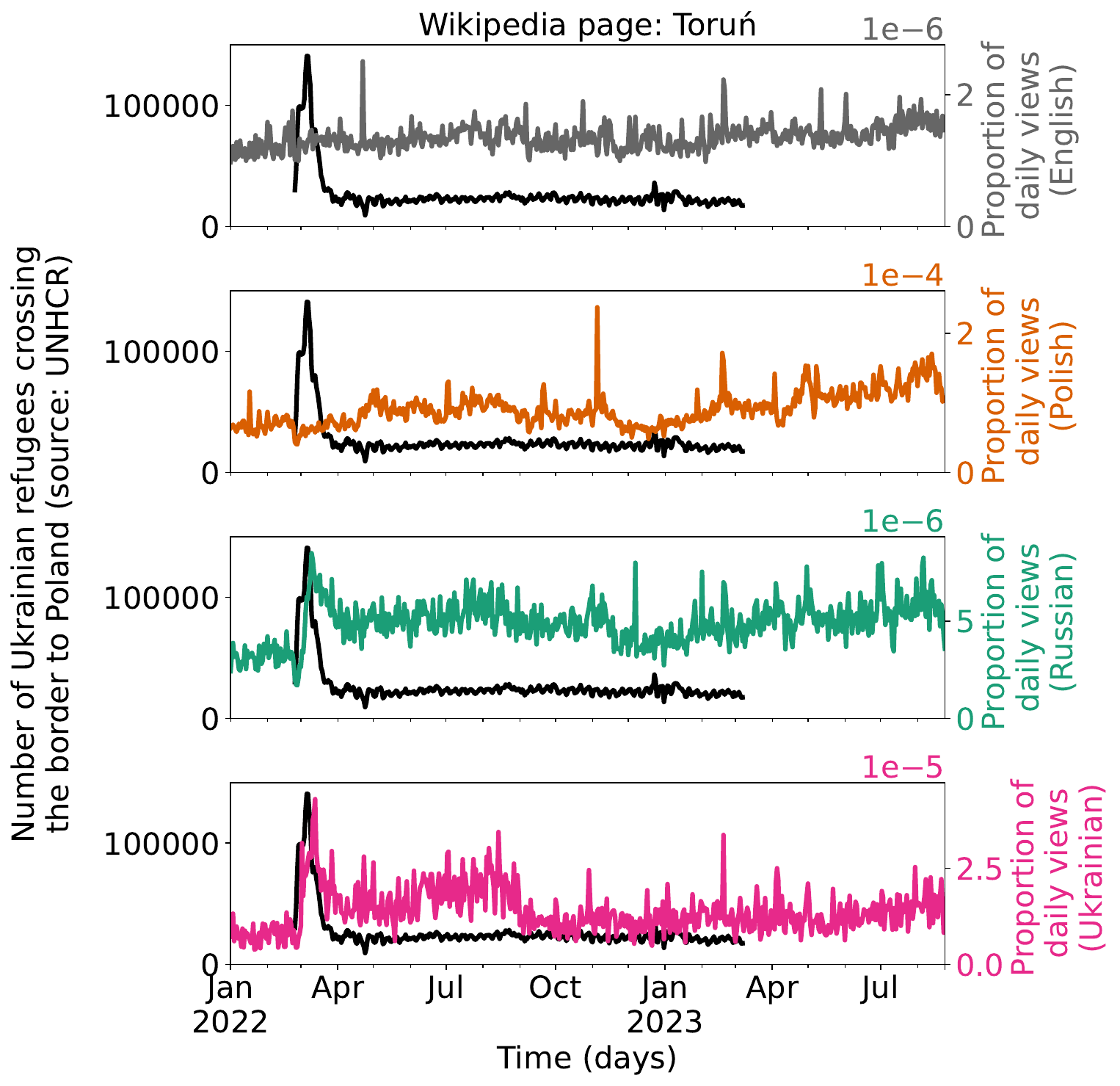}
        \caption{\textit{Toruń}}
    \end{subfigure}
    \begin{subfigure}[h]{0.19\textwidth}
        \centering
        \includegraphics[width=\linewidth]{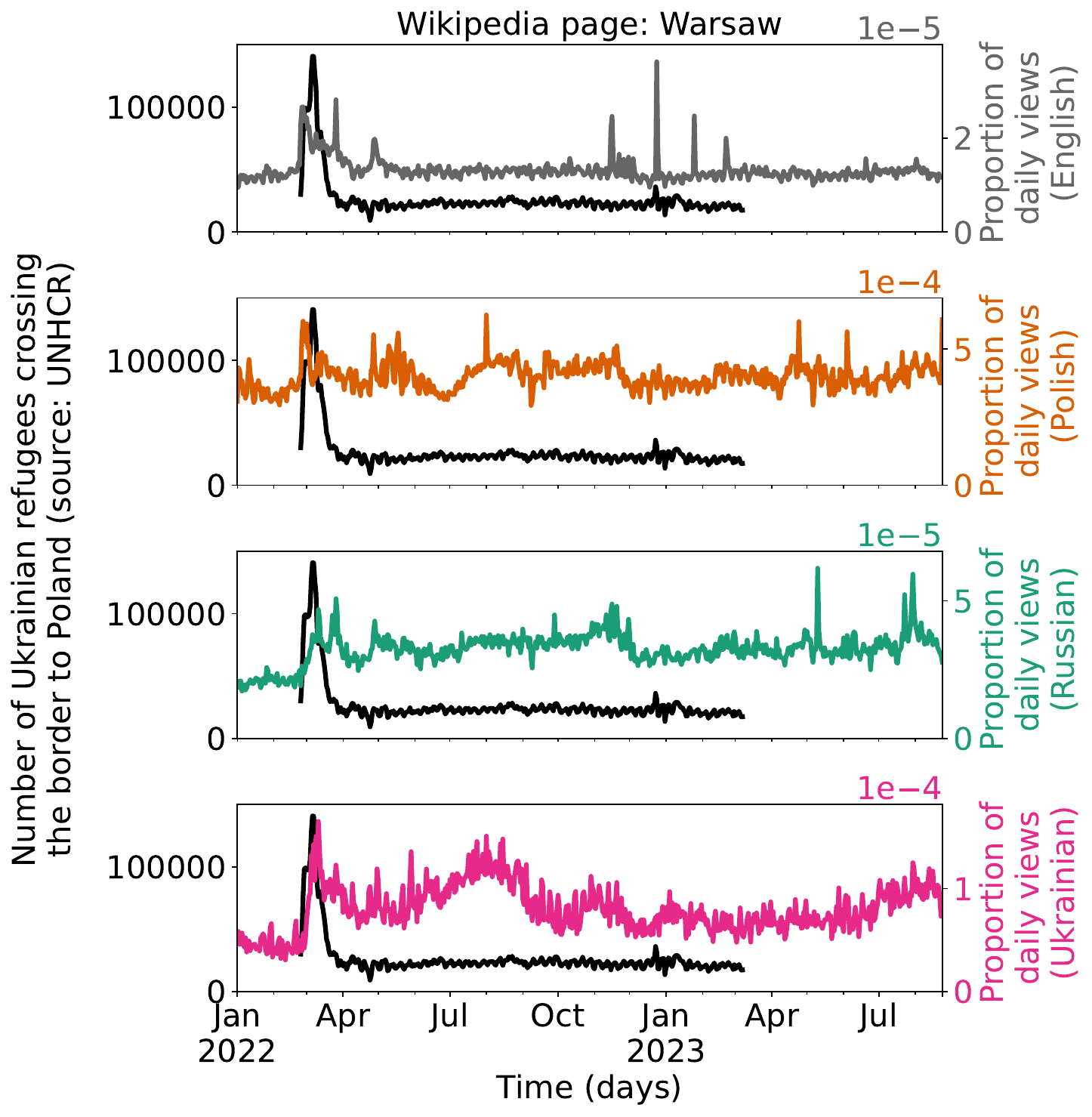}
        \caption{\textit{Warsaw}}
    \end{subfigure}
    \begin{subfigure}[h]{0.19\textwidth}
        \centering
        \includegraphics[width=\linewidth]{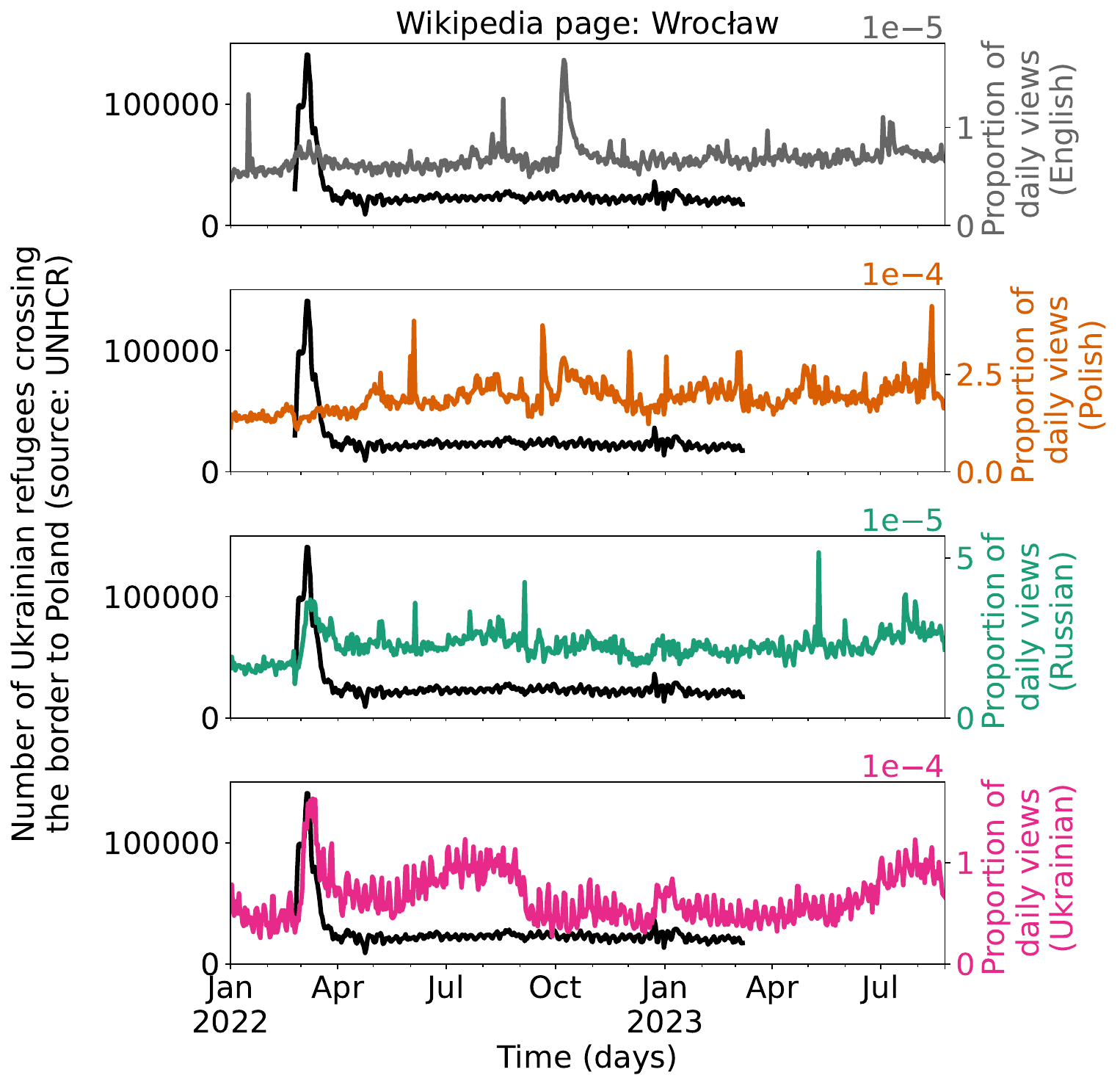}
        \caption{\textit{Wrocław}}
    \end{subfigure}
    \caption{Time series of the daily number of Ukrainian refugees crossing the border from Ukraine to Poland (from February 24, 2022 to March 3, 2023) and the proportion of daily views of Wikipedia articles about the 19 most populous Polish cities in four languages (English, Polish, Russian, and Ukrainian).}
    \label{fig:series-appendix}
\end{figure*}

\newpage
\begin{table*}[ht!]
    \caption*{\textbf{Granger causality}}
    \footnotesize
    \begin{tabular}{llrl}
    \toprule
    \textbf{City} & \textbf{Relationship} & \textbf{Optimal lag} & \textbf{F-statistic (p-value)} \\
    \midrule
    \multirow[t]{3}{*}{Białystok} & Ukrainian refugees in Poland (UNHCR) $\rightarrow$ Wikipedia views in English & 8 & 1.98 (p = 0.0484) \\
     & Ukrainian refugees in Poland (UNHCR) $\rightarrow$ Wikipedia views in Ukrainian & 8 & 8.45 (p = 0.0000) \\
     & Wikipedia views in English $\rightarrow$ Ukrainian refugees in Poland (UNHCR) & 8 & 2.83 (p = 0.0046) \\
    \cline{1-4}
    \multirow[t]{3}{*}{Bydgoszcz} & Ukrainian refugees in Poland (UNHCR) $\rightarrow$ Wikipedia views in English & 8 & 2.00 (p = 0.0462) \\
     & Ukrainian refugees in Poland (UNHCR) $\rightarrow$ Wikipedia views in Russian & 8 & 10.31 (p = 0.0000) \\
     & Ukrainian refugees in Poland (UNHCR) $\rightarrow$ Wikipedia views in Ukrainian & 8 & 10.37 (p = 0.0000) \\
    \cline{1-4}
    \multirow[t]{2}{*}{Częstochowa} & Ukrainian refugees in Poland (UNHCR) $\rightarrow$ Wikipedia views in Russian & 8 & 5.79 (p = 0.0000) \\
     & Ukrainian refugees in Poland (UNHCR) $\rightarrow$ Wikipedia views in Ukrainian & 8 & 5.49 (p = 0.0000) \\
    \cline{1-4}
    \multirow[t]{6}{*}{Gdańsk} & Ukrainian refugees in Poland (UNHCR) $\rightarrow$ Wikipedia views in English & 8 & 2.03 (p = 0.0425) \\
     & Ukrainian refugees in Poland (UNHCR) $\rightarrow$ Wikipedia views in Russian & 23 & 3.97 (p = 0.0000) \\
     & Ukrainian refugees in Poland (UNHCR) $\rightarrow$ Wikipedia views in Ukrainian & 8 & 9.11 (p = 0.0000) \\
     & Wikipedia views in English $\rightarrow$ Ukrainian refugees in Poland (UNHCR) & 8 & 6.06 (p = 0.0000) \\
     & Wikipedia views in Russian $\rightarrow$ Ukrainian refugees in Poland (UNHCR) & 23 & 1.66 (p = 0.0317) \\
     & Wikipedia views in Ukrainian $\rightarrow$ Ukrainian refugees in Poland (UNHCR) & 8 & 2.48 (p = 0.0125) \\
    \cline{1-4}
    \multirow[t]{2}{*}{Gdynia} & Ukrainian refugees in Poland (UNHCR) $\rightarrow$ Wikipedia views in Russian & 8 & 3.09 (p = 0.0022) \\
     & Ukrainian refugees in Poland (UNHCR) $\rightarrow$ Wikipedia views in Ukrainian & 8 & 6.51 (p = 0.0000) \\
    \cline{1-4}
    \multirow[t]{2}{*}{Gliwice} & Ukrainian refugees in Poland (UNHCR) $\rightarrow$ Wikipedia views in Russian & 8 & 2.73 (p = 0.0062) \\
     & Ukrainian refugees in Poland (UNHCR) $\rightarrow$ Wikipedia views in Ukrainian & 8 & 4.32 (p = 0.0001) \\
    \cline{1-4}
    \multirow[t]{3}{*}{Katowice} & Ukrainian refugees in Poland (UNHCR) $\rightarrow$ Wikipedia views in Russian & 8 & 5.99 (p = 0.0000) \\
     & Ukrainian refugees in Poland (UNHCR) $\rightarrow$ Wikipedia views in Ukrainian & 8 & 8.73 (p = 0.0000) \\
     & Wikipedia views in English $\rightarrow$ Ukrainian refugees in Poland (UNHCR) & 8 & 2.56 (p = 0.0101) \\
    \cline{1-4}
    \multirow[t]{2}{*}{Kielce} & Ukrainian refugees in Poland (UNHCR) $\rightarrow$ Wikipedia views in Russian & 8 & 4.19 (p = 0.0001) \\
     & Ukrainian refugees in Poland (UNHCR) $\rightarrow$ Wikipedia views in Ukrainian & 8 & 4.80 (p = 0.0000) \\
    \cline{1-4}
    \multirow[t]{4}{*}{Kraków} & Ukrainian refugees in Poland (UNHCR) $\rightarrow$ Wikipedia views in English & 9 & 4.43 (p = 0.0000) \\
     & Ukrainian refugees in Poland (UNHCR) $\rightarrow$ Wikipedia views in Russian & 8 & 8.01 (p = 0.0000) \\
     & Ukrainian refugees in Poland (UNHCR) $\rightarrow$ Wikipedia views in Ukrainian & 8 & 8.50 (p = 0.0000) \\
     & Wikipedia views in English $\rightarrow$ Ukrainian refugees in Poland (UNHCR) & 9 & 9.67 (p = 0.0000) \\
    \cline{1-4}
    \multirow[t]{3}{*}{Poznań} & Ukrainian refugees in Poland (UNHCR) $\rightarrow$ Wikipedia views in Russian & 8 & 10.24 (p = 0.0000) \\
     & Ukrainian refugees in Poland (UNHCR) $\rightarrow$ Wikipedia views in Ukrainian & 8 & 10.93 (p = 0.0000) \\
     & Wikipedia views in English $\rightarrow$ Ukrainian refugees in Poland (UNHCR) & 8 & 2.65 (p = 0.0078) \\
    \cline{1-4}
    \multirow[t]{3}{*}{Radom} & Ukrainian refugees in Poland (UNHCR) $\rightarrow$ Wikipedia views in Russian & 8 & 5.67 (p = 0.0000) \\
     & Ukrainian refugees in Poland (UNHCR) $\rightarrow$ Wikipedia views in Ukrainian & 15 & 2.48 (p = 0.0018) \\
     & Wikipedia views in Ukrainian $\rightarrow$ Ukrainian refugees in Poland (UNHCR) & 15 & 1.98 (p = 0.0160) \\
    \cline{1-4}
    \multirow[t]{2}{*}{Rzeszów} & Ukrainian refugees in Poland (UNHCR) $\rightarrow$ Wikipedia views in English & 8 & 3.00 (p = 0.0029) \\
     & Ukrainian refugees in Poland (UNHCR) $\rightarrow$ Wikipedia views in Russian & 8 & 2.44 (p = 0.0139) \\
    \cline{1-4}
    \multirow[t]{2}{*}{Sosnowiec} & Ukrainian refugees in Poland (UNHCR) $\rightarrow$ Wikipedia views in Russian & 8 & 3.53 (p = 0.0006) \\
     & Ukrainian refugees in Poland (UNHCR) $\rightarrow$ Wikipedia views in Ukrainian & 8 & 2.86 (p = 0.0043) \\
    \cline{1-4}
    \multirow[t]{3}{*}{Szczecin} & Ukrainian refugees in Poland (UNHCR) $\rightarrow$ Wikipedia views in Russian & 10 & 6.70 (p = 0.0000) \\
     & Ukrainian refugees in Poland (UNHCR) $\rightarrow$ Wikipedia views in Ukrainian & 8 & 10.82 (p = 0.0000) \\
     & Wikipedia views in Russian $\rightarrow$ Ukrainian refugees in Poland (UNHCR) & 10 & 5.32 (p = 0.0000) \\
    \cline{1-4}
    \multirow[t]{3}{*}{Toruń} & Ukrainian refugees in Poland (UNHCR) $\rightarrow$ Wikipedia views in Russian & 8 & 6.21 (p = 0.0000) \\
     & Ukrainian refugees in Poland (UNHCR) $\rightarrow$ Wikipedia views in Ukrainian & 8 & 5.25 (p = 0.0000) \\
     & Wikipedia views in Ukrainian $\rightarrow$ Ukrainian refugees in Poland (UNHCR) & 8 & 2.90 (p = 0.0039) \\
    \cline{1-4}
    \multirow[t]{6}{*}{Warsaw} & Ukrainian refugees in Poland (UNHCR) $\rightarrow$ Wikipedia views in English & 17 & 2.94 (p = 0.0001) \\
     & Ukrainian refugees in Poland (UNHCR) $\rightarrow$ Wikipedia views in Polish & 8 & 2.54 (p = 0.0107) \\
     & Ukrainian refugees in Poland (UNHCR) $\rightarrow$ Wikipedia views in Russian & 23 & 2.06 (p = 0.0035) \\
     & Ukrainian refugees in Poland (UNHCR) $\rightarrow$ Wikipedia views in Ukrainian & 8 & 6.60 (p = 0.0000) \\
     & Wikipedia views in English $\rightarrow$ Ukrainian refugees in Poland (UNHCR) & 17 & 3.81 (p = 0.0000) \\
     & Wikipedia views in Russian $\rightarrow$ Ukrainian refugees in Poland (UNHCR) & 23 & 1.59 (p = 0.0452) \\
    \cline{1-4}
    \multirow[t]{3}{*}{Wrocław} & Ukrainian refugees in Poland (UNHCR) $\rightarrow$ Wikipedia views in Russian & 8 & 5.94 (p = 0.0000) \\
     & Ukrainian refugees in Poland (UNHCR) $\rightarrow$ Wikipedia views in Ukrainian & 17 & 2.83 (p = 0.0002) \\
     & Wikipedia views in Ukrainian $\rightarrow$ Ukrainian refugees in Poland (UNHCR) & 17 & 3.00 (p = 0.0001) \\
    \cline{1-4}
    \multirow[t]{2}{*}{Łódź} & Ukrainian refugees in Poland (UNHCR) $\rightarrow$ Wikipedia views in Russian & 8 & 4.51 (p = 0.0000) \\
     & Ukrainian refugees in Poland (UNHCR) $\rightarrow$ Wikipedia views in Ukrainian & 8 & 2.71 (p = 0.0065) \\
    \cline{1-4}
    \bottomrule
    \end{tabular}
    \caption{Granger causality relationships between Ukrainian refugee flows crossing the border into Poland and the proportion of daily views of Wikipedia articles about the 19 most populous Polish cities in Ukrainian. For each relationship, the table reports the optimal lag length (in days) selected by the model, the associated F-statistic, and the p-value. Only statistically significant relationships (p-value < 0.05) are included.}
    \label{tab:granger}
\end{table*}

\newpage
\onecolumn
\subsection{Comparison between Wikipedia and Google Trends data}
\label{sec:google-appendix}

The growing literature on online information-seeking behavior and migration often relies on the data from online search engines such as Google~\cite{bohme2020searching, sanliturk2024search} and Yandex~\cite{anastasiadou2024war}. Data provided by the Google Trends tool by Google are often preferred due to the widespread use of Google worldwide. While the Google Trends data are proven to be a useful indicator of interest in moving and the possible intention to move, the characteristics of the data pose limitations for advanced statistical analyses. Google does not disclose information on the volume of online searches, but instead produces an index with a range of 0-100 normalized for online search popularity for the given query (keyword), location, and time period. Furthermore, this index is not based on the entire search data for the given parameters, but on a sample large enough to represent the needed search data, yet with a sample size that is unknown to the users. If the query cannot pass a threshold level of interest, which is determined by Google and is unknown to the users, the level of interest is deemed negligible and is reported as zero. Google Trends reports the daily search popularity index for a time period shorter than nine months. For longer periods, researchers may stitch together multiple datasets of nine months and rescale. While the Google Trends Index has proven useful and consistent despite these limitations, it must be acknowledged that the index is representative of the required online search data, but ``might not be a perfect mirror of search activity.''\footnote{\url{https://support.google.com/trends/answer/4365533?hl=en} Accessed in Mar. 2024}

In this study, we introduced the use of Wikipedia data as an indicator of migration and mobility. Wikipedia data have certain advantages over Google Trends data, the most important of which is that Wikipedia provides the absolute number of pageviews instead of a normalized index, which is more suitable for statistical analyses. In Google Trends, distinguishing between two different groups of people, such as host and migrant groups, at the same location is possible if the query words are in different languages or alphabets. Distinguishing by language creates issues when using city or province names as query words because they mostly remain the same across different languages and may leave the alphabetical difference as the only differentiation method. In contrast, Wikipedia pages are available in different languages, which may make it easier to differentiate between two groups. However, Wikipedia provides information only on the language of the viewed page, and not the location of the view. It must be underlined that differentiation by language would also be problematic for languages that are common second languages and/or native languages of multiple countries, such as English, Spanish, French, and Arabic. However, in the context of our case study, pageviews in the Polish and Ukrainian languages may be more easily attributed to the respective countries and their people.

In order to observe and demonstrate the potential advantages of Wikipedia data with respect to Google Trends data, we collected Google Trends data\footnote{Google Trends data were collected using the \texttt{gtrendsR} package in R and \texttt{pytrends} in Python. Both R and Python were used to accelerate the data collection process.} matching the Wikipedia data in our study for a descriptive analysis. Figure \ref{fig:GTI-Wiki_Ukrainian} shows a comparison between the Wikipedia data used in our study and the matching Google Trends data for the 19 most populous Polish cities. We distinguish the online searches for Polish cities made by Ukrainians by setting the location as Ukraine. We take the relevant city as a topic (city) instead of the name of the city as a keyword, because in many cases query by keyword resulted in zero-inflated data or no data at all due to low interest. To enable an easier visual comparison, we normalized the Wikipedia views to the same range as the Google Trends data, i.e., 0-100. We then compare and contrast the changes in searches for information about Polish cities following the Russian invasion of Ukraine. 

Looking at the descriptive analysis of these two data sources, we highlight two main points. First, even using the name of the city as the topic and not as the strict keyword, we can see that Google Trends data reports many zero values and noise in the data. This creates an advantage for Wikipedia data over Google Trends data, which can be observed in the cases of Białystok, Bydgoszcz, Częstochowa, Gdynia, Gliwice, Kielce, Radom, Sosnowiec, and Toruń (Figure \ref{fig:GTI-Wiki_Ukrainian}). Relative to the other cities, for which Google Trends provides good quality data, these cities are less well-known and less populated. Thus, in the case of Wikipedia, the advantage in terms of data quality is more pronounced for smaller cities. Second, for more populated big cities, for which we have better quality (less noisy) Google Trends data, we do not observe important divergences between the patterns of Google Trends data and Wikipedia data. Especially following the start of the Russian invasion of Ukraine, both sources show overlapping increases in interest that are similar in size when good quality data are available from both sources.


\begin{figure*}[ht]
    \centering
    \includegraphics[width=\linewidth]{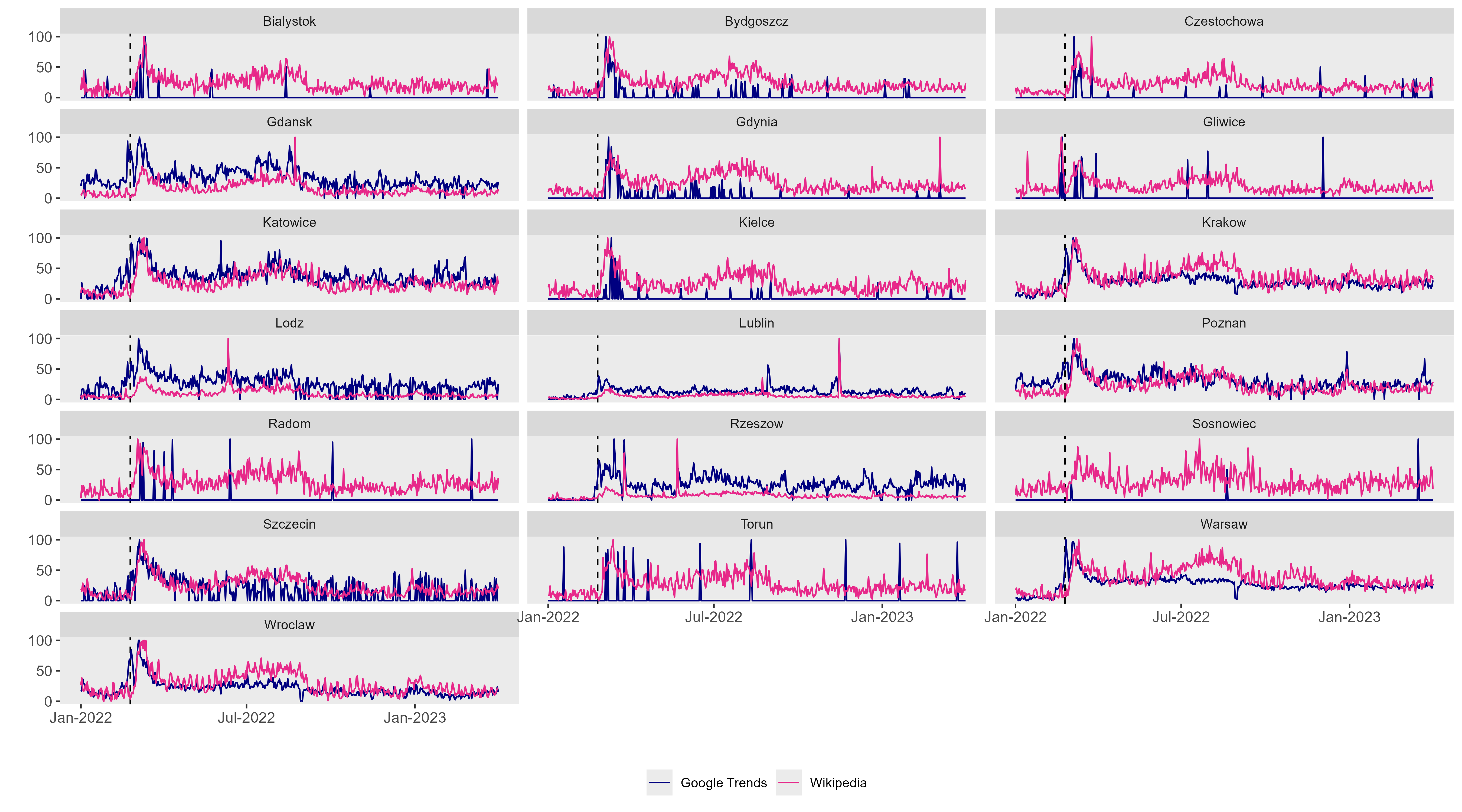}
    \caption{Comparison between the Google Trends Index (GTI) of daily Google searches in Ukraine for Polish cities (as a topic) and the proportion of daily views of the corresponding Wikipedia articles about the 19 most populous Polish cities in Ukrainian. For comparability, Wikipedia views are normalized to the 0–100 range. GTI values are shown in dark blue, and Wikipedia views are shown in pink. The time series cover the period from January 1, 2022, to April 2, 2023, and the vertical dashed line marks the beginning of the Russian invasion of Ukraine (February 24, 2022).}
    \label{fig:GTI-Wiki_Ukrainian}
\end{figure*}

\end{document}